%% file: main.tex
\pdfoutput=1

\documentclass[10pt,twocolumn,letterpaper]{article}

\usepackage[pagenumbers]{cvpr}

\input{preamble}

\usepackage[breaklinks,hidelinks]{hyperref}

\def\confName{CVPR}
\def\confYear{2027}

\title{OpenMAS-GApp: A Comprehensive Benchmark for Domain-Adaptive Multi-Agent Application Construction}

\input{authors}

\begin{document}
\maketitle
\input{sec/0_abstract}

\input{sec/1_intro}
\input{sec/2_problem_statement}
\input{sec/3_benchmark_design}

\input{sec/4_experiments}
\input{sec/5_conclusion}

\clearpage

\makeatletter
\global\@colroom\@colht
\global\vsize\@colht
\makeatother
\bibliographystyle{ieeenat_fullname}
\bibliography{main}

\input{sec/X_suppl}

\end{document}

%% file: preamble.tex
\usepackage{multirow}

\usepackage{flafter}

\input{math_commands.tex}

%% file: math_commands.tex
\usepackage{amsmath,amsfonts,bm}

\def\eqref#1{equation~\ref{#1}}

\def\1{\bm{1}}

\DeclareMathAlphabet{\mathsfit}{\encodingdefault}{\sfdefault}{m}{sl}
\SetMathAlphabet{\mathsfit}{bold}{\encodingdefault}{\sfdefault}{bx}{n}

%% file: authors.tex
\author{
\normalsize Xunkai Li$^{*}$\quad Kairui Yang$^{*}$\quad Jiaxi Li\quad Yuecong Mao\\
\normalsize Zekai Chen\quad Xinle Zhang\quad Rong-Hua Li\quad Guoren Wang\\[5pt]
\normalsize Beijing Institute of Technology\\[3pt]
\small $^{*}$Equal contribution.
}

%% file: sec/0_abstract.tex
\begin{abstract}
Constructing multi-agent system (MAS) applications across domains requires translating diverse requirements into executable systems, involving task decomposition, agent and tool selection, and execution organization. Existing studies have explored individual aspects of MAS construction, including agent generation, architecture design, workflow generation, and component selection. However, these methods are often developed and evaluated independently, using different task specifications, component resources, and execution settings. This makes it difficult to compare their construction capabilities or determine how they can be applied to different domains. To address this problem, we propose \textbf{OpenMAS-GApp}, a One-for-All benchmark for MAS application construction across diverse domains. It standardizes task inputs, component resources, construction interfaces, and execution environments to support MAS construction under consistent conditions. Within OpenMAS-GApp, we provide \textbf{Graph Harness} as a reference construction method that derives task dependencies, selects compatible agents and tools, and organizes their interactions into executable graphs. We evaluate construction methods across 19 benchmark datasets under 20 evaluation settings spanning four application types and four professional domains.
\end{abstract}

%% file: sec/1_intro.tex
\begin{figure}[!t]
    \centering
    \includegraphics[width=\columnwidth]{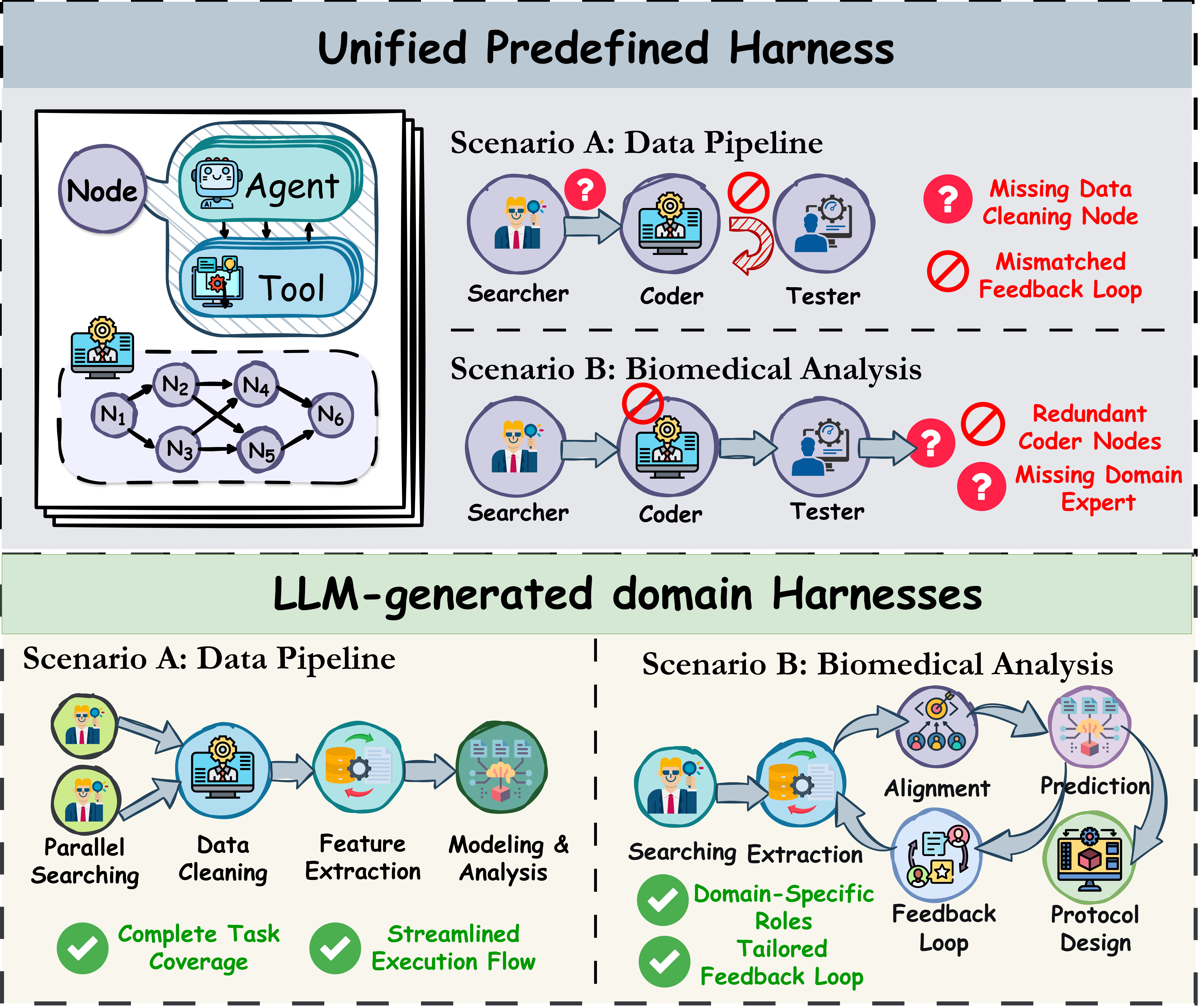}
    \caption{Illustration of Harness construction across domains.}
    \label{fig:domain_harness_motivation}
\end{figure}

\section{Introduction}
\label{sec:introduction}

Pattern recognition techniques have been widely applied in domains such as biomedicine, remote sensing, industrial inspection, and autonomous driving. Applications in these domains often consist of multiple interdependent stages, each requiring specific expertise and processing capabilities. Multi-Agent Systems (MAS) address these requirements by assigning different tasks to specialized agents and equipping them with appropriate tools~\cite{hong2024metagpt,wu2023autogen}. These agents must exchange intermediate results and coordinate their execution to complete the overall application~\cite{chen2024agentverse,qian2024chatdev}. Since the tasks, required capabilities, and dependencies vary across applications, the agents, tools, and execution workflows must be configured accordingly. This calls for methods that automatically construct MAS applications based on domain-specific requirements.

To support automated MAS construction, existing studies have explored agent generation~\cite{hu2025automated}, architecture design~\cite{zhang2024g,zhuge2024language}, workflow generation~\cite{zhang2025aflow}, and component composition~\cite{yuan2026automated,shang2025agentsquare}. These studies have introduced different methods for developing and assembling MAS, but have largely evolved along separate research directions. In particular, they adopt different task specifications, component registries, and construction interfaces, with each method typically evaluated within its own experimental setting. These differences mean that MAS applications are constructed according to method-specific procedures, rather than a shared standard that applies across application domains. Existing benchmarks assess agents through interactive task execution~\cite{liu2024agentbench,ma2024agentboard}, general assistant questions~\cite{mialon2024gaia}, and enterprise web tasks~\cite{drouin2024workarena}. Consequently, there is no consistent basis for specifying application requirements, selecting available components, organizing their execution, and evaluating the resulting systems. This calls for a benchmark that establishes a unified standard for MAS application construction and evaluation across domains.

\looseness=-1
To this end, we propose \textbf{OpenMAS-GApp}, a One-for-All benchmark for MAS application construction across diverse domains. Specifically, OpenMAS-GApp integrates 19 benchmark datasets under 20 evaluation settings spanning four application types and four professional domains, a shared registry of agents and tools, and domain-specific resources. It further establishes standardized application requirements and construction interfaces, allowing different methods to build MAS applications using the same inputs and available components. A unified execution environment and evaluation protocol enable the resulting applications to be assessed under consistent conditions.

\looseness=-1
Alongside the benchmark, we provide \textbf{Graph Harness} as a reference method for constructing executable MAS applications. A Harness specifies how selected agents and tools are organized to perform application tasks, including their operations, data exchanges, and execution relations. Since these requirements vary across applications, a predefined Harness may not cover all required operations or provide appropriate data and execution connections, as illustrated in Fig.~\ref{fig:domain_harness_motivation}. Graph Harness addresses this issue by constructing a separate Harness for each application through three stages: Task Structure Derivation identifies the required tasks and their dependencies; Component Selection retrieves suitable agents and tools based on their capabilities and interface compatibility; and Graph Harness Construction organizes the selected components into an executable graph with explicit data connections and control relations.

\noindent\textbf{Our Contributions.} (1) \emph{Unified Benchmark.} We introduce OpenMAS-GApp, a One-for-All benchmark for cross-domain MAS application construction and evaluation. It integrates 19 benchmark datasets under 20 evaluation settings spanning four application types and four professional domains, with a shared component registry, standard construction interfaces, and unified execution and evaluation. (2) \emph{Reference Construction Method.} We provide Graph Harness as a reference construction method within OpenMAS-GApp. It derives task structures, selects agents and tools based on capability and interface compatibility, and organizes them into executable graphs specifying task dependencies, data connections, and execution control. (3) \emph{Comprehensive Evaluation.} We evaluate construction methods across 19 benchmark datasets under 20 settings, analyzing stage contributions, registry scale, resource consumption, and robustness to perturbations.

%% file: sec/2_problem_statement.tex
\section{Problem Statement}
\label{sec:problem_statement}
\label{sec:problem_formulation}

We study cross-domain MAS application construction and evaluation. A construction method receives application requirements and available agents and tools, generates an executable application, and is evaluated through the outputs it produces for each individual problem.

\subsection{MAS Application Construction}

For evaluation entry $b$ in domain $d$, let $R_{bi}$ denote the application requirement of problem $i$. Each entry specifies its source variant, problem list, input conditions, and primary evaluation metric. The available component registry $\mathcal{C}$ contains components $c_j$ with capability descriptions $p_j$ and interfaces $I_j^c$ and $O_j^c$ for inputs and outputs, respectively.
\begin{equation}
\mathcal{C}=\{c_j=(p_j,I_j^c,O_j^c)\mid j\in\{1,\ldots,M\}\}.
\label{eq:registry}
\end{equation}

All methods share domain resources $K_d$, an execution environment $\Omega_d$, permissions $\mathcal{P}_d$, and resource budgets $\mathcal{B}_d$. For construction method $F_m$ and outer run $u$ with random seed $\xi_u$, the generated application is
\begin{equation}
\mathcal{A}_{biu}^{(m)}=F_m(R_{bi},\mathcal{C},K_d,\Omega_d,\mathcal{P}_d,\mathcal{B}_d;\xi_u),
\label{eq:construction}
\end{equation}
with $\mathcal{A}_{biu}^{(m)}=(\mathcal{U}_{biu}^{(m)},\mathcal{T}_{biu}^{(m)},\mathcal{D}_{biu}^{(m)},\mathcal{X}_{biu}^{(m)})$. The four entries specify callable agents and tools, task assignments, data connections, and execution relations, respectively.

\subsection{Application Execution and Evaluation}

Let $\mathcal{D}_b=\{(x_{bi},y_{bi})\}_{i=1}^{N_b}$ denote the fixed evaluation set, where $x_{bi}$ and $y_{bi}$ are the input and reference record of problem $i$, respectively. Its reference records form $\mathbf{y}_b=\{y_{bi}\}_{i=1}^{N_b}$. In each outer run, the common executor produces an output and an execution record for each problem
\begin{equation}
(o_{biu}^{(m)},r_{biu}^{(m)})
=
\operatorname{Execute}(\mathcal{A}_{biu}^{(m)},x_{bi},\xi_u;\Omega_d).
\label{eq:execution}
\end{equation}
The complete collections are $\mathbf{o}_{bu}^{(m)}=\{o_{biu}^{(m)}\}_{i=1}^{N_b}$ for outputs and $\mathbf{r}_{bu}^{(m)}=\{r_{biu}^{(m)}\}_{i=1}^{N_b}$ for execution records.

The benchmark evaluates the complete prediction set using the source parser and evaluator. Over $R$ complete runs, we report the mean task score
\begin{equation}
\overline{S}_b^{(m)}=\frac{100}{R}\sum_{u=1}^{R}\mathcal{M}_b(\mathbf{o}_{bu}^{(m)},\mathbf{y}_b),
\label{eq:evaluation}
\end{equation}
where $\mathcal{M}_b$ returns a normalized score in $[0,1]$ using entry $b$'s parser and evaluator on the complete prediction set. We report construction and execution costs separately.

%% file: sec/3_benchmark_design.tex
\section{Benchmark Design}
\label{sec:benchmark_design}

\input{sec/3_1_domain_application_suite}

\begin{figure*}[t]
    \centering
    \includegraphics[width=\textwidth]{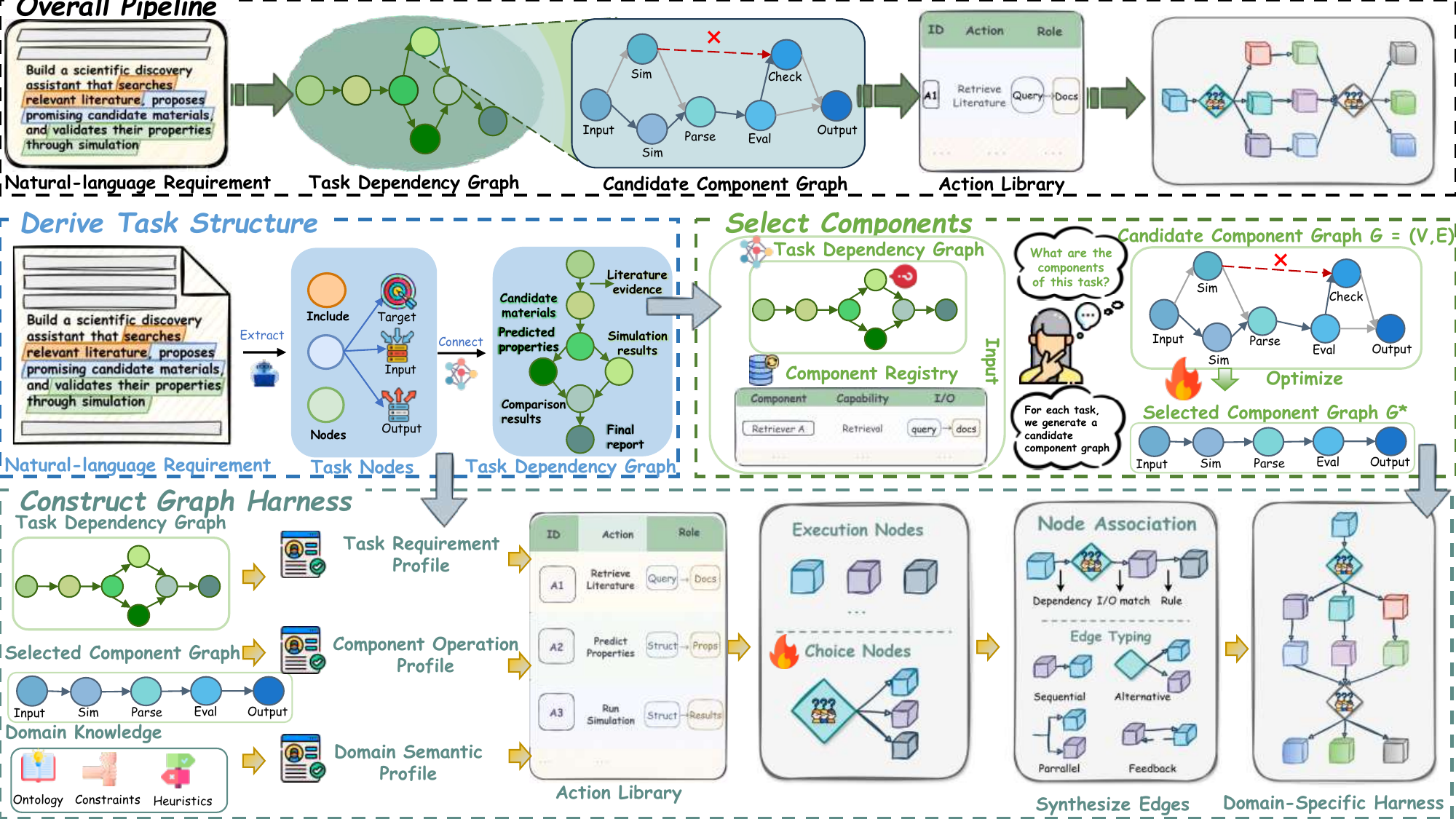}
    \caption{Overview of the domain-adaptive Graph Harness Construction Framework. We derive task structures from application requirements, select agents and tools from a component registry, and construct execution nodes and state transition rules.}
    \label{fig:framework}
\end{figure*}

\input{sec/3_2_reference_framework}

\begin{figure*}[!t]
\centering
\begin{minipage}[t]{\dimexpr(\textwidth-\columnsep)/2\relax}
\vspace{0pt}
\centering
\includegraphics[width=\linewidth,height=168pt,keepaspectratio]{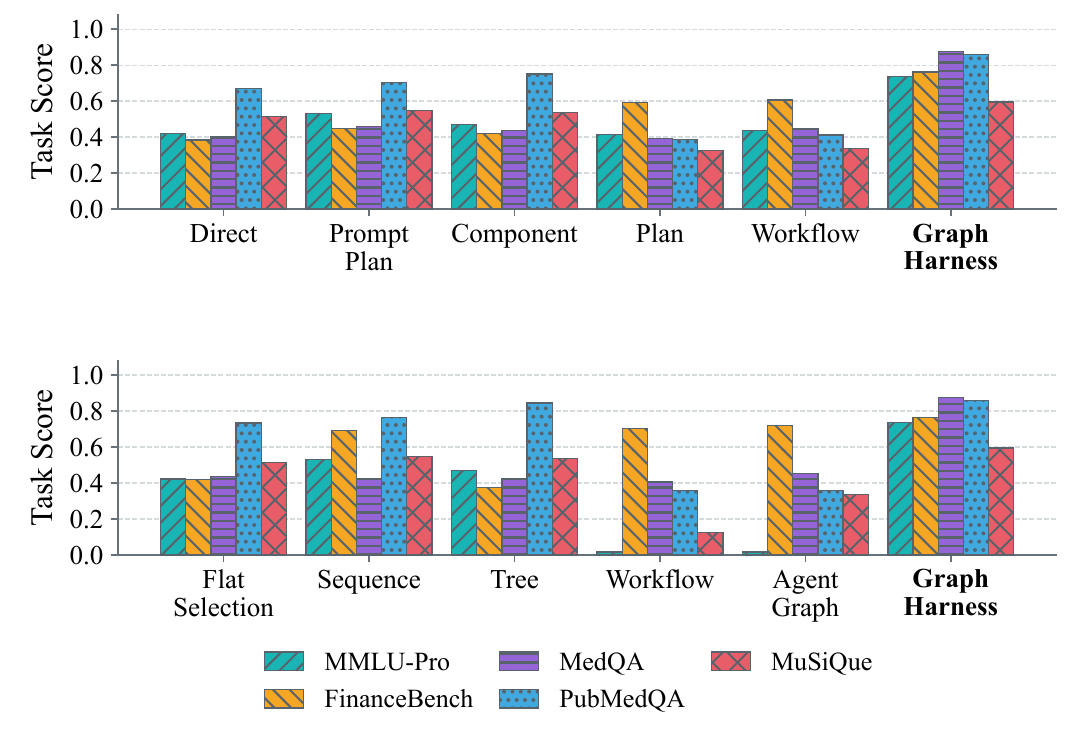}
\caption{Task scores of Graph Harness and the competing construction paradigms and orchestration representations.}
\label{fig:q1_q3_combined}
\end{minipage}\hfill
\begin{minipage}[t]{\dimexpr(\textwidth-\columnsep)/2\relax}
\vspace{0pt}
\centering
\includegraphics[width=\linewidth]{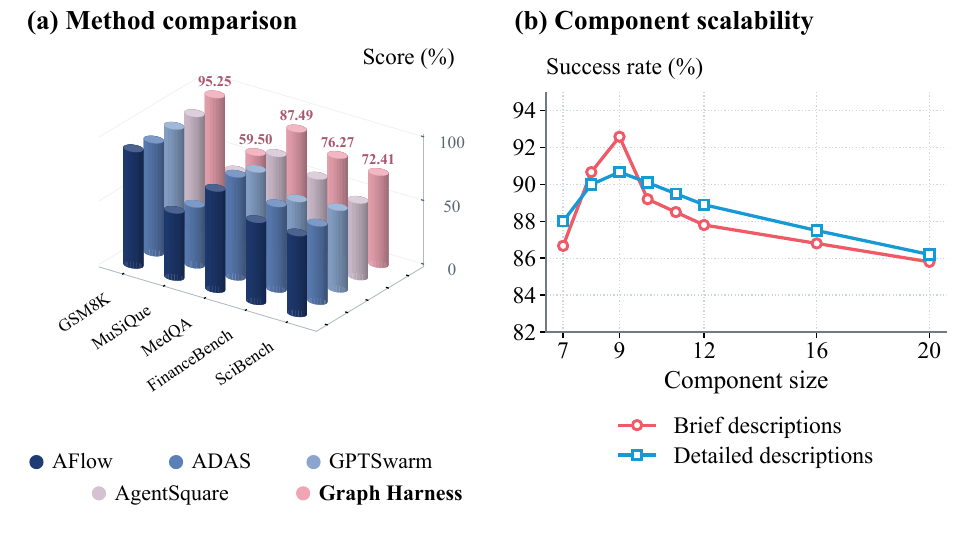}
\caption{Construction method comparison and component scalability. Panel (a) compares AFlow, ADAS, GPTSwarm, AgentSquare, and Graph Harness. Panel (b) compares brief and detailed interface descriptions with the same callable operations and interface schemas.}
\label{fig:component_scale_success_trend.pdf}
\end{minipage}
\end{figure*}

\input{sec/3_3_experiments_evaluations}

%% file: sec/3_1_domain_application_suite.tex
\subsection{Domains and Application Types}
\label{sec:domain_application_suite}

OpenMAS-GApp covers four general application types and four professional domains through 19 benchmark datasets organized into 20 evaluation settings. The application types include general reasoning, mathematical reasoning, reading and question answering, and code generation. The professional domains are medicine, finance, science, and law. BIG-Bench Hard contributes two evaluation settings, BBH-Boolean and BBH-Mixed. We evaluate how construction methods address the knowledge requirements, reasoning steps, and objectives of these domains and application types. Appendix~\ref{sec:reported_scope} gives the comparison scope.

Tables~\ref{tab:domain_datasets} and~\ref{tab:application_type_datasets} list the datasets for professional domains and general application types, respectively. Each dataset provides individual problems with their input data and evaluation information. We use each problem to define an application requirement specifying the objective and permitted resources. We construct, execute, and evaluate one MAS application per problem. We retain source passages, tables, answer choices, and programming specifications.

The resulting application suite covers knowledge reasoning, mathematical problem solving, multi-hop question answering, code generation, and domain-specific tasks in medicine, finance, science, and law. These tasks differ in their required capabilities, input resources, and expected outputs. We evaluate a separate MAS application per problem to assess how construction methods organize components for diverse requirements.

\begin{table}[!t]
    \centering
    \small
    \setlength{\tabcolsep}{4pt}
    \renewcommand{\arraystretch}{1.15}
    \caption{Professional domains and their datasets.}
    \label{tab:domain_datasets}
    \begin{tabular}{p{0.34\columnwidth} p{0.58\columnwidth}}
        \toprule
        \textbf{Domain} & \textbf{Datasets} \\
        \midrule
        Medicine & MedQA~\cite{jin2021disease}, PubMedQA~\cite{jin2019pubmedqa} \\
        Finance & FinQA~\cite{chen2021finqa}, FinanceBench~\cite{islam2023financebench} \\
        Science & SciQ~\cite{welbl2017crowdsourcing}, SciBench~\cite{wang2023scibench} \\
        Law & CaseHOLD~\cite{zheng2021does} \\
        \bottomrule
    \end{tabular}
\end{table}

\begin{table}[!t]
    \centering
    \small
    \setlength{\tabcolsep}{4pt}
    \renewcommand{\arraystretch}{1.15}
    \caption{General application types and their datasets.}
    \label{tab:application_type_datasets}
    \begin{tabular}{p{0.34\columnwidth} p{0.58\columnwidth}}
        \toprule
        \textbf{Application Type} & \textbf{Datasets and Settings} \\
        \midrule
        General Reasoning
        & MMLU~\cite{hendrycks2020measuring}, MMLU-Pro~\cite{wang2024mmlu}, BBH-Boolean, BBH-Mixed~\cite{suzgun2023challenging}, ARC~\cite{clark2018think}, StrategyQA~\cite{geva2021did} \\
        Mathematical\newline Reasoning
        & GSM8K~\cite{cobbe2021training}, MATH-500 \\
        Reading and QA
        & HotpotQA~\cite{yang2018hotpotqa}, MuSiQue~\cite{trivedi2022musique}, DROP~\cite{dua2019drop} \\
        Code Generation
        & HumanEval~\cite{chen2021evaluating}, MBPP~\cite{austin2021program} \\
        \bottomrule
    \end{tabular}
\end{table}

%% file: sec/3_2_reference_framework.tex
\subsection{Reference Construction Framework}
\label{sec:methodology}
\label{sec:reference_construction_framework}

\noindent\textbf{Graph Harness Construction Framework.}
Figure~\ref{fig:framework} shows our three-stage framework. Task Structure Derivation produces a Task Dependency Graph from application requirements. Component Selection retrieves compatible agents and tools and forms Selected Component Graphs. Graph Harness Construction combines these graphs with domain knowledge into execution nodes and transitions specifying component calls, data exchanges, and control.

For the construction below, we fix entry $b$, individual problem $i_0$, method $m$, and outer run $u$. We suppress these fixed indices on intermediate graphs and use local indices for tasks and components. Domain subscript $d$ identifies the resources used for this individual problem.

\noindent\textbf{Task Structure Derivation.}
Given the current problem's application requirement $R_{bi_0}$, we employ an LLM to extract atomic requirements $U_d=\{u_k\}_{k=1}^{K}$ and generate $N$ candidate tasks $\tilde{\tau}_i=(z_i,g_i,I_i,O_i)$, where $z_i$ indicates task selection, $g_i$ specifies its target and operation, and $I_i$ and $O_i$ denote its required inputs and expected outputs. Let $A_{ki}$ indicate whether candidate task $\tilde{\tau}_i$ fulfills requirement $u_k$, and let $b_k$ indicate whether the requirement is mandatory. These indicators are binary. Let $\mathbf{A}_d=(A_{ki})\in\{0,1\}^{K\times N}$, $\mathbf{z}=(z_i)$, and $\mathbf{b}=(b_k)$. We require element-wise

\begin{equation}
\mathbf{A}_d\mathbf{z}\succeq\mathbf{b},
\quad \mathbf{z}\in\{0,1\}^{N}.
\label{eq:task_coverage}
\end{equation}

After task extraction, we identify dependencies between selected tasks. We distinguish explicit execution prerequisites, indicated by binary $c_{ij}$, from data dependencies established through validated output-to-input bindings, indicated by binary $b_{ij}$. Their joint waiting relation is

\begin{equation}
a_{ij}=1-(1-c_{ij})(1-b_{ij}),\quad i\neq j,
\label{eq:task_dependency}
\end{equation}

which defines the dependency edges $D_d$ of the Task Dependency Graph $G_d^T=(T_d,D_d)$. This graph specifies which tasks must be completed before subsequent tasks can execute and which outputs are required by their successors.

\noindent\textbf{Component Selection.}
Given $G_d^T$ and the component registry $\mathcal{C}=\{c_j\}_{j=1}^{M}$, where each component $c_j=(p_j,I_j^c,O_j^c)$ describes its capability and input/output interfaces, we retrieve candidate components for each task $\tau_i$. Their possible combinations are represented as a Candidate Component Graph $G_i^C=(V_i^C,E_i^C)$, with candidate execution paths $\mathcal{P}_i^C$.

For each of the $R_i$ capability requirements $k_{ir}$, we score component $c_j$ by $q_{ijr}=v_{ijr}(1+\cos(\mathbf{e}_{p_j},\mathbf{e}_{k_{ir}}))/2$. Binary $v_{ijr}$ indicates whether the component supports this requirement, and $\mathbf{e}_{p_j}$ and $\mathbf{e}_{k_{ir}}$ denote their nonzero semantic embeddings. The capability coverage of a path is

\begin{equation}
Q_i(p)=\mathbf{w}_i^{\top}
\left[\max_{c_j\in p}q_{ijr}\right]_{r=1}^{R_i},
\label{eq:path_coverage}
\end{equation}

where $\mathbf{w}_i=(w_{ir})_{r=1}^{R_i}$ is a column vector of nonnegative capability weights summing to one. The bracketed vector contains the per-requirement maxima as a column.

After capability assessment, we evaluate interface compatibility to determine whether source outputs can supply required target inputs. For fields $f=(n,t,s)$, where $n$, $t$, and $s$ denote the field name, data type, and semantic description, respectively, compatibility is defined as

\begin{equation}
\chi(f_o,f_r)=\mathbb{I}[t_o\preceq t_r]h_f(f_o,f_r)s_f(f_o,f_r).
\label{eq:interface_compatibility}
\end{equation}

Here, $t_o\preceq t_r$ permits assignment or a registered conversion, binary $h_f$ records semantic acceptance, and $s_f\in[0,1]$ scores accepted matches. Every mandatory input requires a binding passing both hard checks. The weighted coverage of required input fields determines edge compatibility $\eta_e$, while $\Gamma_i(p)$ is the mean edge compatibility along path $p$.

We validate complete paths and freeze the hard-feasible family $\mathcal{P}_i^{\mathrm{feasible}}$ before normalizing invocation and communication costs. With fixed extrema, path utility is

\begin{equation}
U_i(p)=
\boldsymbol{\lambda}^{\top}
[Q_i(p),\Gamma_i(p),-\bar{C}_i(p)]^{\top},
\label{eq:path_utility}
\end{equation}

where $\boldsymbol{\lambda}=[\alpha,\beta,\gamma]^{\top}$ contains nonnegative coefficients summing to one. Paths passing $Q_i(p)\geq\delta_i$ and $\eta_e\geq\epsilon_i$ for every edge form the admissible family $\mathcal{P}_i^{\mathrm{adm}}$. We retain $k_i=2$ paths and let $\mathcal{F}_i$ contain all two-element subsets of this family. The selected set is

\begin{equation}
S_i^*\in\arg\max_{S_i\in\mathcal{F}_i}
\sum_{p\in S_i}U_i(p).
\label{eq:path_selection}
\end{equation}

Fewer than two admissible paths returns a construction failure. Otherwise, we merge the selected paths into a Selected Component Graph

\begin{equation}
G_i^{C*}=
\left(\bigcup_{p\in S_i^*}V(p),
\bigcup_{p\in S_i^*}E(p)\right),
\label{eq:selected_component_graph}
\end{equation}

which preserves suitable component combinations for subsequent Harness construction. Here, $V(p)$ and $E(p)$ are the vertices and edges of path $p$.

\noindent\textbf{Graph Harness Construction.}
Given the Task Dependency Graph, Selected Component Graphs, and domain knowledge $K_d$, we first construct three profiles

\begin{equation}
P_d=\Psi(G_d^T,\{G_i^{C*}\}_{\tau_i\in T_d},K_d),
\label{eq:profile_construction}
\end{equation}

where $\Psi$ produces $P_d=(P_d^T,P_d^C,P_d^D)$, comprising the Task Requirement Profile, Component Operation Profile, and Domain Semantic Profile. They record task objectives and dependencies, component operations and interfaces, and domain constraints.

The synthesizer proposes nodes from these profiles

\begin{equation}
v_i=(s_i,\ell_i,I_i^v,O_i^v,\rho_i),
\label{eq:node_construction}
\end{equation}

where $s_i$ is the execution state, $\ell_i$ the operation, $I_i^v$ and $O_i^v$ the input and output, and $\rho_i\in\{\mathrm{Exec},\mathrm{Choice}\}$ the node type. Each execution activation invokes one registered operation on an associated component after contract checks. Execution nodes form $V_d^{\mathrm{Exec}}$ and satisfy

\begin{equation}
\emptyset\neq\mu_d(v_i)\subseteq C_i,
\quad v_i\in V_d^{\mathrm{Exec}}.
\label{eq:node_association}
\end{equation}

The validator checks task coverage, registry membership, required bindings, preserved waiting relations, and typed control conditions. Source task obligations and component contracts remain fixed during synthesis. Each $C_i$ contains compatible components from the selected graphs.

Each transition retains its type, activation condition, and passed data

\begin{equation}
e_{ij}=(v_i,\tau_{ij},\gamma_{ij},v_j,m_{ij}),
\label{eq:edge_construction}
\end{equation}

where $\tau_{ij}\in\{\mathrm{Seq},\mathrm{Alt},\mathrm{Par},\mathrm{Fb}\}$ denotes sequential, alternative, parallel, and feedback control, respectively. The condition $\gamma_{ij}$ governs activation, and $m_{ij}$ specifies the transferred information. The synthesized nodes and transitions form the final domain-specific Graph Harness $H_d=(V_d,E_d)$. Appendix~\ref{app:reference_construction} details validation and execution.

%% file: sec/3_3_experiments_evaluations.tex
\subsection{Experiments and Evaluations}
\label{sec:benchmark_evaluations}

All methods share application requirements, the component registry, domain resources, and an LLM backbone. We execute each constructed application and score its output with the source dataset protocol. Appendix~\ref{sec:reported_scope} lists the datasets and protocols for each comparison.

\noindent\textbf{Construction Effectiveness.}
We compare construction methods across domains and application types. Ablations examine task structure derivation, component selection, graph construction, and interface constraints. Orchestration comparisons hold the remaining construction process fixed to assess how execution organization affects application performance under shared resource budgets.

\noindent\textbf{Scalability and Efficiency.}
We vary registry size and interface-description detail to assess scalability, and compare task performance against mean construction tokens to evaluate efficiency for each constructed application.

\noindent\textbf{Robustness and Domain Adaptation.}
We test requirement ambiguity, component metadata errors, distractor components, and resource failures at two intensity levels. Medical and legal cases examine how domain requirements shape component assignments, resource connections, and execution structures of their generated Harnesses.

%% file: sec/4_experiments.tex
\section{Experiments}

\looseness=-1
We evaluate application construction on OpenMAS-GApp through the following research questions:
\textbf{Q1 (Performance).} How does Graph Harness compare with existing construction methods across tasks and domains?
\textbf{Q2 (Harness Necessity):} To what extent is a harness-based orchestration paradigm necessary for MAS application construction
\textbf{Q3 (Construction Ablation).} How do task structure derivation, component selection, graph construction, and interface constraints affect the quality of the constructed MAS applications
\textbf{Q4 (Graph Organization):} What role does graph-based orchestration play in representing and coordinating application capabilities, dependencies, and constraints
\textbf{Q5 (Scalability):} How well does Graph Harness scale as the component ecosystem becomes larger and more diverse
\textbf{Q6 (Efficiency).} How does Graph Harness balance task performance and mean construction-token consumption
\textbf{Q7 (Robustness):} When construction conditions are incomplete, noisy, or conflicting, does Graph Harness maintain stable application quality
\textbf{Q8 (Cross-Domain Applicability).} How does Graph Harness construct different Harnesses for individual problems across professional domains

\subsection{Experimental Setup}

\paragraph{Datasets.}
We turn source questions and permitted contexts into construction instances with fixed components, domain resources, and budgets. We execute each returned application and score its output with the source evaluator. Direct answering provides a task-level reference.
For general knowledge and structured reasoning, we use MMLU, MMLU-Pro,
BBH-Boolean, BBH-Mixed, ARC, and StrategyQA. For mathematical reasoning, we include GSM8K
and MATH-500, a subset of MATH~\cite{hendrycks2021measuring}. Multi-hop reading and question answering are evaluated on
HotpotQA, MuSiQue, and DROP. For code generation, we use HumanEval and MBPP.
Our domain-specific benchmarks comprise MedQA and PubMedQA for medicine,
FinQA and FinanceBench for finance, and SciQ and SciBench for science, with CaseHOLD for legal tasks.
Together, these datasets cover tasks with different knowledge requirements,
reasoning structures, computational demands, and domain constraints.
BBH-Boolean and BBH-Mixed are two evaluation settings of the same BIG-Bench Hard benchmark. They use different candidate pools and construction templates, as detailed in Appendix~\ref{app:benchmark_sources}.
Appendix~\ref{app:experimental_settings} specifies dataset manifests and scoring records.

\paragraph{Baselines.}
We compare construction adapters and a direct-answering reference in
four groups. (1) \textbf{Direct answering and planning}
includes Direct LLM Answering, Prompt-only MAS Planning, and Plan-based
Construction, which produce an answer or MAS directly from the requirement
or through an intermediate task plan. (2) \textbf{Component and
process-based construction} includes Component-based Assembly, Flat
Component Selection, Sequence, Tree, and Workflow, which organize selected
components using flat, sequential, hierarchical, or predefined process
structures. (3) \textbf{Agent-centric graph construction} includes Agent
Graph, which models communication dependencies among agents without an
application-level representation of capabilities, resources, and
constraints. (4) \textbf{Automated agent and workflow design} includes
AFlow~\cite{zhang2025aflow}, which searches over executable workflows,
ADAS~\cite{hu2025automated}, which generates agent designs in code,
GPTSwarm~\cite{zhuge2024language}, which optimizes node prompts and graph connections,
and AgentSquare~\cite{shang2025agentsquare}, which searches through module evolution and recombination.
All methods use DeepSeek-V4-flash as the LLM backbone and share the same
component ecosystem, input requirements, and execution environment.
Appendix~\ref{app:baseline_settings} describes the baseline implementations.

\begin{table*}[!t]
\centering
\caption{Ablation scores across eleven evaluation configurations. Values are mean $\pm$ std (\%) over three runs for the two BBH settings and five runs for the other settings, with the highest means in bold.}
\label{tab:ablation}

\footnotesize
\setlength{\tabcolsep}{2.5pt}
\renewcommand{\arraystretch}{1.12}
\resizebox{\textwidth}{!}{
\begin{tabular}{@{}l*{11}{c}@{}}
\toprule
\multirow{2}{*}{Variant}
& \multicolumn{2}{c}{Mathematical Reasoning}
& \multicolumn{4}{c}{General/Structured Reasoning}
& \multicolumn{3}{c}{Multi-hop QA}
& \multicolumn{2}{c}{Domain-specific} \\
\cmidrule(lr){2-3}\cmidrule(lr){4-7}\cmidrule(lr){8-10}\cmidrule(lr){11-12}
& GSM8K & MATH-500 & MMLU-Pro & ARC & BBH-Boolean & BBH-Mixed
& HotpotQA & DROP & MuSiQue & FinanceBench & SciBench \\
\midrule
Full & \textbf{95.25}$\pm$0.86 & \textbf{73.24}$\pm$1.73 & \textbf{73.75}$\pm$0.67 & \textbf{91.37}$\pm$2.54 & \textbf{76.67}$\pm$5.77 & \textbf{72.22}$\pm$1.92 & \textbf{56.00}$\pm$3.38 & \textbf{83.32}$\pm$2.21 & \textbf{59.50}$\pm$1.64 & \textbf{76.27}$\pm$3.55 & \textbf{72.41}$\pm$2.96 \\
w/o Task Structure & 90.24$\pm$3.62 & 68.36$\pm$2.87 & 70.43$\pm$2.15 & 88.26$\pm$3.79 & 70.00$\pm$0.00 & 67.78$\pm$3.85 & 48.94$\pm$1.46 & 82.36$\pm$3.94 & 56.73$\pm$4.27 & 72.93$\pm$2.65 & 68.24$\pm$0.92 \\
w/o Component Selection & 91.52$\pm$2.03 & 63.36$\pm$2.76 & 71.36$\pm$2.48 & 87.25$\pm$3.91 & 73.33$\pm$5.77 & 68.89$\pm$3.85 & 54.56$\pm$2.58 & 82.62$\pm$4.63 & 51.17$\pm$3.07 & 70.27$\pm$3.45 & 69.28$\pm$2.89 \\
w/o Graph Construction & 91.93$\pm$2.74 & 66.76$\pm$3.28 & 72.73$\pm$2.53 & 89.35$\pm$4.08 & 73.33$\pm$5.77 & 68.89$\pm$3.85 & 53.67$\pm$2.29 & 81.73$\pm$3.83 & 53.36$\pm$2.95 & 71.87$\pm$3.69 & 65.41$\pm$2.78 \\
w/o I/O Constraints & 90.45$\pm$3.69 & 66.72$\pm$3.05 & 71.87$\pm$1.83 & 90.56$\pm$4.47 & 70.00$\pm$0.00 & 68.89$\pm$3.85 & 50.94$\pm$1.94 & 82.75$\pm$4.25 & 58.06$\pm$2.37 & 73.33$\pm$3.77 & 68.83$\pm$4.89 \\
\bottomrule
\end{tabular}
}

\medskip
\centering
    \includegraphics[width=\linewidth]{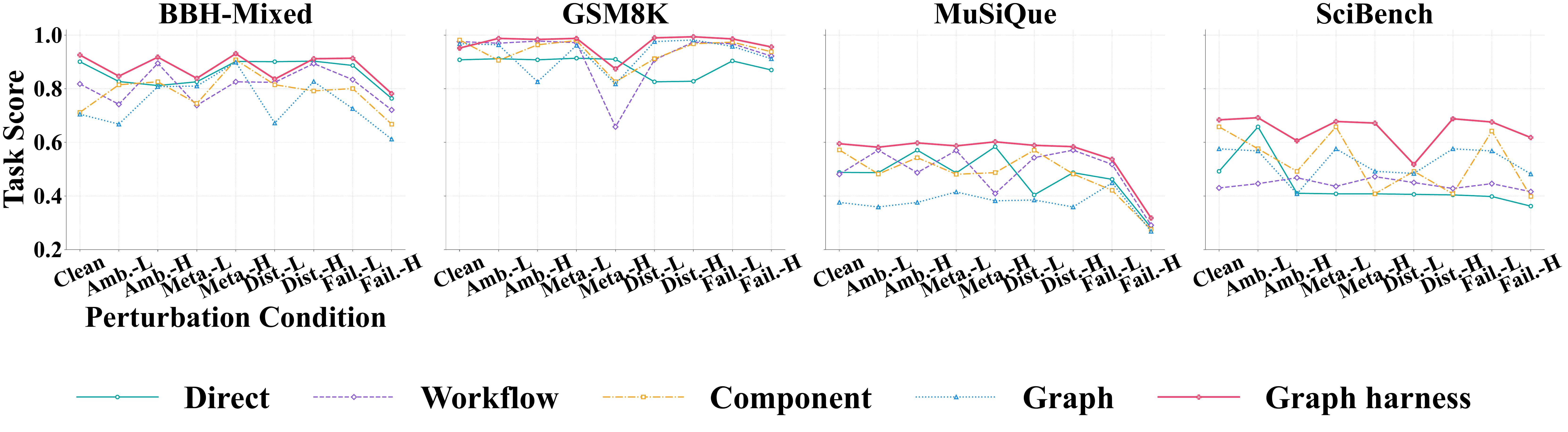}
    \captionof{figure}{Robustness under clean and perturbed construction conditions.
Amb., Meta., Dist., and Fail. denote requirement ambiguity, metadata errors,
distractor components, and resource failures, respectively; L and H indicate
low and high perturbation levels.}
    \label{fig:q6_robustness}
\end{table*}

\subsection{Overall Performance (Q1)}
\label{sec:performance}
To answer \textbf{Q1}, we compare Graph Harness with AFlow, ADAS, GPTSwarm, and AgentSquare across 20 evaluation settings. Figure~\ref{fig:component_scale_success_trend.pdf}(a) presents five representative datasets, and Table~\ref{tab:extended_official_baselines} reports the complete results. Graph Harness achieves the highest score in 15 settings and an average score of \textbf{77.52\%}, with equal weight assigned to each setting. Compared with AFlow, the strongest baseline by average score at $69.13\%$, we obtain an average improvement of \textbf{8.39 percentage points}. Across the four professional domains, Graph Harness leads on all seven datasets. On MedQA and CaseHOLD, its scores reach $87.49\%$ and $79.00\%$, exceeding the strongest competitor, ADAS, by 6.56 and 11.00 percentage points, respectively. Competing methods obtain higher scores on MMLU-Pro, the two BBH settings, HotpotQA, and HumanEval. These comparisons demonstrate strong overall application performance and consistent gains across the evaluated professional domains under the shared benchmark protocol.

\subsection{Construction Paradigms (Q2)}
\looseness=-1
To answer \textbf{Q2}, we compare Graph Harness with construction paradigms and a direct-answering reference, including prompt-only planning, component-based assembly, plan-based construction, and workflow-based construction. As shown in Figure~\ref{fig:q1_q3_combined}, Graph Harness achieves the highest score on all five datasets, reaching $73.75\%$, $76.27\%$, $87.49\%$, $85.88\%$, and $59.50\%$ on MMLU-Pro, FinanceBench, MedQA, PubMedQA, and MuSiQue, respectively. Compared with the strongest alternative on each dataset, it improves by 20.65, 15.60, 41.69, 10.80, and 4.60 percentage points, with an average score of $76.58\%$. Direct answering and prompt-only planning fluctuate substantially across domains, while plan- and workflow-based methods perform well on selected datasets but degrade on others. These scores support using an explicit Harness layer to organize requirements, tasks, capabilities, components, resources, and constraints before MAS execution.

\subsection{Ablation Study (Q3)}
\looseness=-1
To answer \textbf{Q3}, we evaluate four ablated variants of Graph Harness. As shown in Table~\ref{tab:ablation}, the complete framework achieves the highest score on all eleven evaluation settings. Removing task derivation reduces the mean by \textbf{4.16 percentage points}. Replacing component selection with generated configurations gives the largest mean decrease of \textbf{4.22 percentage points}, including losses of 9.88 percentage points on MATH-500 and 8.33 percentage points on MuSiQue. With the task graph and selected components retained, sequential execution reduces the average score by \textbf{3.72 percentage points}. Removing I/O constraints causes an average decrease of \textbf{3.42 percentage points}. These results support the contributions of task derivation, component selection, graph-based execution organization, and \mbox{interface compatibility checks}.

\begin{figure*}[!t]
\centering
    \includegraphics[width=\textwidth]{figures/domain_case_studies_v3.png}
    \caption{Medical and legal Harnesses generated for different domain requirements.}
    \label{fig:domain_case_studies}

\medskip
\centering
    \includegraphics[width=\textwidth]{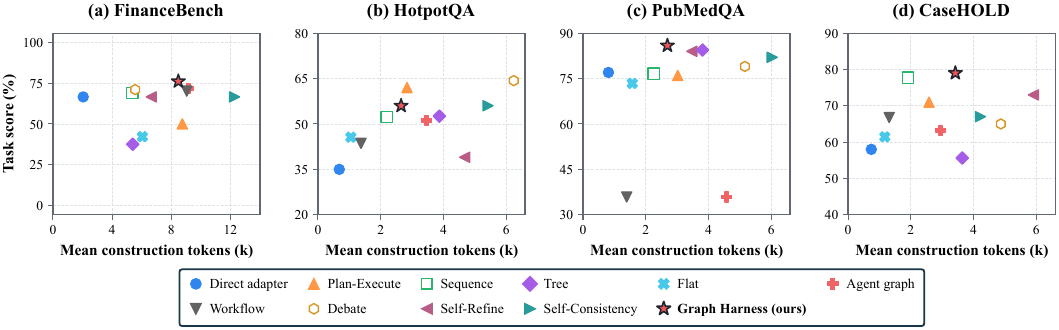}
    \caption{Task scores and mean construction tokens across four datasets.}
    \label{fig:efficiency_preview}
\end{figure*}

\subsection{Orchestration Representation Analysis (Q4)}
\looseness=-1
To answer \textbf{Q4}, we fix the Harness-level construction process and vary only its internal orchestration representation, comparing Graph Harness with flat component selection, sequence-based orchestration, tree-based planning, workflow-based templates, and agent graph orchestration. As shown in Figure~\ref{fig:q1_q3_combined}, Graph Harness obtains the highest score on all five datasets, reaching $73.75\%$ on MMLU-Pro, $76.27\%$ on FinanceBench, $87.49\%$ on MedQA, $85.88\%$ on PubMedQA, and $59.50\%$ on MuSiQue. Relative to the strongest alternative for each dataset, the improvements are 20.65, 4.40, 42.20, 1.48, and 4.60 percentage points, respectively. Sequence- and tree-based representations perform well on selected tasks but are less reliable across domains, whereas workflow and agent-graph representations show sharp drops on MMLU-Pro and MuSiQue. These results support graph-based organization for representing tasks, capabilities, components, resources, constraints, and the dependencies among them within a MAS application.

\subsection{Component Scalability (Q5)}
To answer \textbf{Q5}, we increase the component pool size and compare brief
and detailed interface descriptions. Both settings retain the same callable
operations, endpoints, and interface schemas, so the paired comparison varies
the information available during component selection. As shown in
Figure~\ref{fig:component_scale_success_trend.pdf}, task success peaks near nine
components at approximately $92.6\%$ with brief descriptions and $90.7\%$ with
detailed descriptions. As the pool expands to 20 components, the scores
decline to $85.8\%$ and $86.2\%$, respectively. We obtain higher scores with
detailed descriptions at the larger evaluated pool sizes. These results show that
adding components does not consistently improve task performance and that
interface descriptions and pool size should be evaluated jointly.

\subsection{Efficiency Analysis (Q6)}
To answer \textbf{Q6}, we compare task performance and construction costs across four datasets in Figure~\ref{fig:efficiency_preview}. Graph Harness achieves scores of $76.27\%$, $56.00\%$, $85.88\%$, and $79.00\%$ on FinanceBench, HotpotQA, PubMedQA, and CaseHOLD, respectively. Direct denotes a construction adapter here, distinct from the direct-answering reference in Q2. Graph Harness exceeds Self-Consistency on three datasets and matches its score on HotpotQA, while using $19.0\%$ to $55.4\%$ fewer construction tokens across the four datasets. On PubMedQA, its score is $85.88\%$ with 2.69k tokens, compared with $82\%$ and 6.03k tokens for Self-Consistency. On HotpotQA, both methods score $56\%$, while Graph Harness uses 2.65k tokens compared with 5.42k, reducing construction tokens by $51.1\%$. On CaseHOLD, we gain 12 percentage points over Self-Consistency and reduce tokens by $19.0\%$.

\subsection{Robustness Analysis (Q7)}

\looseness=-1
To answer \textbf{Q7}, we evaluate Graph Harness under clean inputs and
four types of construction perturbations, including ambiguous requirements,
metadata errors, distractor components, and resource failures at two
intensity levels. The comparison covers BBH-Mixed, GSM8K, MuSiQue, and
SciBench. As shown in Figure~\ref{fig:q6_robustness}, Graph Harness obtains
the highest score in 33 of 36 dataset-condition pairs. The exceptions are
BBH-Mixed with low distractor intensity and GSM8K under clean inputs and
high metadata-error intensity. Relative to its clean score, Graph Harness
shows its largest decline on MuSiQue under severe resource failure, from
$59.5\%$ to $31.8\%$. Low-intensity perturbations can also reduce
performance, as SciBench drops from $68.4\%$ to $51.8\%$ with distractor
components. Thus, Graph Harness leads most comparisons, while its
robustness depends on the dataset and the type of construction perturbation.

\subsection{Cross-Domain Case Studies (Q8)}
\looseness=-1
To answer \textbf{Q8}, we examine cross-domain applicability through the medical and legal Harnesses in Figure~\ref{fig:domain_case_studies}. For the author-specified medical query about progressive retrosternal burning, the medical Harness connects retrieval on systemic sclerosis and esophageal involvement to reasoning, safety assessment, automated review, and a human-approval gate. For the dismissal query, the legal Harness connects retrieval and citation matching to legal databases, checks precedents and contractual obligations, and routes the draft through professional review before reporting. These differences in agent assignments, resource connections, and execution dependencies illustrate how the same Graph Harness framework generates different Harnesses for the requirements of each domain and individual problem.

%% file: sec/5_conclusion.tex
\section{Conclusion}
\label{sec:conclusion}

We present OpenMAS-GApp, a benchmark for domain-adaptive MAS application construction, and a Graph Harness Construction Framework with three stages, namely Task Structure Derivation, Component Selection, and Graph Harness Construction. Graph Harness achieves the highest score in 15 of 20 evaluation settings and leads the five-dataset construction and orchestration comparisons. Ablations support the contributions of all three stages and interface constraints. We further examine registry scale, construction efficiency, and robustness. The medical and legal cases illustrate how the same framework generates distinct Harnesses for different domain requirements. Through this benchmark, we provide a common construction task and evaluation protocol for executable MAS applications across domains. In future work, we will examine dynamic component ecosystems and more complex applications.

%% file: sec/X_suppl.tex
\clearpage
\appendix
\input{sec/A_benchmark}
\clearpage
\input{sec/B_theory}
\clearpage
\input{sec/C_implementation}
\clearpage
\input{sec/D_settings}
\clearpage
\input{sec/E_additional}

\clearpage
\input{sec/F_extended_evaluation_tables}

%% file: sec/A_benchmark.tex
\section{OpenMAS-GApp Benchmark}
\label{app:benchmark}

OpenMAS-GApp is a meta-level benchmark for domain-adaptive MAS application construction. For problem $i$ in evaluation entry $b$ from domain $d$, the constructor receives requirement $R_{bi}$, component registry $\mathcal{C}$, domain resources $K_d$, environment $\Omega_d$, permissions $\mathcal{P}_d$, and resource budgets $\mathcal{B}_d$. In outer run $u$, method $F_m$ returns application $\mathcal A_{biu}^{(m)}$ under seed $\xi_u$. The common executor runs that application on input $x_{bi}$. We apply $\mathcal{M}_b$ to the complete predictions from each run and then average the configured run scores, following Sec.~\ref{sec:problem_statement}. Each output and its construction and execution records retain the entry, problem, and run identifiers.

\subsection{Domains, Application Types, and Data Sources}
\label{app:benchmark_sources}

\noindent\textbf{Domains, application types, and individual problems.}
We distinguish professional domains, application types, and individual problems. A professional domain identifies a body of specialized knowledge, such as medicine or law. An application type groups tasks with related objectives and input/output requirements, such as medical question answering or legal holding selection. An individual problem provides the concrete question and materials for one construction run. Problems of the same application type can share structural patterns while requiring different operations for their specific inputs and requested outputs.

\noindent\textbf{Task coverage.}
Table~\ref{tab:benchmark_sources} groups the data sources by domain or application type. General reasoning and mathematics provide questions with different requirements for knowledge use, calculation, and verification. Reading and question answering further require the use of supplied information, while code generation evaluates programs produced through agent and tool execution. Medical, financial, scientific, and legal problems introduce professional concepts and task-specific output requirements. Together, these sources provide application requirements with different inputs, operations, and expected outputs.

\begin{table*}[!t]
\centering
\caption{Domains, application types, and data sources in the evaluation suite. Dataset labels follow the main experiments.}
\label{tab:benchmark_sources}
\small
\setlength{\tabcolsep}{5pt}
\renewcommand{\arraystretch}{1.12}
\begin{tabular}{@{}p{0.23\textwidth}p{0.69\textwidth}@{}}
\toprule
Domain or application type & Data sources \\
\midrule
General reasoning & MMLU, MMLU-Pro, BBH-Boolean, BBH-Mixed, ARC, StrategyQA \\
Mathematical reasoning & GSM8K, MATH-500 \\
Reading and QA & HotpotQA, MuSiQue, DROP \\
Code generation & HumanEval, MBPP \\
Medicine & MedQA, PubMedQA \\
Finance & FinQA, FinanceBench \\
Science & SciQ, SciBench \\
Law & CaseHOLD \\
\bottomrule
\end{tabular}
\end{table*}

\noindent\textbf{Dataset records.}
We distinguish the application type, source dataset, source split, and individual problem identifier. Application-type counts describe the range of construction tasks, while problem counts determine the number of individual construction runs. We preserve the partitions released by each source. Table~\ref{tab:benchmark_manifest} identifies source partitions and the two local BBH settings. The experiment manifest records the source revision, problem membership, and submitted count for each reported comparison.

Partition configuration. Use the named evaluation partition for each source and variant. The two local BBH settings use experiment-specific cohorts drawn from their candidate pools. Preserve the released training, development, and test memberships and the original problem identifiers. The split choices in Table~\ref{tab:benchmark_manifest} are evaluation selections. Identify a development evaluation by its released split name. Freeze the source revision and use the same problem list for every method and outer run in a comparison.

Manifest records. Save the dataset revision, variant, split, task, problem identifier, permitted input fields, input digest, submitted count, and scorer identifier. Source collection sizes describe the released partitions. The submitted count and problem membership for each reported comparison come from its experiment manifest. Keep tuning problems separate from evaluation problems. Link the same problem and repeat to one complete-method run across Q2, Q3, and Q4. Record the BBH-Boolean and BBH-Mixed task lists and their score-to-run associations under Configuration D04.

\begin{table*}[!t]
\centering
\caption{Source partitions and local evaluation settings. Actual evaluated membership follows the corresponding experiment records.}
\label{tab:benchmark_manifest}
\small
\setlength{\tabcolsep}{4pt}
\renewcommand{\arraystretch}{1.12}
\begin{tabular}{@{}p{0.30\textwidth}p{0.62\textwidth}@{}}
\toprule
Data sources & Evaluation partition \\
\midrule
\href{https://github.com/hendrycks/test}{MMLU}~\cite{hendrycks2020measuring} and \href{https://huggingface.co/datasets/TIGER-Lab/MMLU-Pro}{MMLU-Pro}~\cite{wang2024mmlu} & Official test split of each source \\
\href{https://huggingface.co/datasets/allenai/ai2_arc}{ARC-Challenge}~\cite{clark2018think} & Official Challenge test split of 1,172 problems \\
\href{https://github.com/suzgunmirac/BIG-Bench-Hard}{BBH-Boolean and BBH-Mixed}~\cite{suzgun2023challenging} & Boolean-only and mixed-configuration settings from the same benchmark. Candidate pools contain 100 and 240 problems, respectively. The main comparisons and ablations use the first 10 Boolean and first 30 Mixed problems, respectively. \\
\href{https://aclanthology.org/2021.tacl-1.21/}{StrategyQA}~\cite{geva2021did} & Official test partition through the source evaluation \\
\href{https://github.com/openai/grade-school-math}{GSM8K}~\cite{cobbe2021training} & Released test split \\
\href{https://github.com/openai/simple-evals/blob/main/math_eval.py}{MATH-500} & Released 500-problem evaluation subset of MATH~\cite{hendrycks2021measuring} \\
\href{https://hotpotqa.github.io/}{HotpotQA}~\cite{yang2018hotpotqa} & Official distractor development split \\
\href{https://github.com/StonyBrookNLP/musique}{MuSiQue}~\cite{trivedi2022musique} & Official MuSiQue-Ans development split, version 1.0 \\
\href{https://aclanthology.org/N19-1246/}{DROP}~\cite{dua2019drop} & Released development split \\
\href{https://github.com/openai/human-eval}{HumanEval}~\cite{chen2021evaluating} & Complete released evaluation set \\
\href{https://github.com/google-research/google-research/blob/master/mbpp/README.md}{MBPP}~\cite{austin2021program} & Original test partition with task IDs 11 to 510 \\
\href{https://github.com/jind11/MedQA}{MedQA}~\cite{jin2021disease} & Official English four-option test split \\
\href{https://github.com/pubmedqa/pubmedqa}{PubMedQA}~\cite{jin2019pubmedqa} & Expert-labeled PQA-L test membership from the source release \\
\href{https://github.com/czyssrs/FinQA}{FinQA}~\cite{chen2021finqa} & Released public test partition in test.json \\
\href{https://github.com/patronus-ai/financebench}{FinanceBench}~\cite{islam2023financebench} & Complete public annotated sample of 150 problems \\
\href{https://huggingface.co/datasets/allenai/sciq}{SciQ}~\cite{welbl2017crowdsourcing} & Official test split \\
\href{https://github.com/mandyyyyii/scibench}{SciBench}~\cite{wang2023scibench} & The ten main-branch text evaluation files, totaling 580 problems \\
\href{https://huggingface.co/datasets/casehold/casehold}{CaseHOLD}~\cite{zheng2021does} & Official test split of the all configuration \\
\bottomrule
\end{tabular}
\end{table*}

Split interpretation. HotpotQA releases distractor and fullwiki development sets, while MuSiQue provides separate Ans and Full variants. The table fixes the variant used for each local evaluation. StrategyQA's authors identify their derived train/dev files as a split of the official training data\footnote{\href{https://github.com/eladsegal/strategyqa}{StrategyQA source split description}.}. Use its official test partition for source test evaluation. Record each BBH setting's task membership and aggregation weights in its manifest.

\noindent\textbf{Fixed source versions.}
ARC denotes ARC-Challenge with all 1,172 test problems. We use the publisher's \texttt{allenai/ai2\_arc} release at revision \texttt{210d026f}. SciBench uses the ten text evaluation files from main commit \texttt{e14e0ca3}, totaling 580 problems\footnote{\href{https://github.com/mandyyyyii/scibench/tree/e14e0ca3d3db0f02928a18d435b9c3d15656a7c5/dataset/original}{SciBench fixed text release}.}. Files with suffix \texttt{\_sol} contain worked examples and are excluded from this evaluation set. Use textbook file, row, source ID, and input digest to distinguish repeated SciBench IDs.

The BIG-Bench Hard source is commit \texttt{9ee07bd4} of the authors' repository. Its complete release contains 6,511 problems in 27 files representing 23 conceptual tasks\footnote{\href{https://aclanthology.org/2023.findings-acl.824.pdf}{BBH source tasks and original evaluation}.}. We use two local evaluation settings from this benchmark, named BBH-Boolean and BBH-Mixed. For BBH-Boolean, the preparation script reads examples in a preset task order and stops after 100 records, yielding the first 100 examples of \texttt{boolean\_expressions}. For BBH-Mixed, the script follows a fixed order of 27 task configurations and reads one example from each configuration per pass until 240 records are collected, giving eight or nine examples per configuration. These numbers describe candidate pools. The main comparisons and ablations use the first ten Boolean problems and first 30 Mixed problems. The latter cover all 27 configurations, with two problems for each Logical Deduction configuration and one for every other configuration. Both cohorts are selected deterministically and use seeds 11, 22, and 33. Identify each problem by its task name and within-task index because Mixed source identifiers repeat across tasks. Boolean uses a sequential construction template and basic exact matching. Mixed uses a multi-branch template and answer normalization before exact matching. Average correctness over the submitted problems within each run. These cohort sizes and repeat counts apply to the main comparisons and ablations. The former Full label indicated broad configuration coverage. The two settings also use different construction templates, so their score difference combines differences in task composition, sampling, and construction conditions. Full source revisions and candidate-pool provenance are recorded in \texttt{protocol/official\_source\_versions.json} and \texttt{protocol/bbh\_setting\_provenance.json}.

\subsection{Application Requirement Construction}
\label{app:benchmark_requirements}

\noindent\textbf{Original questions as requirements.}
We use the original question directly as the application requirement. The wording of the problem specifies the objective from which the constructor derives tasks and dependencies. For datasets with multiple-choice questions, reading passages, tables, or programming specifications, the corresponding source fields define the problem inputs and requested output. We retain the association between each question and its source record throughout construction and evaluation.

\noindent\textbf{An example requirement.}
The second record of the released GSM8K test file provides a concrete example\footnote{\href{https://github.com/openai/grade-school-math/blob/master/grade_school_math/data/test.jsonl}{GSM8K source test file}.}. Its original question is reproduced below.
\begin{quote}
\raggedright
A robe takes 2 bolts of blue fiber and half that much white fiber. How many bolts \mbox{in total does it take}
\end{quote}
We use this question as the requirement for an individual construction run. The constructor determines the operations and their dependencies from the quantities and requested total in the question.

\noindent\textbf{Construction and execution inputs.}
Table~\ref{tab:benchmark_fields} separates construction inputs, evaluator records, and application outputs. Every constructor receives the permitted problem fields, component registry, domain resources, and execution environment. For our Graph Harness Construction Framework, we derive a Task Dependency Graph and Selected Component Graphs as intermediate records. We use these records to synthesize the final Harness and its component associations. The benchmark executor runs the returned application to obtain the task prediction.

\begin{table}[!t]
\centering
\caption{Instance information and Graph Harness outputs.}
\label{tab:benchmark_fields}
\small
\setlength{\tabcolsep}{3pt}
\renewcommand{\arraystretch}{1.14}
\begin{tabular}{@{}p{0.35\columnwidth}p{0.59\columnwidth}@{}}
\toprule
Item & Content \\
\midrule
Problem identity & Source dataset and problem identifier \\
Requirement & Objective, supplied inputs, output requirements, and application constraints \\
Execution inputs & Materials and resources available to the application \\
Component registry & Available agents and tools with capability and interface descriptions \\
Domain resources & Supplied domain concepts, constraints, and heuristics \\
Execution environment & Common runtime, tool permissions, and resource budgets \\
Evaluator records & Reference output and scoring assets retained by the evaluator \\
Generated graph & The generated Harness and its component associations \\
\bottomrule
\end{tabular}
\end{table}

\noindent\textbf{Instance-specific construction.}
We construct a Harness separately for every problem. Within one application type, related questions can require similar operations and therefore exhibit common structural patterns. At the same time, the supplied materials and requested outputs determine the operations and dependencies relevant to each question. The medical and legal examples in Fig.~\ref{fig:domain_case_studies} illustrate domain differences in component roles, resource connections, and review operations. The benchmark evaluates the applications obtained from these individual construction runs.

Input-access policy. Supply the original question and its permitted source context, choices, tables, or function signature to construction and execution. Store reference answers, worked solutions, and held-out scoring tests with the evaluator. For MBPP, preserve the distinction between example tests allowed by the source prompt and tests used for scoring. The data loader applies these field selections before preparing the construction request.

\subsection{Component Registry and Domain Resources}
\label{app:benchmark_registry}

\noindent\textbf{Component descriptions.}
The registry contains callable agents and tools. Each component has a capability description, an input interface, and an output interface, as defined in Sec.~\ref{sec:problem_formulation}. Capability descriptions record declared operations. Interface descriptions specify the information accepted and returned by the callable entry. Each construction method receives these records through the same registry interface. In Graph Harness construction, we use the descriptions to rank candidates and use the declared contracts to validate bindings and component invocations.

\noindent\textbf{Domain information.}
The framework also uses domain ontologies, constraints, and heuristics when constructing the Domain Semantic Profile. These resources describe the professional concepts and operation rules supplied for the target application. Following Fig.~\ref{fig:framework}, we organize the current tasks into the Task Requirement Profile, component operations into the Component Operation Profile, and domain information into the Domain Semantic Profile. We combine the three profiles when constructing execution nodes and their transitions. For each problem, the constructor uses this information to determine its execution organization.

\noindent\textbf{Resources across experiments.}
The main comparisons use the same component ecosystem and execution environment across methods. The scalability experiment changes the size of that ecosystem, while the perturbation experiment changes requirements, component information, or resource availability. Each of these settings therefore needs an explicit record of the components and resources available in that condition. The detailed baseline configurations and perturbation operations belong to the experimental settings, while the benchmark description identifies \mbox{the resources to which those operations apply}.

Registry composition. Use a pool of 20 components comprising six general agents, eight domain agents, and six tools. General agents perform retrieval, reasoning, parsing, verification, code generation, and reporting. Domain agents cover analysis and review in medicine, finance, science, and law. Tools provide document search, document reading, table extraction, arithmetic calculation, Python execution, and unit normalization. Each entry records an identifier, capability description, required input fields, output fields, callable endpoint, and resource dependencies. 

Domain resources. Associate each domain with a versioned resource bundle containing domain terminology, operation constraints, and review criteria. Record the document or knowledge-source identifiers, permitted retrieval scope, and version of every bundle. Use the same pool and resource bundle across compared methods for a given problem. For scalability and perturbation settings, record the exact modified component and resource sets. 

\subsection{Evaluation Protocol}
\label{app:benchmark_evaluation}

\noindent\textbf{Construction and execution.}
For each problem, a construction method returns an application and its callable associations under the common input and resource contract. We check its schema, task assignments, required bindings, and control definitions before execution. We then run the admitted application through the common executor and score its prediction with the source evaluator. Different task and component arrangements can satisfy this contract. We evaluate each admitted application through its own execution output. The evaluation configuration fixes the dataset variant, parser, and scorer revision.

\noindent\textbf{Evaluation entries and score scale.}
An entry $b$ fixes the source revision, variant, problem list, input conditions, primary metric, parser, and failure policy. Its ordered reference records $\mathbf y_b$ are identical across methods and outer runs. The parser maps each final output to its source answer format or the abstention symbol $\bot$. We evaluate all $N_b$ predictions from one run together. The adapter expresses the source metric as $\mathcal{M}_b\in[0,1]$, dividing percentage outputs by 100 when necessary. Equation~\ref{eq:evaluation} converts the resulting run metrics to the reported percentage scale exactly once.

\noindent\textbf{Metrics defined per problem.}
Let $g_b(\hat y,y)\in[0,1]$ be the normalized source score for a parsed prediction $\hat y$ and reference $y$. For accuracy, $g_b$ is one for an accepted correct answer and zero otherwise. For answer F1, $g_b$ is the source evaluator's score after its normalization, reference matching, and span alignment. Set $g_b(\bot,y)=0$. With run indices suppressed, the dataset metric is
\begin{equation}
\mathcal{M}_b(\mathbf o,\mathbf y)=\frac{1}{N_b}\sum_{i=1}^{N_b}g_b(P_b(o_i),y_i),
\label{eq:decomposable_metric}
\end{equation}
where $P_b$ is the fixed output parser. This definition covers accuracy, answer F1, and correctness of one submitted program. Every frozen problem contributes one term. Failed construction or execution contributes an abstention.

\noindent\textbf{Class macro-F1.}
For CaseHOLD, let $L_b=\{0,1,2,3,4\}$ be the five source labels. For each outer run, compute $\mathrm{TP}_c$, $\mathrm{FP}_c$, and $\mathrm{FN}_c$ from the complete predictions for class $c$. The metric is
\begin{equation}
\mathcal{M}_b(\mathbf o,\mathbf y)=\frac{1}{5}\sum_{c\in L_b}\frac{2\mathrm{TP}_c}{2\mathrm{TP}_c+\mathrm{FP}_c+\mathrm{FN}_c},
\label{eq:class_macro_f1}
\end{equation}
with a zero contribution when a class denominator is zero. An abstention adds a false negative to its true class and no false positive to the five scored classes. We recompute these counts separately for every run before averaging its dataset scores. The same construction applies to secondary class macro-F1 metrics with their own fixed label sets.

\noindent\textbf{Choice and label prediction.}
For multiple-choice tasks, accuracy compares the selected option with the source label. PubMedQA additionally provides macro-F1 over its answer labels\footnote{\href{https://github.com/pubmedqa/pubmedqa/blob/master/evaluation.py}{PubMedQA evaluation code}.}. CaseHOLD uses macro-F1 over its five option labels in the official implementation\footnote{\href{https://github.com/reglab/casehold/blob/main/multiple_choice/run_multiple_choice.py}{CaseHOLD evaluation code}.}. Both BBH settings retain task-specific extraction when aggregating outputs.

\noindent\textbf{BIG-Bench Hard task aggregation.}
The BBH main comparisons and ablations average problem correctness within each run. For a separately configured conceptual-task summary, give represented conceptual tasks equal weight. Let $G_b$ contain the conceptual tasks represented in entry $b$, and let $J_{bg}$ contain their included task files. Each set is nonempty and fixed by the evaluation manifest. If $a_{bj}$ is the exact-match accuracy on the selected problems from file $j$, the normalized score for one run is
\begin{equation}
\mathcal{M}_b=\frac{1}{|G_b|}\sum_{g\in G_b}\frac{1}{|J_{bg}|}\sum_{j\in J_{bg}}a_{bj}.
\label{eq:bbh_group_score}
\end{equation}
In the full release, Logical Deduction and Tracking Shuffled Objects each contain three files. Their six file weights are $1/69$, and each of the remaining 21 file weights is $1/23$. For an evaluation subset, we retain its represented groups and files across every method, run, and perturbation condition. Each file accuracy uses all selected problems from that file, including abstentions as incorrect answers.

\noindent\textbf{Reading and mathematical answers.}
HotpotQA evaluates answer exact match and token F1, with separate supporting-fact and joint measures\footnote{\href{https://github.com/hotpotqa/hotpot/blob/master/hotpot_evaluate_v1.py}{HotpotQA evaluation code}.}. MuSiQue provides answer and supporting-context scores, while its Full variant adds answerability-related evaluation\footnote{\href{https://github.com/StonyBrookNLP/musique}{MuSiQue evaluation instructions}.}. DROP computes exact match and its span-aligned F1 using the released evaluator\footnote{\href{https://github.com/allenai/allennlp-reading-comprehension/blob/master/allennlp_rc/eval/drop_eval.py}{DROP evaluation code}.}. These output fields identify what the application must return for each metric. For GSM8K, we score the extracted numerical answer. For MATH-500, we evaluate the final mathematical answer using the selected \mbox{MATH evaluator and its equivalence rules}.

\noindent\textbf{Scientific and financial answers.}
For SciBench, the released checker applies \texttt{math.isclose} with relative tolerance 0.05 and absolute tolerance zero\footnote{\href{https://github.com/mandyyyyii/scibench/blob/e14e0ca3d3db0f02928a18d435b9c3d15656a7c5/eval/eval_zero.py}{SciBench numerical evaluation}.}. For finite predicted and target values $a$ and $y$, it accepts
\begin{equation}
|a-y|\leq 0.05\max(|a|,|y|).
\label{eq:scibench_numeric_match}
\end{equation}
We retain the source answer extraction and scientific-notation preprocessing. The dataset score is the fraction correct over all 580 evaluation problems, which weights textbooks by their problem counts. FinQA reports execution accuracy and program accuracy\footnote{\href{https://github.com/czyssrs/FinQA}{FinQA evaluation instructions}.}. Execution accuracy evaluates the numerical result of the predicted program, whereas program accuracy also evaluates the generated program. The output contract must therefore specify whether the application \mbox{returns a program, its result, or both}.

FinanceBench uses the answer-assessment rubric in its original study\footnote{\href{https://arxiv.org/html/2311.11944v1}{FinanceBench}, Sec.~4.1.}. Responses are assigned to correct, incorrect, or failure-to-answer categories. The rubric permits equivalent units and small rounding differences, and considers whether the stated calculation or reasoning agrees with the reference material. We use this rubric when judging the final response. Its application requires the question, reference answer, and \mbox{relevant report material at the evaluation stage}.

\noindent\textbf{Program correctness.}
For HumanEval and MBPP, we submit one final program per problem to the held-out tests. HumanEval uses its functional-correctness evaluator\footnote{\href{https://github.com/openai/human-eval}{HumanEval evaluation instructions}.}. MBPP retains its source prompt format and the example tests permitted in that prompt\footnote{\href{https://github.com/google-research/google-research/tree/master/mbpp}{MBPP evaluation instructions}.}. A run's pass@1 is the fraction of submitted final programs that pass their tests. Candidate programs and revisions generated within an application count toward its execution budget. Only the declared final program is submitted for task scoring.

\noindent\textbf{Aggregation and incomplete runs.}
We record construction outcome, execution outcome, and application performance separately. Construction succeeds when the returned application passes the declared schema and static checks. Execution terminates with completed or failed status, together with any timeout, interface error, or exhausted budget. We score the declared final output recorded within the allowed budget. A missing, invalid, or unscorable output becomes $\bot$. Each submitted problem therefore retains a prediction slot and its construction and execution outcomes.

Reporting policy. Evaluate the same frozen problems in three complete runs for the BBH main comparisons and ablations and five complete runs for the other settings, repeating construction and execution for every problem. Compute one normalized dataset metric from each run's full prediction set, then report its percentage mean and sample standard deviation. Construction and execution completion rates use separate denominators. For a cross-entry summary, list the included entries and give each distinct evaluation configuration equal weight. Two aliases with the same evaluation manifest and scoring contract identify one entry. Describe the aggregate as a mean of task scores. Use the output adapters in Configuration D05 and the repeat aggregation in Configuration D10.

\noindent\textbf{Efficiency measurements.}
The efficiency comparison measures Harness construction through task derivation, component selection, and Harness synthesis. For each submitted problem, we sum model input and output tokens from these stages and report their mean across problems. Runtime choice planning and the subsequent agent and tool operations belong to application execution and are recorded separately. Each dataset retains its task-specific performance metric when plotted against construction cost.

Measurement configuration. Start construction when the constructor receives the original problem and end when it returns an accepted Harness or a construction failure. Count model input and output tokens from task derivation, component selection, and Harness synthesis, including validation, repairs, and retries within these stages. For each problem, add its construction input and output tokens, then average these totals over the common set of submitted problems. Retain the construction costs of failed attempts in that mean. Record application execution, evaluator, embedding, and local-tool costs separately. Attribute model calls made inside a construction tool to the originating construction stage. Configuration D09 specifies caching, missing counters, and failure handling, with successful-construction summaries retaining their stated conditional denominator.

%% file: sec/B_theory.tex
\section{Theoretical Details and Analysis}
\label{app:theory}

We specify the reference construction algorithm for Sec.~\ref{sec:methodology} and analyze its task constraints, component contracts, and execution rules. We take the extracted requirements, coverage indicators, component descriptions, and accepted bindings as given inputs. We first give the construction details, then establish properties of the task graph and component scores and analyze contract preservation and finite execution. The execution rules state the implementation conditions for these properties.

\input{sec/B_reference_construction}

\newcounter{harnessclaim}[section]
\renewcommand{\theharnessclaim}{\thesection.\arabic{harnessclaim}}
\newcommand{\harnessclaim}[1]{\par\smallskip\refstepcounter{harnessclaim}\noindent\textbf{#1~\theharnessclaim.}}

\subsection{Task Graph Formulation}
\label{app:task_formulation}

\noindent\textbf{Requirements and input bindings.}
Let \mbox{$U_d^{\mathrm{req}}$} contain the extracted requirements with \mbox{$b_k=1$}. A selected task covers requirement \mbox{$u_k$} when its declared operation satisfies the predicate \mbox{$A_{ki}=1$}. For each mandatory input field, we also specify a binding to an external input or an output field of a predecessor task. A binding records the source field, target field, accepted type conversion, and semantic compatibility predicate. Multiple sources may jointly supply a structured input when their declared merge operation \mbox{produces all required fields}.

We distinguish explicit prerequisites \mbox{$D_d^{\mathrm{pre}}$} from selected producer relations \mbox{$D_d^{\mathrm{data}}$}. The indicators \mbox{$c_{ij}$} and \mbox{$b_{ij}$} identify these relations, respectively. An output supplying a mandatory input through an accepted \mbox{$\phi_{ij}$} binding creates a data relation. A task pair can carry both labels. Their union \mbox{$D_d$} defines the waiting graph through Eq.~\ref{eq:dependency}. External inputs are available before their consumers start.

All structural statements below take the extracted requirements, coverage indicators, selected task declarations, and accepted bindings as given. For input availability, assume that external inputs are available and predecessor tasks complete successfully with the outputs specified by their contracts. We use declared type, semantic, and domain predicates to validate those bindings. The source task scorer evaluates the final application output.

\harnessclaim{Proposition} \textit{Coverage and ordering of the extracted task graph.}
\label{prop:task_order}
Suppose that the selected tasks satisfy Eq.~\ref{eq:appendix_task_coverage}, every mandatory input has an accepted external or predecessor binding, and each dependency between selected tasks satisfies Eq.~\ref{eq:topological} with integer ranks in \mbox{$[1,N]$}. Then every requirement in \mbox{$U_d^{\mathrm{req}}$} has a selected covering task. The union waiting graph is acyclic, and a topological order respects its explicit prerequisites and producer dependencies. Under the availability and successful-completion assumptions above, each task has its declared inputs before execution.

\noindent\textbf{Proof.}
For a mandatory requirement, row $k$ of Eq.~\ref{eq:appendix_task_coverage} has $b_k=1$. Its nonnegative binary terms therefore include a selected task with \mbox{$A_{ki}=1$}. Next, every dependency edge imposes a rank increase of at least one. If a directed cycle contained \mbox{$q$} edges, adding its rank inequalities would give \mbox{$h_i\geq h_i+q$} for its starting task, contradicting \mbox{$q\geq1$}. Self-dependencies are excluded when constructing the graph.

Order the selected tasks topologically in the union waiting graph. A first task receives its mandatory inputs from the environment. Inductively, suppose that all earlier tasks have completed under their contracts. Each explicit prerequisite of the next task has then completed, and each selected producer has returned its declared outputs. The external inputs and those outputs supply the next task through its validated bindings. Repeating this argument establishes \mbox{the stated execution order.~$\square$}

\noindent\textbf{Task and control graphs.}
The topological ranks apply to the task waiting graph \mbox{$G_d^T$}. In the final Harness, a task can involve several execution nodes, while one node can contribute to several tasks through \mbox{$\kappa$}. We record task completion after its assigned operations produce the required outputs. Choice nodes carry an empty task assignment and route control. Budgeted feedback repeats operations after input and completion checks for the current versions.

\subsection{Component Compatibility and Selection}
\label{app:component_analysis}

\noindent\textbf{Scores and normalization.}
For a fixed task, abbreviate \mbox{$q_{ijr}$} as \mbox{$q_{jr}$} and write \mbox{$w_r=w_{ir}$}. Nonzero embedding vectors and a binary support indicator give \mbox{$q_{jr}\in[0,1]$}. Use nonnegative capability weights with \mbox{$\sum_r w_r=1$}. The coverage in Eq.~\ref{eq:coverage} then lies in \mbox{$[0,1]$}, giving all tasks a common scale.

We extend the coverage score to a component set \mbox{$A$} by taking the maximum score for each capability over components in \mbox{$A$}, and set the coverage of the empty set to zero. For the marginal calculation, let \mbox{$b_r(A)$} denote that maximum, with \mbox{$b_r(\varnothing)=0$}, and let \mbox{$[x]_+=\max(x,0)$}. Adding component \mbox{$c$} changes coverage by
\begin{equation}
\Delta_c(A)=\sum_r w_r[q_{cr}-b_r(A)]_+.
\label{eq:coverage_marginal}
\end{equation}

\harnessclaim{Lemma} \textit{Coverage over component sets.}
\label{prop:coverage_sets}
The extended coverage score is bounded, monotone, and submodular. In particular, for component sets \mbox{$A\subseteq B$} and a component \mbox{$c\notin B$}, the marginal gain satisfies \mbox{$\Delta_c(A)\geq\Delta_c(B)\geq0$}.

\noindent\textbf{Proof.}
Each capability maximum lies in \mbox{$[0,1]$}, which gives the stated bounds after multiplication by nonnegative weights and summation. Adding a component can only increase each maximum, proving monotonicity. For \mbox{$A\subseteq B$}, we have \mbox{$b_r(A)\leq b_r(B)$} for every capability. Therefore every positive-part term in Eq.~\ref{eq:coverage_marginal} for \mbox{$A$} is at least the corresponding term for \mbox{$B$}. Summing these inequalities proves diminishing marginal gains and hence submodularity.~\mbox{$\square$}

The lemma characterizes capability coverage over unordered component sets. For ordered paths, we check callable entries, required inputs, intermediate bindings, task outputs, and domain predicates. Each mandatory input has an accepted source from the task input or an earlier component output. Multiple sources require a declared deterministic merge. These hard checks define \mbox{$\mathcal{P}_i^{\mathrm{feasible}}$}. Score thresholds then define admissible paths, and a fixed-cardinality objective selects among them.

\noindent\textbf{Interface scores and costs.}
In Eq.~\ref{eq:field_compatibility}, \mbox{$h_f(f_o,f_r)\in\{0,1\}$} records the semantic field-acceptance decision. The continuous score \mbox{$s_f(f_o,f_r)\in[0,1]$} is mapped cosine similarity between field descriptions. Type compatibility permits direct assignment or a registered deterministic conversion. A binding passes when its type check and hard acceptance both succeed. Its continuous score ranks accepted matches.

For each required target field, take the maximum accepted \mbox{$\chi$} score and average these maxima with nonnegative unit-sum field weights to obtain \mbox{$\eta_e$}. Each mandatory field must also have a binding passing the hard checks. For a target with no required fields, set its score to one. The path mean \mbox{$\Gamma_i$} uses its scored edges and the single-component convention in Appendix~\ref{sec:component_selection}.

After full-path validation, freeze the finite family \mbox{$\mathcal{P}_i^{\mathrm{feasible}}$}. An empty family returns construction failure. Otherwise, let \mbox{$c_i^-$} and \mbox{$c_i^+$} be its minimum and maximum path costs. When they differ, normalize cost by
\begin{equation}
\bar C_i(p)=\frac{C_i(p)-c_i^-}{c_i^+-c_i^-}.
\label{eq:cost_scaling}
\end{equation}
When all candidate costs are equal, set normalized costs to zero. We compute utilities with these fixed extrema, then apply the capability and interface thresholds. Thus all compared paths use the same cost scale during a selection step. Define \mbox{$\mathcal{P}_i^{\mathrm{adm}}$} as the paths in \mbox{$\mathcal{P}_i^{\mathrm{feasible}}$} satisfying the capability threshold and every edge threshold. Since coverage, interface means, and normalized cost lie in \mbox{$[0,1]$}, the utility satisfies \mbox{$-\gamma\leq U_i(p)\leq\alpha+\beta$}.

\noindent\textbf{Fixed-cardinality path selection.}
For one task, freeze \mbox{$\mathcal{P}_i^{\mathrm{adm}}$} and let \mbox{$L_i=|\mathcal{P}_i^{\mathrm{adm}}|$}. Choose a fixed integer \mbox{$k_i$} with \mbox{$1\leq k_i\leq L_i$}. The family \mbox{$\mathcal F_i$} contains every \mbox{$k_i$}-element subset of these paths. Utilities are fixed independently of the selected subset. Write their utilities in descending order as \mbox{$u_{(j)}$} for \mbox{$1\leq j\leq L_i$}. Configuration C05 retains \mbox{$k_i=2$} admissible paths and fails if fewer are available.

\harnessclaim{Proposition} \textit{Optimal fixed-cardinality path selection.}
\label{prop:path_selection}
For this fixed task and admissible family, any subset of the \mbox{$k_i$} highest-utility paths satisfies the selection rule in Eq.~\ref{eq:path_optimization}. The maximum total utility is
\begin{equation}
U_i^\star=\sum_{j=1}^{k_i}u_{(j)}.
\label{eq:top_path_mean}
\end{equation}

\noindent\textbf{Proof.}
Take any admissible subset of size \mbox{$k_i$}, and let \mbox{$v_{(j)}$} be its \mbox{$j$}th largest utility. The \mbox{$j$}th largest value in a subset is at most the \mbox{$j$}th largest value in the entire family, so \mbox{$v_{(j)}\leq u_{(j)}$}. Summing over the selected positions gives \mbox{$\sum_{j=1}^{k_i}v_{(j)}\leq U_i^\star$}. The top \mbox{$k_i$} paths attain this bound, with ties permitting several optimal subsets.~\mbox{$\square$}

The fixed count retains a small set of validated alternatives, which may share components. The proposition concerns one task's fixed admissible family and cardinality constraint. During Harness assembly and runtime admission, we enforce shared component budgets and check resource conflicts across tasks.

\subsection{Graph Harness Execution Semantics}
\label{app:execution_semantics}

\noindent\textbf{Global state.}
To describe concurrent execution, we distinguish a node's control state \mbox{$s_i$} from the global runtime state
\begin{equation}
\sigma_t=(A_t,X_t,M_t,J_t,b_t,\xi_t).
\label{eq:runtime_state}
\end{equation}
Here, \mbox{$A_t$} is the finite multiset of active control tokens, \mbox{$X_t$} stores typed data with producer and version identifiers, and \mbox{$M_t$} contains current task-completion records. Tokens identify nodes, bindings, and branch scopes. Task records specify consumed input versions and accepted outputs. Records \mbox{$J_t$} distinguish fork invocations. The vector \mbox{$b_t$} contains the persistent per-edge counters \mbox{$b_e$} and the total invocation counter \mbox{$b_\Sigma$}. The status \mbox{$\xi_t$} is running, completed, or failed.

\noindent\textbf{Node admission.}
An execution node is enabled when a token reaches it in the correct branch scope, the required predecessor records are current, its mandatory inputs have accepted bindings, and its domain predicates hold. Each activation invokes one registered operation on a component in \mbox{$\mu_d(v_i)$} that passes the current operation and input checks. Before committing a return, the executor rechecks consumed input versions and validates the output schema, types, and declared domain predicates. A stale return follows the recovery policy. Accepted outputs enter \mbox{$X_t$}, and replacing a value invalidates records that depend on its previous version. A task enters \mbox{$M_t$} when all required operations have completed, output checks pass, and its consumed input versions remain current. These updates form one committed state transition. A failed invocation follows an explicit failure or recovery transition.

\noindent\textbf{Routes in a merged component graph.}
The graph union in Eq.~\ref{eq:selected_graph} retains the selected paths. For each path, we preserve a validation record specifying its component order, mandatory bindings, intermediate interfaces, and task output contract. Recombining edges can produce an additional route, which must pass the same full-path checks before admission. A change in route or input bindings requires an updated validation record for the remaining operations.

\noindent\textbf{Sequential and alternative transitions.}
A sequential transition forwards its control token and data after its condition is satisfied. The target then passes the prerequisite and binding checks before invocation. At a choice node, \mbox{$N_{\mathrm{valid}}(\sigma_t)$} contains successors whose transition, input, and route checks hold. The planner takes an argmax in Eq.~\ref{eq:choice}, resolving equal scores by node identifier, and activates one successor. With an empty eligible set, the executor follows an enabled recovery or failure transition. A running configuration fails if work and enabled transitions are both absent.

\noindent\textbf{Parallel transitions.}
A fork creates a finite set of branch tokens with a shared invocation identifier. Its join accepts completion records only from those branches and that identifier. The completion product in Eq.~\ref{eq:parallel} enables the join when all branch indicators are one. The join consumes its branch records once and publishes the merged outputs. Concurrent branches use separate output names or a declared deterministic merge operation. The merge validates the required fields before the subsequent node is admitted. Failed branches follow the specified \mbox{recovery or failure policy}.

\noindent\textbf{Feedback transitions.}
Initialize every \mbox{$b_e$} and the total counter \mbox{$b_\Sigma$} to zero once for the complete Harness invocation. Let \mbox{$N_e^{\max}$} bound traversals of edge \mbox{$e$}, and let \mbox{$N_\Sigma^{\max}$} bound all feedback traversals in that invocation. A traversal requires both checks
\begin{equation}
b_e<N_e^{\max},\qquad b_\Sigma<N_\Sigma^{\max}.
\label{eq:feedback_budget}
\end{equation}
The stopping predicate \mbox{$g_e(\sigma_t)$} must also be false. Commit both counter increments and control transfer atomically, preserving \mbox{$b_\Sigma=\sum_e b_e$}. Counters persist across repeated entries and fork invocations. When any feedback guard fails, follow the declared exit. Replacing a bound value invalidates task, branch, and join records that depend directly or transitively on its old version. Tokens with stale bindings repeat their admission checks, and a return to a fork creates a fresh branch identifier. Successful operations restore task records with each accepted value's producer and version.

\subsection{Properties of Harness Construction}
\label{app:harness_properties}

\noindent\textbf{Execution contract.}
Assume that external inputs pass their declared schema and type checks and every accepted component return passes its declared schema, type, and domain checks. The executor enforces the admission, route, version, branch-scope, and join rules in Sec.~\ref{app:execution_semantics}. Its task-to-node mapping covers every required task. The completion guard requires current records for every required task and current versions of all requested final output fields.

\harnessclaim{Proposition} \textit{Preservation of execution-contract invariants.}
\label{prop:execution_preservation}
Under this contract, every admitted component invocation satisfies its declared prerequisites and mandatory input bindings. Every current task-completion record refers to valid input versions and outputs that passed the declared checks. A completed Harness has current completion records for all required tasks and every requested final output in the current store.

\noindent\textbf{Proof.}
Proceed by induction over committed runtime transitions. Initially, the store contains external inputs that passed schema and type checks, and no task-completion records. An invocation starts after its admission checks accept the predecessor records, branch scope, and mandatory bindings. Before publishing a return, the executor rechecks consumed input versions and validates output schemas, types, and domain predicates. Publishing a replacement invalidates records that depend on the previous value before they can be reused. Task completion is committed after every required operation finishes, its output checks pass, and its consumed input versions remain current.

A sequential or choice transition forwards control through the same target checks. A fork creates scoped tokens, and a join publishes data only after the matching branch records and merge checks succeed. Thus these transitions preserve the input and output properties. A feedback update applies the same invalidation rule to records depending directly or transitively on replaced versions. Failure commits retain the validated store and follow the declared recovery or terminal transition. The induction therefore also applies after repeated execution. Finally, the completed-status guard requires current records for every required task and all final output fields, \mbox{giving the last claim.~$\square$}

The graph-level task mapping assigns required operations to execution nodes. The completion guard checks their fulfillment using current task records. The proposition establishes invariants of the declared schema, type, domain-predicate, and version checks. The benchmark's source task scorer separately evaluates the final application answer.

\begin{table*}[!t]
\centering
\caption{Construction costs under the stated finite-candidate model. Path-generation cost and model inference are accounted for separately.}
\label{tab:construction_complexity}
\small
\setlength{\tabcolsep}{4pt}
\renewcommand{\arraystretch}{1.17}
\begin{tabular}{@{}p{0.35\textwidth}p{0.25\textwidth}p{0.32\textwidth}@{}}
\toprule
Operation & Cost & Conditions \\
\midrule
Requirement coverage checks & \mbox{$O(KN)$} & Given coverage indicators \\
Task dependency checks & \mbox{$O(N^2f^2)$} & Pairwise task and field comparisons \\
Topological ordering & \mbox{$O(N+|D_d|)$} & Given the selected task graph \\
Component capability scoring & \mbox{$O(m_iR_i)$} & Cached cosine and support scores \\
Candidate interface checks & \mbox{$O(a_if^2)$} & Given candidate connections \\
Path scoring & \mbox{$O(\Lambda_i(R_i+1))$} & Cached component and edge scores \\
Fixed-cardinality path selection & \mbox{$O(L_i\log L_i)$} & Proposition~\ref{prop:path_selection} \\
Selected graph union & \mbox{$O(\Lambda_i)$} & Expected time with hash-indexed identifiers \\
Harness assembly & \mbox{$O(|V_d|+|E_d|)$} & Given node associations and transition specifications \\
\bottomrule
\end{tabular}
\end{table*}

\noindent\textbf{Conditions for finite execution.}
Assume a finite Harness, a finite initial token set, and finite fan-out at every fork. Removing budgeted feedback edges leaves an acyclic control graph. Both per-edge and total feedback counters are persistent for the entire invocation, monotone, and bounded by finite nonnegative integers. Admission checks, merges, component calls, and planner calls complete or fail in finite time, with finitely many calls and retries within each activation. Under fair scheduling, every continuously enabled transition or activation is eventually processed, as is each available return. With no pending call or enabled transition, the completion guard determines success or failure.

\harnessclaim{Proposition} \textit{Finite execution.}
\label{prop:finite_execution}
Under these conditions and the transition semantics above, every Harness invocation reaches completed or failed status after finitely many control events and component calls.

\noindent\textbf{Proof.}
Let \mbox{$n$} count control nodes and let \mbox{$E_d^{\mathrm{Fb}}$} contain the feedback edges. Set \mbox{$L=\min\{N_\Sigma^{\max},\sum_{e\in E_d^{\mathrm{Fb}}}N_e^{\max}\}$}. Every traversal increments its edge counter and the total counter once, so both limits bound the total feedback events. At most \mbox{$L$} can occur in the invocation. Along a token lineage, each segment without feedback follows the remaining acyclic graph and visits at most \mbox{$n$} nodes. Its depth is therefore bounded by
\begin{equation}
D\leq(L+1)n+L.
\label{eq:lineage_bound}
\end{equation}

Represent token production by a forest rooted at the initial tokens, assigning each join continuation to one consumed token for counting. Equation~\ref{eq:lineage_bound} bounds its depth, and finite fork fan-out bounds the number of children at each event. A forest with finite roots, depth, and branching has finitely many token events. Scoped joins consume their records once. Each activation has finitely many calls and checks, all ending in finite time. The fair scheduler processes the finite work and returns its terminal status.~\mbox{$\square$}

The remaining acyclic graph bounds progress between feedback traversals, and persistent budgets bound their repetition. The terminal status can be completed or failed. A completed invocation additionally satisfies the declared contracts in Proposition~\ref{prop:execution_preservation}.

\subsection{Computational Complexity}
\label{app:construction_complexity}

\noindent\textbf{Cost model.}
We separate graph operations from model inference and external tool calls. Let \mbox{$K$} and \mbox{$N$} count extracted requirements and candidate tasks. For task \mbox{$i$}, let \mbox{$m_i$} count candidate components, \mbox{$a_i$} the directed connections actually checked, and \mbox{$R_i$} capability requirements. Each task or component interface has at most \mbox{$f$} input fields and \mbox{$f$} output fields. Let \mbox{$L_i^{\mathrm{gen}}$} count generated paths and \mbox{$\Lambda_i$} sum their component counts. The \mbox{$L_i=|\mathcal{P}_i^{\mathrm{adm}}|$} admissible paths are sorted, then the first \mbox{$k_i$} are retained. Table~\ref{tab:construction_complexity} counts compatibility tests and cached score lookups at unit cost.

\noindent\textbf{Candidate graphs and path generation.}
Candidate connections satisfy \mbox{$a_i\leq m_i(m_i-1)$}. Their interface checks cost \mbox{$O(a_if^2)$}, giving \mbox{$O(m_i^2f^2)$} for all ordered pairs. Path generation has a separate cost determined by the search procedure. An acyclic graph can have exponentially many unrestricted paths, as layered two-way branches illustrate. Configuration C05 bounds each generated path to four components, complete candidates to 64, and prefix expansions to 512. The corresponding search cost is counted separately from cached path scoring and sorting.

\noindent\textbf{Validation and inference costs.}
Linear Harness assembly assumes that node associations, bindings, merges, and transition conditions are given. Full-route and completion validation incur the costs of their binding checks and verification procedures. Constructing embeddings and evaluating semantic predicates may require model calls. With embedding dimension \mbox{$d$} and cached norms, computing cosine scores costs \mbox{$O(m_iR_id)$}. The table's capability row instead uses cached cosine values. We account separately for registry retrieval, model inference, and external tools. Runtime work depends on the selected routes, parallel branches, and feedback traversals. The efficiency experiments report \mbox{mean construction tokens per application}.

%% file: sec/B_reference_construction.tex
\subsection{Reference Construction Algorithm}
\label{app:reference_construction}

For the construction below, we fix entry $b$, individual problem $i_0$, method $m$, and outer run $u$. We suppress these fixed indices on intermediate graphs and use local indices for tasks and components. Domain subscript $d$ identifies the resources used for this individual problem.

\subsubsection{Task Structure Derivation}
\label{sec:task_structure_derivation}

\noindent\textbf{Structured Task Extraction.}
Given the current problem's application requirement \mbox{$R_{bi_0}$}, we use an LLM to decompose it into atomic requirements \mbox{$U_d=\{u_k\}_{k=1}^{K}$} and extract \mbox{$N$} candidate tasks. Each candidate task is represented as \mbox{$\tilde{\tau}_i=(z_i,g_i,I_i,O_i)$}, where \mbox{$z_i\in\{0,1\}$} indicates task selection, \mbox{$g_i$} specifies its target and operation, and \mbox{$I_i$} and \mbox{$O_i$} denote its required inputs and expected outputs. Let \mbox{$A_{ki}\in\{0,1\}$} indicate whether task \mbox{$\tilde{\tau}_i$} fulfills requirement \mbox{$u_k$}. Let $\mathbf{z}=(z_i)$, $\mathbf{b}=(b_k)$, and $\mathbf{A}_d=(A_{ki})$. We select tasks under the element-wise coverage constraint
\begin{equation}
\mathbf{z}\in\{0,1\}^{N},\quad\mathbf{A}_d\mathbf{z}\succeq\mathbf{b},\quad\mathbf{A}_d\in\{0,1\}^{K\times N},
\label{eq:appendix_task_coverage}
\end{equation}
where \mbox{$b_k\in\{0,1\}$} marks mandatory extracted requirements. The model supplies \mbox{$U_d$}, coverage judgments \mbox{$A_{ki}$}, and task declarations. We validate coverage of the extracted mandatory requirements and form the selected task set \mbox{$T_d$}.

\noindent\textbf{Dependency Modeling and Validation.}
For distinct selected tasks \mbox{$\tau_i,\tau_j\in T_d$}, define binary indicators \mbox{$c_{ij}$}, \mbox{$b_{ij}$}, and \mbox{$\phi_{ij}(o,r)$}. Set \mbox{$c_{ij}=1$} for an execution-order prerequisite. For required input \mbox{$r\in I_j$}, \mbox{$\phi_{ij}(o,r)=1$} marks its selected producer \mbox{$o\in O_i$} after type and semantic checks. Set \mbox{$b_{ij}=1$} exactly when at least one such binding is adopted. Their joint waiting relation is
\begin{equation}
a_{ij}=1-(1-c_{ij})(1-b_{ij}),\quad i\neq j,
\label{eq:dependency}
\end{equation}
with \mbox{$D_d=D_d^{\mathrm{pre}}\cup D_d^{\mathrm{data}}$}. A pair belongs to \mbox{$D_d^{\mathrm{pre}}$} exactly when \mbox{$c_{ij}=1$}, and to \mbox{$D_d^{\mathrm{data}}$} exactly when \mbox{$b_{ij}=1$}. Both labels are retained when both indicators equal one.

Every mandatory input must have a validated external or predecessor binding. For pairs with \mbox{$z_i=z_j=1$}, assign integer ranks \mbox{$h_i\in[1,N]$} and impose
\begin{equation}
h_j \geq h_i+1-N(1-a_{ij}).
\label{eq:topological}
\end{equation}
These ranks make the task waiting graph \mbox{$G_d^T=(T_d,D_d)$} acyclic. The final Harness separately permits bounded feedback among control nodes.

\subsubsection{Component Selection}
\label{sec:component_selection}

\noindent\textbf{Capability Matching.}
Given \mbox{$G_d^T$}, we retrieve candidate components from the registry \mbox{$\mathcal{C}=\{c_j\}_{j=1}^{M}$}, where each component \mbox{$c_j=(p_j,I_j^c,O_j^c)$} specifies its capability and input/output interfaces. For each task \mbox{$\tau_i$}, we construct a candidate component graph \mbox{$G_i^C=(V_i^C,E_i^C)$}, whose vertices represent retrieved components. Let \mbox{$\mathcal{P}_i^C$} be a finite generated family of candidate paths. Each component is evaluated against the task capability requirements \mbox{$K_i=\{k_{ir}\}_{r=1}^{R_i}$}. Its matching score is
\begin{equation}
q_{ijr}=v_{ijr}
\frac{1+\cos(\mathbf{e}_{p_j},\mathbf{e}_{k_{ir}})}{2},
\label{eq:matching}
\end{equation}
where \mbox{$v_{ijr}\in\{0,1\}$} is a semantic support judgment. Embeddings \mbox{$\mathbf{e}_{p_j}$} and \mbox{$\mathbf{e}_{k_{ir}}$} rank eligible components by their descriptions. Invocation validity follows the interface checks below. A path's capability coverage is
\begin{equation}
Q_i(p)=\mathbf{w}_i^{\top}\left[\max_{c_j\in p}q_{ijr}\right]_{r=1}^{R_i},
\label{eq:coverage}
\end{equation}
where $\mathbf{w}_i=(w_{ir})_{r=1}^{R_i}$ is a column vector of nonnegative weights summing to one. The bracketed vector contains per-requirement maxima as a column.

\noindent\textbf{Interface Compatibility and Path Optimization.}
For interface field \mbox{$f=(n,t,s)$} with name, type, and semantic description, define
\begin{equation}
\chi(f_o,f_r)=
\mathbb{I}[t_o\preceq t_r]\,h_f(f_o,f_r)\,s_f(f_o,f_r),
\label{eq:field_compatibility}
\end{equation}
where \mbox{$t_o\preceq t_r$} permits assignment or a registered conversion. Binary \mbox{$h_f$} records semantic acceptance, and \mbox{$s_f\in[0,1]$} ranks accepted matches. Each mandatory field requires an accepted source or validated deterministic merge. Weighted field scores give \mbox{$\eta_e$}. Their mean over path edges is \mbox{$\Gamma_i(p)$}, set to one for a single component. Callable, binding, task-output, and domain checks define \mbox{$\mathcal{P}_i^{\mathrm{feasible}}\subseteq \mathcal{P}_i^C$}. Invocation and communication costs \mbox{$C_i(p)$} use extrema fixed on this family before threshold filtering.

For each task, we define path utility
\begin{equation}
U_i(p)=\boldsymbol{\lambda}^{\top}[Q_i(p),\Gamma_i(p),-\bar{C}_i(p)]^{\top},
\label{eq:appendix_path_utility}
\end{equation}
where $\boldsymbol{\lambda}=[\alpha,\beta,\gamma]^{\top}$, \mbox{$\alpha,\beta,\gamma\geq0$} and \mbox{$\alpha+\beta+\gamma=1$}. Let \mbox{$\mathcal{P}_i^{\mathrm{adm}}$} contain feasible paths passing the capability and edge thresholds. We retain \mbox{$k_i=2$} paths and define \mbox{$\mathcal F_i$} as all \mbox{$k_i$}-element subsets of \mbox{$\mathcal{P}_i^{\mathrm{adm}}$}. Admissibility requires
\begin{equation}
Q_i(p)\geq\delta_i,\quad\eta_e\geq\epsilon_i,\quad |S|=k_i.
\label{eq:path_constraints}
\end{equation}
We select an admissible set by maximizing
\begin{equation}
S_i^*\in\arg\max_{S\in\mathcal{F}_i}\sum_{p\in S}U_i(p).
\label{eq:path_optimization}
\end{equation}
Construction fails if fewer than \mbox{$k_i$} paths are admissible. The selected paths form the Selected Component Graph
\begin{equation}
G_i^{C*}=\left(\bigcup_{p\in S_i^*}V(p),\bigcup_{p\in S_i^*}E(p)\right).
\label{eq:selected_graph}
\end{equation}
Retained paths preserve validated bindings. Recombined routes pass the same full-path checks before admission. Assembly and execution enforce shared resource limits.

\subsubsection{Graph Harness Construction}
\label{sec:graph_harness_construction}

\noindent\textbf{Profile Construction.}
We use a model-assisted structured transformation to form \mbox{$P_d=(P_d^T,P_d^C,P_d^D)$}. The Task Requirement Profile stores objectives, input/output requirements, dependencies, and producer bindings. The Component Operation Profile stores selected operations, interface contracts, and retained paths. We collect domain rules and constraints in the Domain Semantic Profile, giving
\begin{equation}
P_d=\Psi\left(G_d^T,\{G_i^{C*}\}_{\tau_i\in T_d},K_d\right),
\label{eq:profiles}
\end{equation}
where \mbox{$\Psi$} preserves each source identifier.

\noindent\textbf{Execution Node Construction and Association.}
Based on these profiles, the synthesizer proposes control nodes
\begin{equation}
v_i=(s_i,\ell_i,I_i^v,O_i^v,\rho_i),
\end{equation}
where \mbox{$s_i$} is the execution state, \mbox{$\ell_i$} the operation, \mbox{$I_i^v$} and \mbox{$O_i^v$} the input and output, and \mbox{$\rho_i\in\{\mathrm{Exec},\mathrm{Choice}\}$} the node type. Each execution activation invokes one registered operation on an associated component after contract checks. Choice nodes route control and have empty task assignments. The associations satisfy
\begin{equation}
\emptyset\neq\mu_d(v_i)\subseteq C_i,
\quad v_i\in V_d^{\mathrm{Exec}}.
\label{eq:appendix_node_association}
\end{equation}
The validator checks task coverage, registry membership, required bindings, preserved waiting relations, and typed control conditions. Source task obligations and component contracts remain fixed during synthesis. Each \mbox{$C_i$} contains compatible components retained in the selected graphs.

\noindent\textbf{Edge Typing and Harness Synthesis.}
Each transition retains its type, activation condition, and passed data
\begin{equation}
e_{ij}=(v_i,\tau_{ij},\gamma_{ij},v_j,m_{ij}),
\label{eq:edge}
\end{equation}
\looseness=-1
Here, \mbox{$\tau_{ij}\in\{\mathrm{Seq},\mathrm{Alt},\mathrm{Par},\mathrm{Fb}\}$} specifies control type, \mbox{$\gamma_{ij}$} activation, and \mbox{$m_{ij}$} transferred data.

Sequential transitions require current predecessor completions and valid inputs. For an alternative transition in global state \mbox{$\sigma_t$}, the eligible set \mbox{$N_t$} contains successors passing condition, binding, and route checks. The planner chooses the successor through
\begin{equation}
v_{t+1}\in
\arg\max_{v_j\in N_t}\pi_d(v_j\mid\sigma_t,N_t),
\label{eq:choice}
\end{equation}
where \mbox{$N_t=N_{\mathrm{valid}}(\sigma_t)$} and \mbox{$\pi_d$} scores eligible successors. Identifier-based ties make selection deterministic.

For a fork invocation with \mbox{$B$} branches, the join guard is
\begin{equation}
\gamma_{\mathrm{join}}(t)=
\prod_{b=1}^{B}d_b(t),
\label{eq:parallel}
\end{equation}
where \mbox{$d_b(t)\in\{0,1\}$} marks current branch completion. At \mbox{$\gamma_{\mathrm{join}}(t)=1$}, the join consumes that invocation's records once and validates the merged outputs.

Each feedback traversal requires remaining edge and invocation budgets \mbox{$N_e^{\max}$} and \mbox{$N_\Sigma^{\max}$}, with \mbox{$g_e(\sigma_t)$} false. It atomically consumes one unit of both budgets. Otherwise, control follows the exit. Updated bindings transitively invalidate completion records depending on earlier versions.

Finally, we synthesize all execution nodes and typed transitions into the domain-specific Graph Harness \mbox{$H_d=(V_d,E_d)$}, which specifies execution operations, component associations, and runtime control flow.

%% file: sec/C_implementation.tex
\section{Implementation Details}
\label{app:implementation}

This appendix gives the procedures connecting Sec.~\ref{sec:methodology} to the execution contract in Appendix~\ref{app:theory}. Construct a separate Harness for each original problem and retain its inputs, intermediate outputs, validation order, and runtime updates. Save model versions, component implementations, search settings, and prompts with these records.

\subsection{Data Records and Component Interfaces}
\label{app:implementation_records}

\noindent\textbf{Problem records.}
A construction request identifies the source dataset, source split, problem identifier, original question, and permitted input resources. Associated passages, tables, answer choices, and programming specifications retain their source identifiers. The input adapter selects fields according to the access policy of Appendix~\ref{app:benchmark_requirements}. Reference answers and scoring assets belong to the evaluator record. The adapter policy is \mbox{specified in Configuration C01}.

\par\smallskip\noindent\textbf{Configuration C01.} 
Use a fixed adapter that retains the original question, source identifier, permitted context, and output format. For choice tasks, include every option in the source order and keep the correct-option indicator with the evaluator. For SciQ, assemble and shuffle the correct option and distractors with the data seed, then remove the correct-option indicator from the request. For HotpotQA, MuSiQue, DROP, and PubMedQA, retain the permitted passages or abstracts and their identifiers. Supporting-fact labels, answerability labels, worked solutions, and reference answers remain in the evaluator record.

For FinQA, include the question, table, and associated report text. For FinanceBench, expose the question and the permitted report collection through document identifiers. HumanEval receives the source prompt and entry point. MBPP receives its source prompt and allowed example tests. Evaluation tests are stored separately. Save the source-to-adapter field map, option ordering, resource versions, and input digest for each problem.
\par\smallskip

\noindent\textbf{Task records.}
For each candidate task, store its identifier, operation, required inputs, expected outputs, selection flag, and links to extracted requirements. Each input field records its name, type, required status, and intended meaning. Store prerequisite and data-production labels separately for every task pair, preserving both when they apply. Each adopted binding identifies the source field, target field, accepted conversion, and semantic predicate result. Their union records the tasks required before the consumer starts.

\noindent\textbf{Component and path records.}
A registry entry contains a stable component identifier, a capability description, typed input and output fields, and a callable descriptor. Implementation and resource versions distinguish entries that expose the same capability through different interfaces. Candidate paths store ordered component identifiers, accepted bindings, capability scores, edge scores, and estimated costs. Record membership in the generated, structurally feasible, and admissible path families separately. The selected paths retain their identifiers and validation records when merged into the Selected Component Graph. The registry and domain-resource records are \mbox{specified in Configuration C02}.

\par\smallskip\noindent\textbf{Configuration C02.} 
Instantiate the 20-entry registry specified in Appendix~\ref{app:benchmark_registry}. Its six general agents perform retrieval, reasoning, parsing, verification, code generation, and reporting. Its eight domain agents provide analysis and review for medicine, finance, science, and law. Its six tools provide document search, document reading, structured-table access, arithmetic, Python execution, and unit conversion. All agent roles use the backbone in Configuration D01. The table tool reads supplied structured tables, and document tools use the problem's frozen collection.

Assign each entry a unique identifier and semantic version. Store its capability statements, required fields, returned fields, callable entry point, resource identifiers, and invocation budget. A domain agent can declare several operations, each with its own schema. Separate safety and review nodes can thus invoke declared operations of one medical-review component. Use UTF-8 files for domain terminology, operation rules, and review criteria. Hash their contents together with the registry manifest. Any human-review node uses an external callback that records the reviewer decision and its time. A model-generated review retains the identity of its model component.
\par\smallskip

\noindent\textbf{Harness records.}
An execution node records its registered operation, component associations, task assignments, input bindings, and output declarations. Its task assignment \mbox{$\kappa(v_i)$} can overlap those of other execution nodes. Thus several nodes may jointly support one task, and one node may support several tasks. Choice nodes have empty task assignments and registry associations. Their planner selects among eligible successors. Each execution activation invokes one registered operation on one associated component that passes its current contract checks.

A transition records its source, target, type, activation condition, and transferred data. Fork and join records specify branch identifiers and the merge operation. Feedback records specify a stopping predicate, a per-edge budget, and an exit transition. The invocation record additionally stores the total feedback budget. A serialized Harness identifies its entry configuration, final output fields, and completion condition. These records connect the node assignments to the data and control checks used during execution.

\noindent\textbf{Invocation records.}
For each call, the executor records the problem identifier, node activation identifier, component version, branch scope, input versions, start event, end event, and outcome. For each execution activation, also record the configured component-selection policy identifier, the eligible associated components, and the component actually invoked. The selected component belongs to \mbox{$\mu_d(v_i)$} and passes its operation and binding checks for the current inputs. Accepted outputs receive producer and version identifiers. A task-completion record links its consumed inputs to the outputs that satisfy its contract. Model-call and resource counters retain the activation that incurred them. Serialization formats and the runtime backend are \mbox{specified in Configuration C03}.

\par\smallskip\noindent\textbf{Configuration C03.} 
Serialize problem records, registry entries, component paths, and Harnesses as UTF-8 JSON. Store events in append-only JSONL files. Use a Python 3.11 event loop and a local SQLite transaction log, with at most four external calls in flight. Serve ready transitions and node activations through a FIFO queue, using identifier order for simultaneous arrivals. Dispatch calls outside the state transaction. On return, commit the input-version recheck, output publication, and completion update in one transaction. Index events by run, activation, branch scope, and call identifier. Measure durations with a monotonic clock and align events by UTC timestamps.
\par\smallskip

\subsection{Construction Procedure}
\label{app:construction_procedure}

\noindent\textbf{Procedure C.1.}
Given a problem request, registry, domain resources, and finite budgets, this procedure returns a validated Harness or records construction failure.

\begin{enumerate}
\item Read the original question and allowed resources. Initialize a new construction record for this problem.
\item Extract atomic requirements and candidate tasks. Record requirement coverage and task selection.
\item Derive and label prerequisite relations and adopted input bindings. Check coverage, producer availability, type compatibility, and a topological ordering of their union.
\item For each selected task, retrieve candidate components and construct connections with their interface records.
\item Generate finite candidate paths. Check their complete input and output contracts, freeze the hard-feasible family, and compute its cost extrema and utility scores.
\item Apply capability and interface thresholds to form \mbox{$\mathcal{P}_i^{\mathrm{adm}}$}. Retain the \mbox{$k_i=2$} highest-utility admissible paths for each task. If fewer than two remain, return construction failure. Keep their identifiers when merging graphs and validate each additional route before admission.
\item Form the task, component, and domain profiles through a model-assisted transformation that preserves source identifiers. Propose control nodes, their component associations, and typed transitions.
\item Check task coverage, registry membership, interfaces, dependencies, input bindings, branch scopes, merges, feedback budgets, and completion guards.
\item Return the accepted Harness. A failed check may trigger another proposal within the remaining construction budget. Exhausting that budget returns a construction failure with its failed-check record.
\end{enumerate}

\noindent\textbf{Task extraction and validation.}
The extraction output follows the task schema above. The validator resolves every task and resource identifier, checks coverage of the extracted requirements, and examines each mandatory input binding. For every dependency from Eq.~\ref{eq:dependency}, it retains the explicit prerequisite, adopted producer binding, or both. It computes a topological ordering of this union. Any unresolved producer, incompatible field, or cycle becomes a validation result for the next permitted proposal. The coverage indicators and extracted requirements are the inputs to the structural checks in Appendix~\ref{app:task_formulation}. Their relation to the retained dependency labels is \mbox{specified in Configuration C09}.

\par\smallskip\noindent\textbf{Configuration C09.} 
Record explicit prerequisites through \mbox{$c_{ij}$}. For each mandatory input, list sources with compatible object, role, unit, and type, then select an accepted binding. Set \mbox{$\phi_{ij}(o,r)=1$} for an adopted binding from output \mbox{$o$} to required input \mbox{$r$} that passes these checks. Set \mbox{$b_{ij}=1$} when task \mbox{$i$} supplies at least one such field to task \mbox{$j$}. Use \mbox{$a_{ij}$} for their union in Eq.~\ref{eq:dependency}, retaining both labels. External bindings create no task edge. For multi-source inputs, record all producers and a deterministic merge.

Use the proposed task order to restrict sources to external inputs and earlier tasks. A missing producer or dependency cycle triggers task repair followed by complete revalidation. Preserve required bindings when repairing the order. Component alternatives implementing the same task belong to that task's component graph. 
\par\smallskip

\begin{samepage}
\noindent\textbf{Candidate generation and selection.}
The path generator uses a declared finite search space or search budget and records the order and bindings of each candidate route. First check callable entries, mandatory inputs, intermediate interfaces, task outputs, and domain predicates. An empty structurally feasible family returns construction failure. Otherwise freeze \mbox{$\mathcal{P}_i^{\mathrm{feasible}}$} and compute its cost extrema, normalized costs, and utilities. Equal path costs give normalized cost zero. Apply capability and edge thresholds to form \mbox{$\mathcal{P}_i^{\mathrm{adm}}$}, retaining the same extrema. A task with fewer than \mbox{$k_i=2$} admissible paths returns construction failure. Capability and field weights follow Appendix~\ref{app:component_analysis}. The encoder, match predicates, and thresholds are \mbox{specified in Configuration C04}.\par
\end{samepage}

\par\smallskip\noindent\textbf{Configuration C04.} 
Encode capability statements and field descriptions with \texttt{text-embedding-3-large} using 3072 dimensions\footnote{\href{https://developers.openai.com/api/docs/guides/embeddings}{Embedding model documentation}.}. Normalize nonzero vectors and cache them by text digest and encoder identifier. Use uniform, unit-sum weights for capability requirements and mandatory interface fields. The backbone evaluates each capability against the component's declared operation and returns a binary support flag. For a field pair, it returns semantic acceptance \mbox{$h_f(f_o,f_r)\in\{0,1\}$} and the matched object, data role, unit, and granularity. Type validation accepts direct assignment or an allowed registered deterministic converter. A failed type check rejects the binding.

Map the cosine of field-description embeddings to the unit interval to obtain \mbox{$s_f(f_o,f_r)$}. Equation~\ref{eq:field_compatibility} combines this continuous score with the type and semantic acceptance gates. A rejected type or semantic check gives \mbox{$\chi=0$}. For every mandatory input, require a producer or external binding that passes both hard checks. Multiple producers require a declared deterministic merge. Cache judgments by descriptions, encoder, judge, and prompt version. Set minimum capability coverage to 0.70 and minimum edge compatibility to 0.80. Use utility weights of 0.50 for capability coverage, 0.30 for interface compatibility, and 0.20 for the normalized cost penalty. Each mandatory binding is checked individually before the aggregate edge score is used. An edge supplying only part of the target input can fail the edge threshold. If a wrapper forwards the question or context references, declare those fields in its output schema. For path ranking, assign a planning cost of 1 per model component, 0.1 per local tool, and 0.01 per KiB transferred. Record actual tokens and latency \mbox{separately in the efficiency evaluation}.
\par\smallskip

For the fixed-cardinality objective in Proposition~\ref{prop:path_selection}, sort \mbox{$\mathcal{P}_i^{\mathrm{adm}}$} by utility and retain the first \mbox{$k_i=2$} paths. Resolve equal utility by a fixed order on path identifiers. Shared component resources and global concurrency budgets are checked during Harness assembly and execution. Candidate search rules, path length, weights, and proposal budgets are \mbox{specified in Configuration C05}.

\par\smallskip\noindent\textbf{Configuration C05.} 
Retrieve at most eight components per task. Generate simple paths containing at most four components, stopping after 64 complete candidates or 512 prefix expansions. Order prefixes by higher current capability coverage, lower raw planning cost, and then component identifier. Check the full input and output contract for each completed candidate. Freeze the nonempty structurally feasible family before computing cost extrema and normalized utilities. Apply capability and edge thresholds to obtain admissible paths, keeping those extrema fixed.

Fix the retained path count at \mbox{$k_i=2$}. Select the two highest-utility admissible paths as in Proposition~\ref{prop:path_selection}. Fewer than two admissible paths produces a construction failure. Each model-assisted construction stage permits one initial proposal and two repairs within the shared budget. Limit a proposal to 24 tasks and 64 control nodes. Record the termination reason and selected path IDs per task.
\par\smallskip

\noindent\textbf{Profile and Harness synthesis.}
To form the three profiles, we use a model-assisted transformation that retains source identifiers. We place objectives, input and output requirements, labeled dependencies, and producer bindings in \mbox{$P_d^T$}. We place selected components, callable descriptors, interface contracts, and retained paths in \mbox{$P_d^C$}, and supplied domain concepts and rules in \mbox{$P_d^D$}. The synthesizer receives these records and returns a Harness proposal. Static validation checks task coverage, registry membership, unchanged component interfaces, mandatory dependencies, required input sources, and typed transition records.

The selected component graph may admit routes assembled from several retained paths. Validate the full input bindings, intermediate interfaces, task output contract, and domain predicates of each added route before admitting it. For feedback, deleting the budgeted feedback edges must leave an acyclic control graph. Record the required tasks and final output fields checked at successful termination. These obligations connect construction to the executor assumptions of Propositions~\ref{prop:execution_preservation} and~\ref{prop:finite_execution}.

\subsection{Execution Procedure}
\label{app:execution_procedure}

\noindent\textbf{Procedure C.2.}
Given an accepted Harness and its permitted inputs, initialize the global state of Eq.~\ref{eq:runtime_state}. Its vector \mbox{$b_t$} includes the per-edge counters \mbox{$b_e$} and the total counter \mbox{$b_\Sigma$}. All branch scopes, completion records, and counters belong to this problem invocation.

\begin{enumerate}
\item Initialize the typed input store and entry tokens. Set every \mbox{$b_e$} and \mbox{$b_\Sigma$} to zero once and activate the invocation.
\item Process returned or timed-out calls. Accept returns after checking input versions and output contracts.
\item Commit the version recheck and accepted-output publication atomically. Record task completion once the assigned operations and output checks pass.
\item Advance ready transitions in FIFO order with identifier-based ties. Apply binding, branch-scope, and budget checks before changing the active token set.
\item For each ready execution token, apply its configured component-selection policy to the associated components passing the operation and route checks. Record the policy identifier, selected component, inputs, and deadline. Atomically mark the activation as in flight and dispatch one registered operation. If none is eligible, follow the declared recovery or failure transition. Consume the activation once on completion and create successor tokens through its transitions.
\item Continue while calls are pending or transitions can advance. With no pending call and no enabled transition, return completed if the final-output and task-completion conditions hold, and failed otherwise.
\end{enumerate}

\noindent\textbf{Runtime component selection.}
At task entry, rank the retained validated routes by the construction utility in Eq.~\ref{eq:path_utility}, resolving ties by route identifier. Select the first route whose required external inputs, predecessor records, and resource reservations are available. Fix that route's component associations for the task activation. Invoke components in the route's declared order and recheck bindings and versions before each call. If no route is eligible, follow the declared recovery or failure transition. A transient call failure follows Configuration C06's retry policy. A route change requires a new task activation through a declared recovery transition and a fresh check of the full route contract. Record the selected route and component at dispatch. Identify this policy as \texttt{ranked-route-v1}.

\noindent\textbf{Versions and concurrent returns.}
A replacement of a bound value invalidates completion records that consume its earlier version, together with dependent task, branch, and join records. Calls already in progress retain their consumed version identifiers. Their returns pass the same version check before publication. A stale return follows the declared recovery policy and retains its resource counters. The executor thus checks output validity while accounting for all dispatched calls.

\noindent\textbf{Branch scopes and feedback.}
Each fork activation creates a new fork-invocation identifier for its branches. A join consumes only the completion records of that activation and publishes outputs through the declared merge operation. For feedback edge \mbox{$e$}, require a false stopping predicate, \mbox{$b_e<N_e^{\max}$}, and \mbox{$b_\Sigma<N_\Sigma^{\max}$}. Commit both budget checks, both counter increments, source-token consumption, and successor-token creation atomically. The counters persist across repeated entries. If either budget is exhausted or the stopping predicate holds, use the declared exit. Returning to a fork creates a fresh branch scope and retains every feedback counter. A recovery transition that creates a cycle is subject to the same bounded-feedback condition.

\noindent\textbf{Termination and recovery records.}
Every planner call, component call, validator, and merge must end or return a failure under its declared execution limit. The scheduler eventually serves every activation that remains ready. Recovery actions identify their target, supplied data, and remaining budget. Record the failure stage and final invocation status with the run. These requirements supply the finite-call and fair-scheduling premises of Proposition~\ref{prop:finite_execution}. Candidate timeouts, counter updates, and \mbox{recovery rules are \mbox{specified in Configuration C06}}.

\par\smallskip\noindent\textbf{Configuration C06.} 
Use the limits in Configuration D02. Set the per-edge limit to \mbox{$N_e^{\max}=3$} and the invocation limit to \mbox{$N_\Sigma^{\max}=16$}. Initialize their counters once per Harness invocation. Every committed feedback traversal increments both counters in the transaction that consumes its source token and creates its successor. A component or planner call has a 60-second limit. Local tools have a 10-second limit, with a five-second limit for a Python test process. A transient service failure permits one retry using the same invocation key. Before retrying a state-changing tool, require an idempotent wrapper or an accepted status query.

Discard stale returns from the accepted data store and retain their resource counters. Recovery follows a declared transition within the remaining global budget. Conflicting branch outputs require a registered deterministic merge operation. Exhausted feedback, call, token, or time limits select an exit or failed terminal state according to the node contract. For reproducibility, save the timeout, retry, merge, and counter policy with the Harness.
\par\smallskip

\subsection{Prompt and Configuration Records}
\label{app:prompt_records}

\noindent\textbf{Structured prompt interfaces.}
Each model-assisted stage receives the original requirement and the records relevant to that stage. The task parser returns atomic requirements, task declarations, coverage links, labeled dependencies, and bindings. Capability and interface evaluators return candidate identifiers, scores, and predicate outcomes. The Harness synthesizer returns nodes, associations, typed edges, and output conditions. At a choice node, the planner receives global state \mbox{$\sigma_t$} and eligible successor identifiers. The executor validates the returned records before dispatch.

\begin{samepage}
\noindent\textbf{Instruction for task extraction.}
Given the original question and permitted input fields, identify the operations needed to produce the requested output. Assign a unique identifier to each extracted requirement and task. For every task, return its objective, mandatory inputs, expected outputs, and covering requirement identifiers. For every dependency, return its prerequisite or producer relation and the corresponding field bindings. Express the result using the supplied task schema.\par
\end{samepage}

\noindent\textbf{Instruction for Harness synthesis.}
Given the task, component, and domain profiles, construct execution nodes and their component associations. Return the sequential, alternative, parallel, and feedback transitions required to connect these operations. For every transition, specify its activation condition and transferred data. For parallel execution, return scoped fork and join records with the merge operation. For feedback, return the stopping predicate, traversal limit, and exit transition. Identify the entry configuration, completion records, and \mbox{required final output fields}.

Response schemas, planner selection, and parsing or repair rules follow \mbox{Configuration C07}. Stage models and version records follow \mbox{Configuration C08}. Save these records alongside the resource manifests for the experimental setup.

\par\smallskip\noindent\textbf{Configuration C07.} 
Use five prompt interfaces. The task parser receives the unchanged question and allowed fields, then returns requirements, tasks, coverage links, producer candidates, prerequisites, and a proposed order. The capability evaluator returns a component identifier, support flag, and supporting capability statement. The field evaluator returns semantic acceptance \mbox{$h_f$}, the proposed converter, object identity, unit, and field-role match. Validate converter registration and calculate the continuous score \mbox{$s_f$} from the cached field embeddings. The synthesizer receives the three profiles and returns the node, association, transition, merge, feedback, and completion records specified above.

At a choice node, supply global state \mbox{$\sigma_t$} and successors that pass transition, input, and route checks. Require one score in the unit interval for each eligible identifier. Subtract the largest score before applying softmax, then select the maximum with identifier-based tie breaking. This is the deterministic argmax policy in Eq.~\ref{eq:choice}. Parse one complete JSON object and validate its schema, references, enums, duplicate identifiers, and global constraints. A repair prompt uses the permitted inputs, rejected output, validation errors, and remaining budget of its originating stage.

Batch capability and field judgments in groups of at most 16 relations, retaining individual identifiers. Charge every batch and repair to its originating budget. A missing or truncated judgment triggers a smaller batch or a recorded failure. Reuse a cached registry-only judgment only when the fields, model, and prompt versions match its saved record.
\par\smallskip

\par\smallskip\noindent\textbf{Configuration C08.} 
Use \texttt{OpenMAS-GApp v0.1} as the interface specification and \texttt{gh.required.v1} as the configuration identifier. Use D01's backbone for model-assisted stages, C04's encoder and dimensions, and D02's temperatures and response limits.

Version the common, task, capability, field, synthesis, and choice templates separately as v0.1. Use record schema v0.3 with these templates. Build each prompt from the common instruction, stage template, schema, and permitted instance fields. Hash the fixed template, schema, and generation settings with SHA-256. For every call, save the digest, configuration identifier, requested and returned model identifiers, and registry version. Store templates and their version map with the configuration.
\par\smallskip

%% file: sec/D_settings.tex
\section{Experimental Settings}
\label{app:experimental_settings}

We organize the experimental settings by the comparison being performed, the resources available to each method, and the unit used for scoring and cost accounting. The benchmark inputs and source evaluation rules are described in Appendix~\ref{app:benchmark}. The construction and execution procedures are described in Appendix~\ref{app:implementation}. These records link results to their configurations and \mbox{evaluated problem identifiers}.

\subsection{Reported Scope}
\label{sec:reported_scope}

Q1 covers all 20 evaluation settings. The Q2 and Q4 summaries cover five datasets, while Q3 covers eleven evaluation settings. The complete suite contains 19 benchmark datasets organized into 20 evaluation settings across domains and application types. Q7 reports a Graph Harness lead in 33 of 36 dataset-condition pairs. The additional comparison in Figure~\ref{fig:component_scale_success_trend.pdf}(a) covers GSM8K, MuSiQue, MedQA, FinanceBench, and SciBench. It compares AFlow, ADAS, GPTSwarm, AgentSquare, and Graph Harness under the settings in Appendix~\ref{app:official_baselines}. Table~\ref{tab:extended_official_baselines} reports the measured scores for these four methods and Graph Harness across the evaluation suite.

\subsection{Models, Environment, and Data Records}
\label{app:environment_settings}

\noindent\textbf{Shared comparison conditions.}
The main comparison design uses the same language-model backbone, component ecosystem, problem requirements, and execution environment across methods. The construction-ablation design further specifies matched decoding, execution budgets, random seeds, and scoring rules. Each run therefore needs a configuration identifier linking those shared settings to its method-specific construction procedure. The model and environment settings are listed in Table~\ref{tab:environment_records}.

\begin{table*}[!t]
\centering
\caption{Environment and model records.}
\label{tab:environment_records}
\small
\setlength{\tabcolsep}{4pt}
\renewcommand{\arraystretch}{1.12}
\begin{tabular}{@{}p{0.28\textwidth}p{0.44\textwidth}p{0.20\textwidth}@{}}
\toprule
Record & Required configuration & Current entry \\
\midrule
Language-model backbone & Provider, model identifier, revision, and serving endpoint & DeepSeek-V4-flash \\
Stage-specific models & Task parser, semantic evaluator, synthesizer, and planner & Shared backbone \\
Decoding & Temperature, sampling parameters, output limits, and seeds & Configuration D02 \\
Embedding and matching & Encoder revision, normalization, predicates, and thresholds & 3072-dimensional encoder \\
Runtime environment & Operating system, hardware, memory, and software versions & 16 vCPU, 64 GiB \\
Serving configuration & Local or remote inference, concurrency, and batching & Remote API, four calls \\
Data manifest & Dataset release, split, selected identifiers, and sample count & Configuration D04 manifests \\
Registry and resources & Component and domain-resource versions for each run & 20 versioned components \\
\bottomrule
\end{tabular}
\end{table*}

\par\smallskip\noindent\textbf{Configuration D01.} 
Use DeepSeek-V4-flash as the shared backbone for extraction, semantic judgments, synthesis, planning, and agent operations. Submit requests through the configured inference endpoint. Log requested and returned model IDs, serving revision, call date, response ID, and any backend fingerprint.
\par\smallskip

\par\smallskip\noindent\textbf{Configuration D02.} 
Set task extraction, semantic judgments, synthesis, and discrete planning to temperature zero. Set agent execution to 0.2 and top-p to one, with frequency and presence penalties zero. Limit extraction and synthesis responses to 4096 tokens, semantic responses to 2048, choice responses to 512, and final agent responses to 2048. Use data seed 42 and root seeds 11, 22, and 33 for the BBH main comparisons and ablations. Use root seeds 11, 22, 33, 44, and 55 for the other reported settings. BBH cohort selection is deterministic and independent of these run seeds. Forward a supported model seed and save the full request.

For each submitted problem, cap construction at 120 seconds, 12 model calls, and 16,000 input-plus-output tokens. Cap execution at 300 seconds, 32 model calls, 64 local tool calls, and 32,000 model tokens. Include repairs and retries in these limits. Apply the per-call, feedback, and recovery limits in Configuration C06. Apply these shared caps and record actual resource consumption.
\par\smallskip

\par\smallskip\noindent\textbf{Configuration D03.} 
Use a client host with Ubuntu 22.04, Python 3.11, 16 virtual CPU cores, and 64 GiB memory. Model inference is remote, with no client GPU allocation. Use the standard event loop and transactional store in Configuration C03. Enforce a global limit of four external calls. For per-problem latency, submit one problem at a time while retaining its allowed within-Harness parallelism. Save the actual machine identifier, operating-system build, dependency lockfile, API-client version, network region, and execution dates in the environment record.
\par\smallskip

\noindent\textbf{Dataset and metric identity.}
A data manifest identifies the exact problems evaluated in each comparison. We follow the official benchmark partitions and evaluation protocols. Table~\ref{tab:experiment_scope} records the datasets used in the main figures and ablation table. Each manifest retains the source release, official split, problem identifiers, and scoring procedure described in Appendix~\ref{app:benchmark_sources}.

The official source scorer determines the task metric, answer parser, and accepted output form. In particular, CaseHOLD provides macro-F1, while FinanceBench uses the assessment rubric described in Appendix~\ref{app:benchmark_evaluation}. Configuration D05 fixes the metrics and output parsing. D04 records the BBH cohorts and instance-mean scoring.

\par\smallskip\noindent\textbf{Configuration D04.} 
Use the official evaluation partition for each benchmark, as specified in Appendix~\ref{app:benchmark_sources}. Preserve its released question identifiers and required input fields. Freeze the source revision, evaluator, and configuration before comparing methods. Give the paradigm, ablation, and orchestration studies identifiers Q2, Q3, and Q4. Share problems and permitted resources across methods.

Construct the complete method once for each problem and outer-run identifier, and reference that same Full run in Q2, Q3, and Q4. Save the task graph and selected components for the Q4 representation adapters and the sequential ablation. Give a manifest its digest over the ordered source revision, split, task, problem identifier, input digest, and scoring contract. A shared run also has the same model, prompt, registry, budget, and repeat identifiers. Reuse an existing Full record only when those identifiers match.

Retain official partition membership for the named sources. For the two local BBH settings, retain the recorded evaluation cohort from the corresponding candidate pool. Save each BBH setting's source-task list and aggregation weights in its manifest. The reported-result register links the 55 Table~\ref{tab:ablation} entries and 44 Figure~\ref{fig:efficiency_preview} coordinates to their source files. Attach the corresponding release identifiers and run records when preparing the final artifact.

ARC uses the complete ARC-Challenge test partition with 1,172 problems. StrategyQA, MBPP, PubMedQA, FinQA, and SciQ use their complete evaluation cohorts of 490, 500, 500, 1,147, and 1,000 problems, respectively. SciBench uses the 580 problems in the ten text evaluation files of main commit \texttt{e14e0ca3}. The BIG-Bench Hard source is commit \texttt{9ee07bd4}. Its configuration manifests retain the selected task files and conceptual groups for Eq.~\ref{eq:bbh_group_score}. BBH-Boolean and BBH-Mixed identify its boolean-only and mixed-configuration settings. Candidate pools contain 100 and 240 records, respectively. The main comparisons and ablations use the first 10 Boolean and first 30 Mixed problems with seeds 11, 22, and 33. The Mixed cohort covers 27 configurations, including two problems for each Logical Deduction configuration and one for every other configuration. Use composite task-name and within-task-index identifiers. Boolean uses a sequential template and Mixed a multi-branch template. Fix source releases and file digests separately from model and runtime settings.
\par\smallskip

\par\smallskip\noindent\textbf{Configuration D05.} 
Use accuracy for MMLU, MMLU-Pro, ARC-Challenge, StrategyQA, MedQA, SciQ, and PubMedQA. Retain PubMedQA macro-F1 as a secondary metric. Use five-class macro-F1 for CaseHOLD and also retain accuracy. For HotpotQA, MuSiQue-Ans, and DROP, use answer F1 as the primary metric and retain the source exact-match score. Score supporting-context outputs separately when the output contract requests them. For the BBH main comparisons and ablations, use basic exact matching for Boolean and BBH answer normalization followed by exact matching for Mixed. Average correctness over the submitted problems within each run. Eq.~\ref{eq:bbh_group_score} defines a separate conceptual-group summary when explicitly requested.

Extract each prediction from its declared final-answer field or final-answer line, using one parser per dataset for all methods. Normalize choice labels and yes/no/maybe labels to their source forms. A missing or conflicting final answer is unscorable. For GSM8K, compare the extracted numerical answer after the source normalization. For SciBench, use Eq.~\ref{eq:scibench_numeric_match} and the fixed release's answer extraction and scientific-notation preprocessing. Average correctness over all evaluated problems. Require a program and its executed value for FinQA, and report execution accuracy with program accuracy retained separately. For HumanEval and MBPP, submit one final program per problem and report pass@1 using held-out tests.

For MATH-500, use the OpenAI simple-evals mathematical-answer evaluator\footnote{\href{https://github.com/openai/simple-evals/blob/main/math_eval.py}{MATH evaluation implementation}.}. Set its internal repeat count to one for each outer run and save the equivalence grader's model identifier and decisions. Charge grader calls to evaluation. For FinanceBench, label answers correct, incorrect, or unanswered using the source rubric, and report the fraction marked correct. One grader scores all answers with method identities hidden. A second grader reviews ambiguous outputs and a fixed 20\% sample chosen before scoring. Save both judgments and the adjudicated decision.

Freeze the parser, scorer repository revision, dependencies, primary metric, and normalization rules before evaluation. Save raw outputs, parsed predictions, references, and scoring outputs under their problem and run identifiers. Each entry's adapter returns $\mathcal{M}_b\in[0,1]$. Calculate CaseHOLD macro-F1 from the complete prediction set of each outer run. Convert normalized run metrics to percentages once when computing the reported mean. The metric assignments and record fields are stored in \texttt{evaluation\_plan.json} and \texttt{record\_schemas.json}.
\par\smallskip

\begin{table*}[!t]
\centering
\caption{Displayed experiment scope and protocol entries. Dataset partitions follow the official benchmark protocols.}
\label{tab:experiment_scope}
\small
\setlength{\tabcolsep}{4pt}
\renewcommand{\arraystretch}{1.13}
\begin{tabular}{@{}p{0.25\textwidth}p{0.67\textwidth}@{}}
\toprule
Comparison & Displayed datasets or applications \\
\midrule
Construction paradigms and orchestration & MMLU-Pro, FinanceBench, MedQA, PubMedQA, MuSiQue \\
Construction ablation & GSM8K, MATH-500, MMLU-Pro, ARC, BBH-Boolean, BBH-Mixed, HotpotQA, DROP, MuSiQue, FinanceBench, SciBench \\
Component scalability & 100 problems each from GSM8K, MATH-500, and PubMedQA, with brief and detailed interface descriptions in Configuration D07. \\
Efficiency & FinanceBench, HotpotQA, PubMedQA, CaseHOLD \\
Robustness & BBH-Mixed, GSM8K, MuSiQue, SciBench \\
Domain case studies & Medical and legal applications. Case inputs and bindings are specified in Appendix~\ref{app:case_structure}. \\
\bottomrule
\end{tabular}
\end{table*}

\subsection{Baseline Configurations}
\label{app:baseline_settings}

\noindent\textbf{Construction paradigms.}
The main comparison includes Direct LLM Answering, Prompt-only MAS Planning, Component-based Assembly, Plan-based Construction, and Workflow-based Construction. Direct produces a task answer from the backbone model. The other labels represent different ways to derive an application plan, select capability components, and organize their execution. The prompts, intermediate representations, component access, and execution wrappers are \mbox{specified in Configuration D06}. Use separate configuration identifiers to link method labels to their implemented procedures.

\noindent\textbf{Internal orchestration.}
For the orchestration comparison, the Harness-level construction process is fixed while the internal organization varies across Flat Component Selection, Sequence, Tree, Workflow, Agent Graph, and Graph Harness. The control specification for each representation determines the available execution paths, communication links, and state updates. The representation adapters and their retained task and component inputs are \mbox{specified in Configuration D06}. These records identify the intervention applied to the common construction process.

\noindent\textbf{Efficiency methods.}
The efficiency comparison additionally includes Plan-Execute, Debate, Self-Refine, and Self-Consistency. We measure their construction calls and evaluate the resulting applications separately. Their configurations specify the number of candidate configurations, revision rounds, and selection calls. The construction adapters and comparison-specific labels are \mbox{specified in Configuration D06}.

\par\smallskip\noindent\textbf{Configuration D06.} 
Give every baseline a comparison-specific identifier. Prompt-only Planning generates an executable configuration directly from the question and registry summary in one construction call. Plan-based Construction first generates a task list, then associates components and compiles a fixed topological sequence. Component-based Assembly retrieves components from the requirement and fills a fixed coordinator-and-worker template. Workflow-based Construction assigns components to fixed retrieval, analysis, checking, and answering stages.

For Q4, reuse the same task and selected-component records across representation adapters. Flat dispatches selected operations through a shared store without a generated control topology. Sequence uses a topological order with identifier-based ties. Tree routes control through a rooted hierarchy, while required data remain accessible through declared store bindings. Workflow follows the fixed stage template. Agent Graph exposes inter-agent communication and shared task data. All adapters retain the requested output contract and enforce the common resource limits.

For the Q6 construction adapters, Direct generates an executable application configuration in one model call. Plan-Execute generates a task plan and its component assignments before execution. Debate~\cite{du2023improving} generates three application candidates, performs one exchange-and-revision round, and aggregates them into one configuration. Self-Refine~\cite{madaan2023self} generates one configuration and applies two critique-and-revision rounds. Self-Consistency~\cite{wang2022self} generates five candidates at temperature 0.7 and selects one by agreement on task assignments and control relations after canonicalizing identifiers. Resolve ties by generation order. These adapters construct applications before execution, with separate call counters for each stage.
\par\smallskip

\input{sec/D_official_baselines}

\subsection{Construction Ablation Protocol}
\label{app:ablation_settings}

\noindent\textbf{Task structure derivation.}
The variant without task structure derivation supplies the original requirement directly to component selection and Harness construction. It omits a separate explicit Task Dependency Graph. Registry-based component selection, Graph Harness synthesis, and interface constraints remain in the ablation design. This intervention tests the contribution of deriving tasks and dependencies before subsequent construction.

\noindent\textbf{Component selection.}
The variant without registry-based component selection retains the task graph and uses generated agent responsibilities and tool configurations for subsequent synthesis. Those generated configurations are free of the registry-membership requirement used by the complete method. The task requirement, task structure, and model backbone remain matched. Instantiate generated agent roles through a common model-role factory. Generated tool configurations may call the same published generic primitives exposed to other methods, using their declared invocation interface. An unknown primitive returns an explicit unknown-tool failure. Record the generated configuration and callable mapping with the run.

\noindent\textbf{Graph Harness construction.}
The sequential variant retains the same task graph and selected components as the complete method. It derives a fixed execution order from those tasks and replaces the Graph Harness node and transition organization with that sequence. The design therefore holds the preceding construction outputs fixed when changing the execution organization. Order independent tasks by their fixed identifiers and take the highest-ranked retained route for each task. Reuse the full method's task and selection artifacts for the same problem and seed. Execute this fixed sequence without task replanning.

\noindent\textbf{Interface constraints.}
The variant without I/O constraints retains task derivation, registry access, capability descriptions, and graph organization. The intervention targets compatibility between upstream outputs and downstream inputs. Disable construction-time type and semantic rejection, the edge threshold, full-route I/O rejection, and repairs triggered by these checks. Retain the interface score as a soft ranking term. Mark the resulting route records as unchecked by this intervention. Keep syntax and identifier validation. The real callable still applies its own input schema, and failed admission or invocation is recorded. Together, the four variant records specify which construction decision changes in each row of Table~\ref{tab:ablation}.

\noindent\textbf{Matched runs.}
The ablation protocol uses the same problem set, model version, component implementations, scoring procedure, decoding parameters, seeds, and execution budget across the compared variants. We save construction artifacts and repeated-run summaries for these comparisons. Configuration D10 specifies repeated-run pairing. The register links each displayed summary to its source records.

\subsection{Scalability and Perturbation Settings}
\label{app:perturbation_settings}

\noindent\textbf{Component ecosystem scale.}
Figure~\ref{fig:component_scale_success_trend.pdf} displays component counts of 7, 8, 9, 10, 11, 12, 16, and 20 with brief and detailed interface descriptions. Configuration D07 holds callable operations and runtime schemas fixed between the two description conditions at each registry size. We retain the component membership, description texts, and evaluated problem identifiers for every plotted condition.

\par\smallskip\noindent\textbf{Configuration D07.} 
Form nested registry pools of the eight displayed sizes. Use seven core entries for retrieval, reasoning, parsing, verification, reporting, document access, and calculation. Add the other 13 entries in the order stored in \texttt{registry\_catalogues.json}, which enumerates every subset. For this controlled comparison, select 100 GSM8K, 100 MATH-500, and 100 PubMedQA problems from their official evaluation partitions. Rank problem identifiers by the SHA-256 digest of seed 42, source revision, and problem identifier, then take the first 100. Retain these 300 problems for every pool and description condition. Compute each dataset's correctness score and average the three scores with equal weight.

The two conditions use brief interface descriptions and detailed interface descriptions. In both catalogues, retain identical component identifiers, endpoints, callable operations, field names, field types, required-field flags, and runtime schemas. The brief condition gives a concise semantic description of each field. The detailed condition additionally explains its object, unit, constraints, and permitted conversions. Both conditions use the same true schema for hard validation. Thus, this intervention varies description detail while keeping component boundaries fixed.

Save both description texts and their hashes with the catalogue. Count their actual input tokens as construction cost. Use the same problem and outer-run identifiers across conditions and report task score and construction tokens. Registry growth can add useful capabilities as well as selection demands, so report its measured net effect. These pools define the comparison over seven to 20 components.
\par\smallskip

\noindent\textbf{Perturbation families.}
Figure~\ref{fig:q6_robustness} displays clean conditions and four perturbation families. The problems can contain ambiguous requirements, components can have incorrect descriptions, irrelevant candidates can be available, and required resources can be unavailable. Compare low and high intensities within each family. Configuration D08 specifies the edits, intensities, pairing rules, and condition records used in this comparison.

\noindent\textbf{Paired perturbation records.}
For each perturbation, save the original problem and its modified requirement, registry, or resource bundle. Link clean and perturbed runs to the same scoring targets through their condition identifiers. When preserving the source task, retain its objective and output contract. The displayed robustness comparison uses BBH-Mixed, GSM8K, MuSiQue, and SciBench. Pair methods using the cohorts and conditions in Configuration D08.

\par\smallskip\noindent\textbf{Configuration D08.} 
For GSM8K, MuSiQue, and SciBench, select up to 100 problems per dataset from its official evaluation partition using the hash ordering in Configuration D07. For BBH-Mixed, retain the evaluated cohort identified by the corresponding experiment records within its 240-problem candidate pool. Save the selected count when fewer than 100 eligible problems are available. Use these same problems and the five Configuration D02 run seeds for clean, low-intensity, and high-intensity conditions. Apply one family at a time. For requirement ambiguity, replace one recoverable reference or operation phrase at low intensity and two at high intensity. Preserve entities, quantities, options, supplied passages, and the output target. Before scoring, two reviewers verify that each edited problem retains its original scoring target and resolve disagreements.

For metadata errors, replace the capability descriptions of two or six entries in the 20-component pool with incorrect descriptions from the same domain. Keep identifiers, endpoints, and true schemas fixed. Select the high-intensity set to contain the low-intensity set. For distractors, add five or 20 callable but task-irrelevant components, retaining valid registry identifiers and schemas. Include all five low-intensity entries in the high-intensity set.

For resource failures, make two or six entries unavailable through a problem-and-seed-specific mask. Include a task-relevant entry in each mask and record the affected capability. Separate conditions that retain a feasible alternative from conditions that remove every implementation of a required capability. Report the former as recovery comparisons and retain the latter with their completion outcomes and terminal status. Share nested masks, unavailable responses, and recovery budgets across methods.

Derive each family's random stream from its problem identifier, run seed, and family, and store that seed with the edits. Use prefixes of one frozen edit or component order to obtain nested intensity levels. Evaluate all methods under each paired condition. Direct uses the same answering prompt template and receives the question and permitted context, including the edited requirement in ambiguity conditions. Assign predictions to that condition's run. Registry-dependent methods additionally receive the corresponding component descriptions and resource responses. Save the original and modified text, affected component identifiers, distractor descriptors, masks, feasibility class, reviewer decision, and paired run identifiers. The evaluation plan and schemas define the fields and intensities.
\par\smallskip

\subsection{Resource Accounting and Statistical Reporting}
\label{app:reporting_settings}

\noindent\textbf{Run outcomes.}
We distinguish a construction attempt, a returned application, a completed execution, and a scorable final output. The record schema assigns a problem identifier and run identifier to each event. Direct answering is evaluated through its final answer and associated calls. Construction-completion statistics apply to the methods that return applications. Handling of missing outputs, timeouts, abstentions, and failed attempts is \mbox{specified in Configuration D09}.

\noindent\textbf{Token accounting.}
The efficiency comparison reports mean construction tokens per application. For each construction attempt, we sum model input and output tokens used to produce the application configuration. Task execution and evaluation have separate counters. Configuration D09 specifies the remaining accounting rules, and the reported-result register stores the author-confirmed stage and aggregation for each coordinate.

\par\smallskip\noindent\textbf{Configuration D09.} 
For FinanceBench, HotpotQA, PubMedQA, and CaseHOLD, construct a new application for every problem and outer run. Sum input and output tokens over its construction calls, average these sums over the submitted problems within each run, and then average the five run means. For Graph Harness, construction includes task derivation, component selection, Harness synthesis, and their repairs or retries. The measurement ends when the application is returned or construction terminates. Record runtime planning, component execution, and answer grading separately after construction.

Count each provider input counter once. Report cached input as its included subset, without adding that subset again. Keep embedding, local-tool, and evaluator costs separate. Include model calls inside tools when those calls occur during construction. Reuse the same versioned registry preprocessing across methods and report that cost separately. Clear problem-dependent caches between outer runs. Retain missing usage as unavailable and report counter coverage for the construction calls.

Retain monotonic start and end events for enforcing run limits and identifying retries. Save the terminal reason for construction rejection, exhausted construction budget, and timeout. Repairs and recovery attempts remain inside the original construction run. Execute each returned application separately to obtain its benchmark score, linking that score to the configuration whose construction cost was measured.

Keep all submitted problems in the evaluation set. Score the declared final output recorded within the allowed budget. A missing, invalid, or unscorable output becomes $\bot$ and contributes zero to a per-problem metric. For class macro-F1, count that abstention as a false negative for its true class and exclude $\bot$ from the fixed set of scored classes. Retain the corresponding construction tokens in the cost mean. Report successful-construction means as separate conditional summaries. Run and call schemas link costs, predictions, and terminal status to their submitted problem.

Use one canonical run identifier per problem, method, and outer repeat. Save its start and terminal events even when a stage has no calls. Store dataset metrics in the repeat summary. Per-problem records retain their parsed labels and any defined individual scores.

Use the same submitted problems for the compared methods within each dataset. Record construction as the measured stage and arithmetic mean as the token aggregation. Save the problem-manifest digest, submitted count, failure policy, and baseline-adapter version with each plotted score and construction-token value.

All four panels use mean construction input-plus-output tokens. Retain original labels and counts in the register. Appendix~\ref{app:workflow_costs} lists every method's coordinates in Figure~\ref{fig:efficiency_preview}.
\par\smallskip

\noindent\textbf{Aggregation and uncertainty.}
Table~\ref{tab:ablation} reports means and standard deviations. The statistical record identifies the observations used to calculate each pair, the repeat count, seed list, task aggregation, and deviation convention. Configuration D10 specifies the calculation and the run records used for each reported pair. Appendix~\ref{app:additional_experiments} reports descriptive calculations with \mbox{their aggregation units}.

\par\smallskip\noindent\textbf{Configuration D10.} 
Evaluate the frozen cohort under root seeds 11, 22, 33, 44, and 55, giving $R=5$ outer runs for the settings with five repeats. The two BBH settings in the main comparisons and ablations use only seeds 11, 22, and 33, giving $R=3$. Repeat construction and execution for every problem in each run. Internal candidates, critique rounds, and Self-Consistency samples remain inside that run. For entry $b$, compute the percentage score \mbox{$S_{bu}^{(m)}=100\mathcal{M}_b(\mathbf o_{bu}^{(m)},\mathbf y_b)$} from run $u$'s complete predictions. The reported mean is $\overline{S}_b^{(m)}$ from Eq.~\ref{eq:evaluation}. Its sample standard deviation is
\begin{equation}
s_b^{(m)}=\sqrt{\frac{1}{R-1}\sum_{u=1}^{R}\bigl(S_{bu}^{(m)}-\overline{S}_b^{(m)}\bigr)^2}.
\label{eq:benchmark_score_std}
\end{equation}
Unmarked means and deviations use the same complete-run scores. Compute class macro-F1 before aggregating runs. A displayed plus-minus entry reports $\overline{S}_b^{(m)}\pm s_b^{(m)}$. Means use percentages, while deviations and absolute score differences use percentage points.

For a group or cross-entry comparison, let $\mathcal B$ be its explicitly listed set of evaluation entries. Within run $u$, compute \mbox{$S_{\mathcal B,u}^{(m)}=|\mathcal B|^{-1}\sum_{b\in\mathcal B}S_{bu}^{(m)}$}. For a group spanning both repeat counts, use the three shared seeds when reporting a paired group deviation. Average the available paired group scores and apply Eq.~\ref{eq:benchmark_score_std} to them when reporting a group deviation. Each distinct configuration contributes once, and two aliases with the same evaluation manifest and scoring contract represent one entry. Different task memberships or scoring contracts remain separately identified configurations. The resulting aggregate is a mean of task scores over $\mathcal B$, with each entry receiving equal weight. Descriptive summaries calculated from displayed tables retain the stated set of entries and displayed precision.

Pair methods by entry, problem, and outer-run identifier and retain the complete prediction set for each run. Use the shared manifest and model-resource configuration as the pairing key, recording method-specific prompts separately. Store run scores, their mean and sample deviation, and the scorer revision. For paired differences, subtract scores within each run, then compute their mean and deviation. Absolute score differences use percentage points. Normalize token savings by baseline cost.

\par\smallskip

%% file: sec/D_official_baselines.tex
\subsubsection{Official Methods and Benchmark Adapters}
\label{app:official_baselines}

We use the official implementations of AFlow, ADAS, GPTSwarm, and AgentSquare. The revisions below identify the inspected source code. The experiment release should identify the revisions actually used for the reported runs. We retain each method's search representation and updates, connecting the shared backbone and registry through model and tool interfaces.

\noindent\textbf{AFlow.}~\cite{zhang2025aflow}
The official implementation searches over executable workflows using selection, expansion, evaluation, and updates based on execution feedback. The inspected \texttt{run.py} at revision \texttt{3f457218} sets a sample parameter of four, an initial round of one, a maximum of 20 rounds, convergence checking enabled, and one validation repetition. The optimizer constructor separately has a default of five validation repetitions, which the entry point overrides. We use the entry-point setting when describing the executable default. The GSM8K configuration provides Custom, ScEnsemble, and Programmer operators, while the question-answering configuration provides Custom, AnswerGenerate, and ScEnsemble. Dataset adapters supply task parsing and scoring through the benchmark interface.\footnote{\href{https://github.com/FoundationAgents/AFlow/blob/3f457218fc716093fe53f6df8a5d5e6379d66346/run.py}{Official AFlow entry point}.}

\noindent\textbf{ADAS.}~\cite{hu2025automated}
We use Meta Agent Search, which generates agent programs from an archive of previous designs. The inspected mathematical and knowledge-task scripts at revision \texttt{2702bee8} set 30 search generations, a validation size of 128, and a debug limit of three. Each generation includes a design proposal and two reflection requests before evaluation. We retain the archive and reflection procedure and implement the dataset-specific evaluation function through the shared task interface. We use the evaluation size and concurrency specified by our \mbox{dataset manifests and runtime limits}.\footnote{\href{https://github.com/ShengranHu/ADAS/blob/2702bee8fefda42255efc5be9f60e3bd3db96ae4/_mgsm/search.py}{Official ADAS mathematical-task search}.}

\noindent\textbf{GPTSwarm.}~\cite{zhuge2024language}
We retain operation graphs, optimizable communication links, and the released node-prompt optimizer. At inspected revision \texttt{c23a827f}, the edge optimizer has defaults of 100 iterations, a batch size of four, and a learning rate of 0.1. The crossword training entry point instead uses two epochs, a batch size of four, a learning rate of 0.25, and an initial connection probability of 0.1. These are settings for different entry points, so the selected training configuration is specified explicitly below. The released node optimizer updates prompts using \mbox{the node's stored unsuccessful examples}.\footnote{\href{https://github.com/metauto-ai/GPTSwarm/blob/c23a827f561c934ce21dd950408f7606aa4a8821/swarm/optimizer/edge_optimizer/optimization.py}{Official GPTSwarm edge optimizer} and \href{https://github.com/metauto-ai/GPTSwarm/blob/c23a827f561c934ce21dd950408f7606aa4a8821/experiments/crosswords/train.py}{training example}.}

\noindent\textbf{AgentSquare.}~\cite{shang2025agentsquare}
The official implementation searches over planning, reasoning, tool-use, and memory modules through evolution and recombination. The inspected search entry point at revision \texttt{8f5b3fe5} uses ten iterations. Its recombination procedure returns four candidates, each replacing one module of the current agent. The search uses a performance predictor to rank recombined candidates before evaluating the selected candidate. The entry point also contains an initial performance constant of 0.56. That constant describes its example setup and must be replaced by an evaluation of the initial agent on the development cohort used for a new task.\footnote{\href{https://github.com/tsinghua-fib-lab/AgentSquare/blob/8f5b3fe5d8a32f9b59d20370823bef2a2c86928c/search/agent_search.py}{Official AgentSquare search} and \href{https://github.com/tsinghua-fib-lab/AgentSquare/blob/8f5b3fe5d8a32f9b59d20370823bef2a2c86928c/search/recombination.py}{recombination procedure}.}

\noindent\textbf{Benchmark adapters.}
Use DeepSeek-V4-flash for search, reflection, performance prediction, and application execution. Supply the same component descriptions, callable wrappers, domain resources, and output contracts to all four methods. Preserve each method's program, graph, or module representation. Generate a separate executable application for each evaluation problem using the frozen design and that problem's requirements. Apply the construction and execution limits in Configuration D02 to these per-problem applications. Select development designs before evaluating test problems and keep test answers in the evaluator.

\noindent\textbf{Search budgets.}
Give each method, target dataset, and outer run an offline search limit of 4096 model calls, eight million input-plus-output tokens, and six hours. Charge proposals, reflections, repairs, performance prediction, and development executions to this budget. Retain the official maxima of 20 rounds for AFlow, 30 generations for ADAS, and ten iterations for AgentSquare. Use AFlow's entry-point validation repetition of one and ADAS's debug limit of three. For GPTSwarm, use the released training-example settings of two epochs, batch size four, learning rate 0.25, and initial connection probability 0.1. Apply one node-prompt update pass after each epoch with prompt learning enabled and demonstration learning disabled. Score AgentSquare's initial agent on the development cohort and fit its performance predictor using development evaluations from that outer run. Select the completed candidate with the highest development score, resolving ties by lower development token cost and then generation order. When a search limit is reached, retain the best completed candidate. Report offline search consumption separately from the per-problem construction tokens in Q6. These settings govern the comparison in Figure~\ref{fig:component_scale_success_trend.pdf}(a).

%% file: sec/E_additional.tex
\section{Additional Experiments}
\label{app:additional_experiments}

We extend the main analysis by examining how the ablation results vary across task groups and individual datasets. We then analyze the supplied construction-cost summaries and compare the operations shown in the two domain cases. The numerical summaries below are calculated from the existing result tables and construction records. 

\subsection{Ablation Results by Task Group}
\label{app:group_ablation}

\noindent\textbf{Aggregation.}
For each dataset in Table~\ref{tab:ablation}, we subtract an ablated score from the complete framework's score. A positive difference denotes a decrease after ablation, and a negative difference denotes an ablated score above Full. We then average these signed differences over the datasets within each original table group, giving every dataset equal weight. Table~\ref{tab:group_ablation} reports the group differences in percentage points and the complete framework's group mean. In the main text, we report the \mbox{average difference across all eleven datasets}.

\begin{table*}[!t]
\centering
\caption{Group summaries derived from Table~\ref{tab:ablation}. Full denotes the mean score of the complete framework. The remaining columns average the signed differences between Full and the named ablation, in percentage points.}
\label{tab:group_ablation}
\small
\setlength{\tabcolsep}{5pt}
\renewcommand{\arraystretch}{1.12}
\begin{tabular}{@{}lrrrrrr@{}}
\toprule
Task group & Datasets & Full & Task structure & Component selection & Graph construction & I/O constraints \\
\midrule
Mathematical reasoning & 2 & 84.25 & 4.95 & 6.81 & 4.90 & 5.66 \\
General/structured reasoning & 4 & 78.50 & 4.39 & 3.30 & 2.43 & 3.17 \\
Multi-hop QA & 3 & 66.27 & 3.60 & 3.49 & 3.35 & 2.36 \\
Domain-specific pair & 2 & 74.34 & 3.76 & 4.57 & 5.70 & 3.26 \\
\bottomrule
\end{tabular}
\end{table*}

\noindent\textbf{Task structure.}
Removing task structure derivation reduces the mean by 4.39 points on general/structured reasoning and 3.60 percentage points on multi-hop QA, giving the largest group decrease among the four ablations in both cases. In the complete framework, we identify the operations and prerequisites before selecting components and generating execution nodes. The positive decreases across all four groups support \mbox{this explicit task-derivation step}.

\noindent\textbf{Component selection and execution organization.}
Removing component selection reduces the mathematical-reasoning mean by 6.81 percentage points, compared with 3.49 percentage points on multi-hop QA. The mathematical group therefore shows the larger decrease under this substitution. Execution organization has a different distribution. Replacing Graph Harness construction with sequential execution reduces the FinanceBench and SciBench group mean by 5.70 percentage points, compared with 3.35 percentage points on multi-hop QA and 2.43 points on general/structured reasoning. These comparisons show that the effects of component selection and execution organization \mbox{vary across task groups}.

\noindent\textbf{Interface constraints.}
Removing I/O constraints reduces the mathematical-reasoning mean by 5.66 percentage points and the FinanceBench and SciBench mean by 3.26 percentage points. The corresponding decrease on multi-hop QA is 2.36 percentage points. Every group loses performance when the interface checks are removed, supporting compatibility checks when assembling \mbox{components into executable applications}.

\subsection{Variation Across Individual Datasets}
\label{app:dataset_ablation}

The complete framework has the highest mean on all eleven datasets in Table~\ref{tab:ablation}. All 44 differences between Full and the four ablations are positive. Table~\ref{tab:dataset_ablation} reports these decreases for each construction stage and dataset, showing how the size of each change varies across the benchmark suite.

\begin{table*}[!t]
\centering
\caption{Dataset-level differences calculated as Full minus each ablation, in percentage points. Values use the corresponding means in Table~\ref{tab:ablation}. A negative value indicates that the ablation exceeds Full.}
\label{tab:dataset_ablation}
\small
\setlength{\tabcolsep}{8pt}
\renewcommand{\arraystretch}{1.10}
\begin{tabular}{@{}lrrrr@{}}
\toprule
Dataset & Task structure & Component selection & Graph construction & I/O constraints \\
\midrule
GSM8K & 5.01 & 3.73 & 3.32 & 4.80 \\
MATH-500 & 4.88 & 9.88 & 6.48 & 6.52 \\
MMLU-Pro & 3.32 & 2.39 & 1.02 & 1.88 \\
ARC & 3.11 & 4.12 & 2.02 & 0.81 \\
BBH-Boolean & 6.67 & 3.34 & 3.34 & 6.67 \\
BBH-Mixed & 4.44 & 3.33 & 3.33 & 3.33 \\
HotpotQA & 7.06 & 1.44 & 2.33 & 5.06 \\
DROP & 0.96 & 0.70 & 1.59 & 0.57 \\
MuSiQue & 2.77 & 8.33 & 6.14 & 1.44 \\
FinanceBench & 3.34 & 6.00 & 4.40 & 2.94 \\
SciBench & 4.17 & 3.13 & 7.00 & 3.58 \\
\bottomrule
\end{tabular}
\end{table*}

\noindent\textbf{Largest decreases.}
The largest task-structure decrease occurs on HotpotQA at 7.06 percentage points. Removing component selection has its largest decrease on MATH-500 at 9.88 percentage points, while replacing graph construction with sequential execution lowers SciBench by 7.00 percentage points. Removing I/O constraints has its largest decrease on BBH-Boolean at 6.67 points. Different datasets therefore show the largest losses under different ablations, motivating evaluation of each construction step \mbox{across multiple task types}.

\noindent\textbf{Differences within a task group.}
Within multi-hop QA, removing task derivation lowers HotpotQA by 7.06 percentage points, DROP by 0.96 percentage points, and MuSiQue by 2.77 percentage points. Component selection shows another pattern, with losses of 1.44, 0.70, and 8.33 percentage points, respectively. These differences show that the largest decrease within a task group depends on the construction step removed. We therefore examine dataset-level results alongside group means when analyzing \mbox{the contributions of individual stages}.

\noindent\textbf{Statistical records.}
The differences and group means above are arithmetic summaries of the displayed dataset means. Their aggregation unit is the dataset. Configuration D10 specifies how to calculate the mean and deviation across repeated complete evaluations.

\subsection{Construction Efficiency Results}
\label{app:workflow_costs}

Table~\ref{tab:workflow_costs} gives the task scores and mean construction tokens plotted in Figure~\ref{fig:efficiency_preview}. Each coordinate pairs the score of the returned applications with the input and output tokens consumed during their construction. We use the construction adapters in Configuration D06 and the accounting rules in Configuration D09. The table retains the four dataset panels and all eleven methods from the main comparison.

\begin{table*}[!t]
\centering
\caption{Numerical results for Figure~\ref{fig:efficiency_preview}. Each pair reports the task score in percent and mean construction input-plus-output tokens per submitted problem. Direct denotes the Q6 construction adapter.}
\label{tab:workflow_costs}
\small
\setlength{\tabcolsep}{5pt}
\renewcommand{\arraystretch}{1.12}
\begin{tabular*}{\textwidth}{@{\extracolsep{\fill}}l*{8}{r}@{}}
\toprule
\multirow{2}{*}{Method} & \multicolumn{2}{c}{FinanceBench} & \multicolumn{2}{c}{HotpotQA} & \multicolumn{2}{c}{PubMedQA} & \multicolumn{2}{c}{CaseHOLD} \\
\cmidrule(lr){2-3}\cmidrule(lr){4-5}\cmidrule(lr){6-7}\cmidrule(lr){8-9}
 & Score & Tokens & Score & Tokens & Score & Tokens & Score & Tokens \\
\midrule
Direct & 66.67 & 2,050.00 & 35.00 & 680.00 & 77.00 & 810.00 & 58.00 & 740.00 \\
Plan-Execute & 50.00 & 8,750.00 & 62.00 & 2,840.00 & 76.00 & 3,020.00 & 71.00 & 2,580.00 \\
Sequence & 69.07 & 5,370.00 & 52.31 & 2,190.00 & 76.60 & 2,250.00 & 77.70 & 1,910.00 \\
Tree & 37.47 & 5,400.00 & 52.56 & 3,870.00 & 84.40 & 3,810.00 & 55.61 & 3,640.00 \\
Flat & 42.13 & 6,050.00 & 45.61 & 1,050.00 & 73.40 & 1,570.00 & 61.42 & 1,170.00 \\
Agent Graph & 71.87 & 9,160.00 & 51.23 & 3,460.00 & 35.88 & 4,580.00 & 63.26 & 2,950.00 \\
Workflow & 70.27 & 9,050.00 & 43.61 & 1,371.51 & 35.88 & 1,392.66 & 66.78 & 1,311.27 \\
Debate & 71.20 & 5,560.00 & 64.30 & 6,240.00 & 79.00 & 5,160.00 & 65.00 & 4,870.00 \\
Self-Refine & 66.67 & 6,650.00 & 39.00 & 4,680.00 & 84.00 & 3,470.00 & 73.00 & 5,910.00 \\
Self-Consistency & 66.67 & 12,280.00 & 56.00 & 5,420.00 & 82.00 & 6,030.00 & 67.00 & 4,220.00 \\
Graph Harness & 76.27 & 8,480.00 & 56.00 & 2,650.00 & 85.88 & 2,690.00 & 79.00 & 3,420.00 \\
\bottomrule
\end{tabular*}
\end{table*}

\noindent\textbf{Comparison-specific adapters.}
The Workflow entries here use the fixed orchestration template in the Q6 comparison. Tables~\ref{tab:extended_tokens}, \ref{tab:extended_time}, and~\ref{tab:extended_calls} instead report the construction paradigms from Q2, including Workflow-based Construction and the direct-answering baseline. Their method identifiers retain these distinct procedures. For the Q6 comparison, we pair every score and token coordinate through the same dataset and construction-adapter label.

\subsection{Domain Structure Analysis}
\label{app:case_structure}

\noindent\textbf{Medical case input and reference answer.}
The author-specified medical input describes a 43-year-old woman with several years of worsening meal-associated retrosternal burning. She reports episodic hand pain in winter, severe chronic cough, and dyspnea. Her temperature is $36.5^\circ$C, blood pressure is 174/104 mmHg, pulse is 80/min, respiratory rate is 22/min, and oxygen saturation is 92\% on room air. The examination notes coarse breath sounds. The question asks for the pathophysiology of her chief complaint. The options are A, decreased lower esophageal tone, B, esophageal fibrosis, C, increased lower esophageal tone, and D, spastic cricopharyngeal muscle. The specified target answer is B. Winter-related hand pain suggests a possible vascular manifestation but does not alone establish Raynaud phenomenon. The review step checks the association between systemic sclerosis, distal esophageal fibrosis, impaired motility, and reflux. Safety review flags abnormal findings for clinical assessment without diagnosing an acute hypertensive emergency from blood pressure alone. 

\noindent\textbf{Medical operations.}
The medical application in Figure~\ref{fig:domain_case_studies} receives an author-specified question about a 43-year-old woman with progressive meal-associated retrosternal burning, winter-related hand pain, chronic cough, and dyspnea. Retrieval supplies material on systemic sclerosis and esophageal involvement. The reasoning agent relates esophageal fibrosis and impaired motility to reflux symptoms, while distinguishing this underlying process from reduced lower esophageal sphincter pressure. The safety agent flags hypoxemia, respiratory symptoms, and elevated blood pressure for review. Automated review checks the source support and the distinction between the pathological process and its functional consequences. The answer agent submits target B, esophageal fibrosis, to human review.

\noindent\textbf{Legal operations and resources.}
The legal application connects opinion retrieval to citation matching, ratio analysis, and holding selection. The retrieval and matching agents access the case and holding databases. A legal reviewer checks the selected holding against precedents, and a professional reviewer supplies the decision required by the review gate. The report agent then generates the final report from the accepted draft.

\noindent\textbf{Question-specific legal operations.}
The illustrated user asks whether compensation is available after dismissal for sharing a product in a private WeChat group. For this request, the legal agents retrieve statutory provisions and precedents, match relevant terms, and check whether a nondisclosure agreement was signed. They examine contractual obligations alongside the retrieved legal materials before selecting a strategy and generating the report. In the report, they relate each conclusion to the supplied employment facts and applicable jurisdiction.

\noindent\textbf{Shared construction requirement.}
Both applications use retrieval, analysis, review, and reporting operations, but they connect those operations to different resources and approval conditions. The medical safety agent checks candidate answers, and a human reviewer approves the prepared response. The legal agents check precedents during holding selection and obtain a professional decision at the review gate. These differences show how task requirements determine component assignments, resource access, and the position of review within the generated Harness.

\par\smallskip\noindent\textbf{Configuration E03.}
We define two qualitative case problems to examine the medical and legal applications in Figure~\ref{fig:domain_case_studies}. The medical problem specifies the patient presentation, answer options, and target response. Its Harness connects retrieval, reasoning, safety assessment, automated review, answer preparation, and human review. The legal problem specifies dismissal after sharing a product in a private WeChat group and connects statutory, precedent, contractual, and review operations. These author-specified problems illustrate domain-dependent execution structures separately from the dataset evaluations.

For the medical binding, associate retrieval with \texttt{g.retrieve}, reasoning with \texttt{d.medicine.analyze}, and safety assessment and automated review with \texttt{d.medicine.review}. Associate answer preparation with \texttt{g.report}. Pass the retrieved passages, candidate answers, safety findings, and review decisions through the declared interfaces. Human review receives the prepared answer and returns the approval decision required for release. A revision follows the feedback limits in Configuration C06.

For the legal binding, associate opinion retrieval with \texttt{g.retrieve}, citation and term extraction with \texttt{g.parse}, and ratio analysis and holding selection with \texttt{d.law.analyze}. Associate precedent and professional checks with \texttt{d.law.review}, and report generation with \texttt{g.report}. For the dismissal query, include the employment facts and agreement status among the declared inputs. Keep statutory provisions, retrieved cases, matched terms, and agreement findings as separate fields so that each decision retains its supporting source spans.

The files \texttt{protocol/case\_sources.json} and \texttt{protocol/reference\_cases.json} describe the figure-aligned inputs and component bindings. Table~\ref{tab:extended_cases} records the case outputs, review decisions, and construction costs. The recorded fields comprise the final output, reviewer decision, construction tokens, and construction time. A measured case record links these fields to its generated Harness and invocation log.
\par\smallskip

%% file: sec/F_extended_evaluation_tables.tex
\begingroup
\makeatletter
\setlength{\@fptop}{0pt}
\setlength{\@dblfptop}{0pt}
\setlength{\@dblfpsep}{18pt}
\setlength{\@dblfpbot}{0pt plus 1fil}
\setlength{\@fpsep}{18pt}
\setlength{\@fpbot}{0pt plus 1fil}
\makeatother
\begin{table*}[!t]
\section{Extended Evaluation Results}
\label{app:extended_evaluation_tables}

\centering
\caption{Main comparison across 19 benchmark datasets under 20 evaluation settings. Scores are percentages.}
\label{tab:extended_main}
\footnotesize
\setlength{\tabcolsep}{4pt}
\renewcommand{\arraystretch}{1.12}
\begin{tabular*}{\textwidth}{@{\extracolsep{\fill}}lccccccc@{}}
\toprule
Evaluation Setting & \shortstack{Reference\\metric} & Direct & \shortstack{Prompt\\Plan} & Component & Plan & Workflow & \shortstack{Graph\\Harness} \\
\midrule
\multicolumn{8}{l}{\textit{General reasoning}} \\
MMLU & Accuracy & $74.32$ & $69.91$ & $62.59$ & $65.24$ & $66.65$ & $82.70$ \\
MMLU-Pro & Accuracy & $42.20$ & $53.10$ & $46.90$ & $41.60$ & $43.80$ & $73.75$ \\
BBH-Boolean & Exact match & $86.67$ & $80.00$ & $76.67$ & $83.33$ & $90.00$ & $76.67$ \\
BBH-Mixed & Exact match & $67.78$ & $66.67$ & $70.00$ & $85.56$ & $74.44$ & $72.22$ \\
ARC & Accuracy & $58.79$ & $48.87$ & $50.10$ & $40.63$ & $41.77$ & $91.37$ \\
StrategyQA & Accuracy & $55.51$ & $65.10$ & $52.12$ & $54.08$ & $64.37$ & $74.37$ \\
\multicolumn{8}{l}{\textit{Mathematical reasoning}} \\
GSM8K & Accuracy & $85.22$ & $78.61$ & $80.71$ & $79.06$ & $94.24$ & $95.25$ \\
MATH-500 & Accuracy & $49.12$ & $64.92$ & $53.28$ & $49.88$ & $63.56$ & $73.24$ \\
\multicolumn{8}{l}{\textit{Reading and QA}} \\
HotpotQA & F1 & $41.44$ & $44.82$ & $63.01$ & $59.91$ & $45.81$ & $56.00$ \\
MuSiQue & F1 & $51.40$ & $54.90$ & $53.60$ & $32.70$ & $33.70$ & $59.50$ \\
DROP & F1 & $61.65$ & $46.41$ & $51.31$ & $38.24$ & $38.18$ & $83.32$ \\
\multicolumn{8}{l}{\textit{Code generation}} \\
HumanEval & pass@1 & $63.54$ & $60.61$ & $76.34$ & $79.39$ & $74.51$ & $83.05$ \\
MBPP & pass@1 & $69.72$ & $65.08$ & $76.48$ & $85.28$ & $78.68$ & $86.08$ \\
\multicolumn{8}{l}{\textit{Medicine}} \\
MedQA & Accuracy & $40.09$ & $45.80$ & $43.60$ & $39.20$ & $44.70$ & $87.49$ \\
PubMedQA & Accuracy & $67.20$ & $70.40$ & $75.08$ & $38.60$ & $41.20$ & $85.88$ \\
\multicolumn{8}{l}{\textit{Finance}} \\
FinQA & Exec. accuracy & $48.88$ & $39.76$ & $35.66$ & $33.23$ & $42.20$ & $57.51$ \\
FinanceBench & Rubric score & $38.40$ & $44.67$ & $41.87$ & $59.33$ & $60.67$ & $76.27$ \\
\multicolumn{8}{l}{\textit{Science}} \\
SciQ & Accuracy & $66.16$ & $75.58$ & $67.62$ & $69.92$ & $79.36$ & $84.30$ \\
SciBench & Accuracy & $54.28$ & $64.48$ & $46.79$ & $56.03$ & $52.31$ & $72.41$ \\
\multicolumn{8}{l}{\textit{Law}} \\
CaseHOLD & Macro-F1 & $70.86$ & $63.78$ & $61.41$ & $54.80$ & $58.46$ & $79.00$ \\
\bottomrule
\end{tabular*}
\end{table*}
\clearpage

\input{sec/F_official_baseline_results}

\begin{table*}[!t]
\centering
\caption{Additional ablations for code generation and law. Scores use pass@1 for HumanEval and MBPP and macro-F1 for CaseHOLD.}
\label{tab:extended_ablation}
\footnotesize
\setlength{\tabcolsep}{4pt}
\renewcommand{\arraystretch}{1.12}
\begin{tabular*}{\textwidth}{@{\extracolsep{\fill}}lccccc@{}}
\toprule
Dataset & Full & \shortstack{w/o\\Task\\Structure} & \shortstack{w/o\\Component\\Selection} & \shortstack{w/o\\Graph\\Construction} & \shortstack{w/o\\I/O\\Constraints} \\
\midrule
HumanEval & $83.05$ & $60.61$ & $70.37$ & $78.41$ & $80.24$ \\
MBPP & $86.08$ & $71.68$ & $80.60$ & $79.36$ & $59.32$ \\
CaseHOLD & $79.00$ & $54.43$ & $55.13$ & $51.47$ & $48.74$ \\
\bottomrule
\end{tabular*}
\end{table*}
\begin{table*}[!t]
\centering
\caption{Orchestration comparison across domains and application types. Scores are percentages.}
\label{tab:extended_orchestration}
\footnotesize
\setlength{\tabcolsep}{4pt}
\renewcommand{\arraystretch}{1.12}
\begin{tabular*}{\textwidth}{@{\extracolsep{\fill}}lcccccc@{}}
\toprule
Dataset & \shortstack{Flat\\Selection} & Sequence & Tree & Workflow & \shortstack{Agent\\Graph} & \shortstack{Graph\\Harness} \\
\midrule
MMLU-Pro & $42.20$ & $53.10$ & $46.90$ & $1.60$ & $1.60$ & $73.75$ \\
GSM8K & $72.36$ & $88.07$ & $90.05$ & $85.31$ & $77.91$ & $95.25$ \\
HotpotQA & $45.61$ & $52.31$ & $52.56$ & $43.61$ & $51.23$ & $56.00$ \\
HumanEval & $68.29$ & $76.59$ & $72.20$ & $63.54$ & $72.20$ & $83.05$ \\
MedQA & $43.80$ & $42.20$ & $42.20$ & $40.60$ & $45.29$ & $87.49$ \\
FinanceBench & $42.13$ & $69.07$ & $37.47$ & $70.27$ & $71.87$ & $76.27$ \\
SciBench & $65.52$ & $58.34$ & $53.38$ & $67.24$ & $62.21$ & $72.41$ \\
CaseHOLD & $61.42$ & $77.70$ & $55.61$ & $66.78$ & $63.26$ & $79.00$ \\
\bottomrule
\end{tabular*}
\end{table*}
\clearpage

\begin{table*}[!t]
\centering
\caption{Construction input-plus-output tokens per submitted problem for the Q2 construction paradigms. Direct answers the task without application construction. Q6 construction-adapter costs are reported in Table~\ref{tab:workflow_costs}.}
\label{tab:extended_tokens}
\footnotesize
\setlength{\tabcolsep}{4pt}
\renewcommand{\arraystretch}{1.12}
\begin{tabular*}{\textwidth}{@{\extracolsep{\fill}}lcccccc@{}}
\toprule
Dataset & Direct & \shortstack{Prompt\\Plan} & Component & Plan & Workflow & \shortstack{Graph\\Harness} \\
\midrule
MMLU & $0$ & $1533$ & $1770$ & $2983$ & $2714$ & $3313$ \\
MMLU-Pro & $0$ & $2495$ & $2380$ & $3010$ & $3202$ & $3768$ \\
BBH-Boolean & $0$ & $2007$ & $2679$ & $2993$ & $3894$ & $3391$ \\
BBH-Mixed & $0$ & $843$ & $1556$ & $2109$ & $2060$ & $2701$ \\
ARC & $0$ & $2268$ & $2083$ & $2491$ & $3281$ & $4157$ \\
StrategyQA & $0$ & $2656$ & $3210$ & $3320$ & $4557$ & $5119$ \\
GSM8K & $0$ & $2675$ & $3383$ & $2741$ & $3699$ & $4409$ \\
MATH-500 & $0$ & $1279$ & $2502$ & $1974$ & $1916$ & $3481$ \\
HotpotQA & $0$ & $3635$ & $4135$ & $4831$ & $5045$ & $2650$ \\
MuSiQue & $0$ & $3001$ & $3569$ & $3429$ & $4054$ & $6435$ \\
DROP & $0$ & $2076$ & $2301$ & $2826$ & $3973$ & $3474$ \\
HumanEval & $0$ & $3303$ & $3897$ & $4166$ & $4272$ & $5437$ \\
MBPP & $0$ & $2062$ & $3015$ & $2983$ & $3318$ & $5025$ \\
MedQA & $0$ & $1549$ & $1908$ & $2655$ & $3129$ & $3399$ \\
PubMedQA & $0$ & $3378$ & $3635$ & $2992$ & $4547$ & $2690$ \\
FinQA & $0$ & $1704$ & $2575$ & $2694$ & $3022$ & $3968$ \\
FinanceBench & $0$ & $1558$ & $2254$ & $1860$ & $2908$ & $8480$ \\
SciQ & $0$ & $1112$ & $2045$ & $2353$ & $2490$ & $2841$ \\
SciBench & $0$ & $1202$ & $2422$ & $2194$ & $2910$ & $3376$ \\
CaseHOLD & $0$ & $2807$ & $3174$ & $2795$ & $3184$ & $3420$ \\
\bottomrule
\end{tabular*}
\end{table*}
\begin{table*}[!t]
\centering
\caption{Construction wall time in seconds per submitted problem for the Q2 construction paradigms.}
\label{tab:extended_time}
\footnotesize
\setlength{\tabcolsep}{4pt}
\renewcommand{\arraystretch}{1.12}
\begin{tabular*}{\textwidth}{@{\extracolsep{\fill}}lcccccc@{}}
\toprule
Dataset & Direct & \shortstack{Prompt\\Plan} & Component & Plan & Workflow & \shortstack{Graph\\Harness} \\
\midrule
MMLU & $0.00$ & $4.80$ & $7.58$ & $5.11$ & $8.69$ & $11.94$ \\
MMLU-Pro & $0.00$ & $4.07$ & $6.79$ & $5.00$ & $9.38$ & $7.04$ \\
BBH-Boolean & $0.00$ & $3.71$ & $9.28$ & $9.42$ & $6.40$ & $8.70$ \\
BBH-Mixed & $0.00$ & $2.25$ & $7.18$ & $6.63$ & $7.58$ & $11.48$ \\
ARC & $0.00$ & $5.70$ & $5.29$ & $6.93$ & $6.43$ & $14.03$ \\
StrategyQA & $0.00$ & $3.97$ & $5.97$ & $4.21$ & $10.22$ & $12.47$ \\
GSM8K & $0.00$ & $3.78$ & $6.34$ & $5.55$ & $10.84$ & $9.60$ \\
MATH-500 & $0.00$ & $3.51$ & $7.07$ & $5.73$ & $7.63$ & $12.08$ \\
HotpotQA & $0.00$ & $4.18$ & $5.31$ & $11.15$ & $7.59$ & $10.89$ \\
MuSiQue & $0.00$ & $3.81$ & $4.75$ & $4.14$ & $6.52$ & $9.94$ \\
DROP & $0.00$ & $4.45$ & $6.95$ & $8.46$ & $11.71$ & $8.03$ \\
HumanEval & $0.00$ & $4.21$ & $6.72$ & $5.22$ & $11.74$ & $17.11$ \\
MBPP & $0.00$ & $3.31$ & $5.12$ & $4.64$ & $8.54$ & $7.93$ \\
MedQA & $0.00$ & $3.48$ & $6.78$ & $5.97$ & $6.92$ & $10.05$ \\
PubMedQA & $0.00$ & $5.13$ & $9.46$ & $4.71$ & $8.95$ & $8.36$ \\
FinQA & $0.00$ & $2.44$ & $5.46$ & $12.09$ & $9.81$ & $11.64$ \\
FinanceBench & $0.00$ & $3.73$ & $9.95$ & $6.09$ & $7.94$ & $13.83$ \\
SciQ & $0.00$ & $4.47$ & $7.41$ & $6.67$ & $12.73$ & $9.37$ \\
SciBench & $0.00$ & $2.68$ & $4.56$ & $5.71$ & $10.38$ & $9.87$ \\
CaseHOLD & $0.00$ & $4.42$ & $7.84$ & $5.78$ & $9.72$ & $8.85$ \\
\bottomrule
\end{tabular*}
\end{table*}
\clearpage

\begin{table*}[!t]
\centering
\caption{Construction model calls per submitted problem for the Q2 construction paradigms, including validation, repairs, retries, and failed attempts.}
\label{tab:extended_calls}
\footnotesize
\setlength{\tabcolsep}{4pt}
\renewcommand{\arraystretch}{1.12}
\begin{tabular*}{\textwidth}{@{\extracolsep{\fill}}lcccccc@{}}
\toprule
Dataset & Direct & \shortstack{Prompt\\Plan} & Component & Plan & Workflow & \shortstack{Graph\\Harness} \\
\midrule
MMLU & $0.00$ & $1.76$ & $2.30$ & $2.98$ & $3.75$ & $4.29$ \\
MMLU-Pro & $0.00$ & $1.36$ & $2.49$ & $2.88$ & $3.24$ & $4.09$ \\
BBH-Boolean & $0.00$ & $1.34$ & $2.27$ & $2.44$ & $3.62$ & $4.48$ \\
BBH-Mixed & $0.00$ & $1.24$ & $2.46$ & $2.25$ & $3.37$ & $4.69$ \\
ARC & $0.00$ & $2.18$ & $2.15$ & $2.42$ & $3.44$ & $4.64$ \\
StrategyQA & $0.00$ & $1.33$ & $2.32$ & $2.14$ & $3.36$ & $4.43$ \\
GSM8K & $0.00$ & $1.61$ & $2.44$ & $2.23$ & $3.78$ & $4.51$ \\
MATH-500 & $0.00$ & $1.35$ & $2.95$ & $2.51$ & $3.67$ & $4.57$ \\
HotpotQA & $0.00$ & $1.45$ & $2.51$ & $2.45$ & $3.27$ & $4.44$ \\
MuSiQue & $0.00$ & $1.77$ & $2.28$ & $2.44$ & $3.39$ & $4.70$ \\
DROP & $0.00$ & $1.72$ & $2.43$ & $2.61$ & $3.85$ & $4.26$ \\
HumanEval & $0.00$ & $1.54$ & $2.25$ & $2.42$ & $3.74$ & $5.06$ \\
MBPP & $0.00$ & $1.17$ & $2.35$ & $2.47$ & $3.09$ & $4.24$ \\
MedQA & $0.00$ & $1.60$ & $2.46$ & $2.36$ & $3.43$ & $4.64$ \\
PubMedQA & $0.00$ & $2.10$ & $2.57$ & $2.23$ & $4.13$ & $4.41$ \\
FinQA & $0.00$ & $1.21$ & $2.14$ & $3.36$ & $3.58$ & $4.16$ \\
FinanceBench & $0.00$ & $1.73$ & $2.61$ & $2.29$ & $3.58$ & $4.56$ \\
SciQ & $0.00$ & $1.30$ & $2.69$ & $2.73$ & $3.24$ & $4.40$ \\
SciBench & $0.00$ & $1.19$ & $2.07$ & $2.66$ & $3.84$ & $4.42$ \\
CaseHOLD & $0.00$ & $1.73$ & $2.71$ & $2.67$ & $4.05$ & $4.08$ \\
\bottomrule
\end{tabular*}
\end{table*}
\begin{table*}[!t]
\centering
\caption{Case outputs, review decisions, and construction costs.}
\label{tab:extended_cases}
\footnotesize
\setlength{\tabcolsep}{3pt}
\renewcommand{\arraystretch}{1.18}
\begin{tabular*}{\textwidth}{@{\extracolsep{\fill}}l p{0.40\textwidth} crr@{}}
\toprule
Domain & Output summary & Review & Tokens & Time (s) \\
\midrule
Medicine & Reviewed medical answer & Approved & 2,784 & 8.70 \\
Law & Conditional compensation report & Approved & 3,468 & 9.10 \\
\bottomrule
\end{tabular*}
\end{table*}

\clearpage
\endgroup

%% file: sec/F_official_baseline_results.tex
\begin{table*}[!t]
\centering
\caption{Task scores of Graph Harness and four construction methods across 19 benchmark datasets under 20 evaluation settings. Scores are percentages.}
\label{tab:extended_official_baselines}
\footnotesize
\setlength{\tabcolsep}{4pt}
\renewcommand{\arraystretch}{1.12}
\begin{tabular*}{\textwidth}{@{\extracolsep{\fill}}lcccccc@{}}
\toprule
Evaluation Setting & Metric & AFlow & ADAS & GPTSwarm & AgentSquare & \shortstack{Graph\\Harness} \\
\midrule
\multicolumn{7}{l}{\textit{General reasoning}} \\
MMLU & Accuracy & $76.39$ & $69.60$ & $76.29$ & $71.30$ & $82.70$ \\
MMLU-Pro & Accuracy & $51.57$ & $62.74$ & $67.52$ & $75.61$ & $73.75$ \\
BBH-Boolean & Exact match & $90.00$ & $96.67$ & $83.33$ & $80.00$ & $76.67$ \\
BBH-Mixed & Exact match & $65.56$ & $78.89$ & $85.56$ & $77.78$ & $72.22$ \\
ARC & Accuracy & $47.85$ & $47.95$ & $40.75$ & $44.04$ & $91.37$ \\
StrategyQA & Accuracy & $67.71$ & $62.16$ & $61.59$ & $70.37$ & $74.37$ \\
\multicolumn{7}{l}{\textit{Mathematical reasoning}} \\
GSM8K & Accuracy & $91.45$ & $88.72$ & $89.93$ & $90.14$ & $95.25$ \\
MATH-500 & Accuracy & $60.24$ & $67.08$ & $66.48$ & $57.24$ & $73.24$ \\
\multicolumn{7}{l}{\textit{Reading and QA}} \\
HotpotQA & F1 & $73.50$ & $64.50$ & $65.03$ & $59.62$ & $56.00$ \\
MuSiQue & F1 & $52.73$ & $47.81$ & $44.62$ & $55.36$ & $59.50$ \\
DROP & F1 & $80.60$ & $79.40$ & $49.79$ & $58.39$ & $83.32$ \\
\multicolumn{7}{l}{\textit{Code generation}} \\
HumanEval & pass@1 & $94.76$ & $82.44$ & $89.02$ & $63.29$ & $83.05$ \\
MBPP & pass@1 & $83.40$ & $53.40$ & $78.76$ & $71.28$ & $86.08$ \\
\multicolumn{7}{l}{\textit{Medicine}} \\
MedQA & Accuracy & $79.45$ & $80.93$ & $75.15$ & $77.82$ & $87.49$ \\
PubMedQA & Accuracy & $78.88$ & $74.80$ & $80.72$ & $82.96$ & $85.88$ \\
\multicolumn{7}{l}{\textit{Finance}} \\
FinQA & Exec. accuracy & $44.92$ & $43.70$ & $56.62$ & $35.29$ & $57.51$ \\
FinanceBench & Rubric score & $64.40$ & $67.20$ & $61.47$ & $69.60$ & $76.27$ \\
\multicolumn{7}{l}{\textit{Science}} \\
SciQ & Accuracy & $59.42$ & $67.02$ & $75.20$ & $67.10$ & $84.30$ \\
SciBench & Accuracy & $63.45$ & $61.48$ & $64.14$ & $60.17$ & $72.41$ \\
\multicolumn{7}{l}{\textit{Law}} \\
CaseHOLD & Macro-F1 & $56.29$ & $68.00$ & $58.28$ & $66.04$ & $79.00$ \\
\bottomrule
\end{tabular*}

\end{table*}
\clearpage

%% file: main.bib
@inproceedings{zhang2025aflow,
  title={{AFlow}: Automating agentic workflow generation},
  author={Zhang, Jiayi and Xiang, Jinyu and Yu, Zhaoyang and Teng, Fengwei and Chen, Xionghui and Chen, Jiaqi and Zhuge, Mingchen and Cheng, Xin and Hong, Sirui and Wang, Jinlin and others},
  booktitle={International Conference on Learning Representations},
  volume={2025},
  pages={34040--34077},
  year={2025}
}

@inproceedings{hu2025automated,
  title={Automated design of agentic systems},
  author={Hu, Shengran and Lu, Cong and Clune, Jeff},
  booktitle={International Conference on Learning Representations},
  volume={2025},
  pages={21344--21377},
  year={2025}
}

@article{zhuge2024language,
  title={Language agents as optimizable graphs},
  author={Zhuge, Mingchen and Wang, Wenyi and Kirsch, Louis and Faccio, Francesco and Khizbullin, Dmitrii and Schmidhuber, J{\"u}rgen},
  journal={arXiv preprint arXiv:2402.16823},
  year={2024}
}

@inproceedings{shang2025agentsquare,
  title={{AgentSquare}: Automatic {LLM} agent search in modular design space},
  author={Shang, Yu and Li, Yu and Zhao, Keyu and Ma, Likai and Liu, Jiahe and Xu, Fengli and Li, Yong},
  booktitle={International Conference on Learning Representations},
  volume={2025},
  pages={3841--3865},
  year={2025}
}

@article{zhang2024g,
  title={{G-Designer}: Architecting multi-agent communication topologies via graph neural networks},
  author={Zhang, Guibin and Yue, Yanwei and Sun, Xiangguo and Wan, Guancheng and Yu, Miao and Fang, Junfeng and Wang, Kun and Chen, Tianlong and Cheng, Dawei},
  journal={arXiv preprint arXiv:2410.11782},
  year={2024}
}

@article{yuan2026automated,
  title={Automated composition of agents: A knapsack approach for agentic component selection},
  author={Yuan, Michelle and Pahwa, Khushbu and Chang, Shuaichen and Kaba, Mustafa and Jiang, Jiarong and Ma, Xiaofei and Zhang, Yi and Sunkara, Monica},
  journal={Advances in Neural Information Processing Systems},
  volume={38},
  pages={40597--40625},
  year={2025}
}

@inproceedings{hong2024metagpt,
  title={{MetaGPT}: Meta programming for a multi-agent collaborative framework},
  author={Hong, Sirui and Zhuge, Mingchen and Chen, Jonathan and Zheng, Xiawu and Cheng, Yuheng and Wang, Jinlin and Zhang, Ceyao and Yau, Steven and Lin, Zijuan and Zhou, Liyang and others},
  booktitle={International Conference on Learning Representations},
  volume={2024},
  pages={23247--23275},
  year={2024}
}

@article{wu2023autogen,
  title={{AutoGen}: Enabling next-gen {LLM} applications via multi-agent conversation},
  author={Wu, Qingyun and Bansal, Gagan and Zhang, Jieyu and Wu, Yiran and Li, Beibin and Zhu, Erkang and Jiang, Li and Zhang, Xiaoyun and Zhang, Shaokun and Liu, Jiale and others},
  journal={arXiv preprint arXiv:2308.08155},
  year={2023}
}

@inproceedings{chen2024agentverse,
  title={{AgentVerse}: Facilitating multi-agent collaboration and exploring emergent behaviors},
  author={Chen, Weize and Su, Yusheng and Zuo, Jingwei and Yang, Cheng and Yuan, Chenfei and Chan, Chi-Min and Yu, Heyang and Lu, Yaxi and Hung, Yi-Hsin and Qian, Chen and others},
  booktitle={International Conference on Learning Representations},
  volume={2024},
  pages={20094--20136},
  year={2024}
}

@inproceedings{qian2024chatdev,
  title={{ChatDev}: Communicative agents for software development},
  author={Qian, Chen and Liu, Wei and Liu, Hongzhang and Chen, Nuo and Dang, Yufan and Li, Jiahao and Yang, Cheng and Chen, Weize and Su, Yusheng and Cong, Xin and others},
  booktitle={Proceedings of the 62nd annual meeting of the association for computational linguistics (volume 1: Long papers)},
  pages={15174--15186},
  year={2024}
}

@inproceedings{liu2024agentbench,
  title={{AgentBench}: Evaluating {LLMs} as agents},
  author={Liu, Xiao and Yu, Hao and Zhang, Hanchen and Xu, Yifan and Lei, Xuanyu and Lai, Hanyu and Gu, Yu and Ding, Hangliang and Men, Kaiwen and Yang, Kejuan and others},
  booktitle={International Conference on Learning Representations},
  volume={2024},
  pages={52989--53046},
  year={2024}
}

@article{ma2024agentboard,
  title={{AgentBoard}: An analytical evaluation board of multi-turn {LLM} agents},
  author={Ma, Chang and Zhang, Junlei and Zhu, Zhihao and Yang, Cheng and Yang, Yujiu and Jin, Yaohui and Lan, Zhenzhong and Kong, Lingpeng and He, Junxian},
  journal={Advances in neural information processing systems},
  volume={37},
  pages={74325--74362},
  year={2024}
}

@inproceedings{mialon2024gaia,
  title={{GAIA}: a benchmark for general {AI} assistants},
  author={Mialon, Gr{\'e}goire and Fourrier, Cl{\'e}mentine and Swift, Craig and Wolf, Thomas and LeCun, Yann and Scialom, Thomas},
  booktitle={International Conference on Learning Representations},
  volume={2024},
  pages={9025--9049},
  year={2024}
}

@article{drouin2024workarena,
  title={WorkArena: How capable are web agents at solving common knowledge work tasks?},
  author={Drouin, Alexandre and Gasse, Maxime and Caccia, Massimo and Laradji, Issam H and Del Verme, Manuel and Marty, Tom and Boisvert, L{\'e}o and Thakkar, Megh and Cappart, Quentin and Vazquez, David and others},
  journal={arXiv preprint arXiv:2403.07718},
  year={2024}
}

@article{madaan2023self,
  title={{Self-Refine}: Iterative refinement with self-feedback},
  author={Madaan, Aman and Tandon, Niket and Gupta, Prakhar and Hallinan, Skyler and Gao, Luyu and Wiegreffe, Sarah and Alon, Uri and Dziri, Nouha and Prabhumoye, Shrimai and Yang, Yiming and others},
  journal={Advances in neural information processing systems},
  volume={36},
  pages={46534--46594},
  year={2023}
}

@article{wang2022self,
  title={Self-consistency improves chain of thought reasoning in language models},
  author={Wang, Xuezhi and Wei, Jason and Schuurmans, Dale and Le, Quoc and Chi, Ed and Narang, Sharan and Chowdhery, Aakanksha and Zhou, Denny},
  journal={arXiv preprint arXiv:2203.11171},
  year={2022}
}

@article{du2023improving,
  title={Improving factuality and reasoning in language models through multiagent debate},
  author={Du, Yilun and Li, Shuang and Torralba, Antonio and Tenenbaum, Joshua B and Mordatch, Igor},
  journal={arXiv preprint arXiv:2305.14325},
  year={2023}
}

@article{hendrycks2020measuring,
  title={Measuring massive multitask language understanding},
  author={Hendrycks, Dan and Burns, Collin and Basart, Steven and Zou, Andy and Mazeika, Mantas and Song, Dawn and Steinhardt, Jacob},
  journal={arXiv preprint arXiv:2009.03300},
  year={2020}
}

@article{wang2024mmlu,
  title={{MMLU-Pro}: A more robust and challenging multi-task language understanding benchmark},
  author={Wang, Yubo and Ma, Xueguang and Zhang, Ge and Ni, Yuansheng and Chandra, Abhranil and Guo, Shiguang and Ren, Weiming and Arulraj, Aaran and He, Xuan and Jiang, Ziyan and others},
  journal={Advances in Neural Information Processing Systems},
  volume={37},
  pages={95266--95290},
  year={2024}
}

@inproceedings{suzgun2023challenging,
  title={Challenging {BIG-Bench} tasks and whether chain-of-thought can solve them},
  author={Suzgun, Mirac and Scales, Nathan and Sch{\"a}rli, Nathanael and Gehrmann, Sebastian and Tay, Yi and Chung, Hyung Won and Chowdhery, Aakanksha and Le, Quoc and Chi, Ed H and Zhou, Denny and others},
  booktitle={Findings of the Association for Computational Linguistics: ACL 2023},
  pages={13003--13051},
  year={2023}
}

@article{clark2018think,
  title={Think you have solved question answering? try {ARC}, the {AI2} reasoning challenge},
  author={Clark, Peter and Cowhey, Isaac and Etzioni, Oren and Khot, Tushar and Sabharwal, Ashish and Schoenick, Carissa and Tafjord, Oyvind},
  journal={arXiv preprint arXiv:1803.05457},
  year={2018}
}

@article{geva2021did,
  title={Did aristotle use a laptop? a question answering benchmark with implicit reasoning strategies},
  author={Geva, Mor and Khashabi, Daniel and Segal, Elad and Khot, Tushar and Roth, Dan and Berant, Jonathan},
  journal={Transactions of the Association for Computational Linguistics},
  volume={9},
  pages={346--361},
  year={2021},
  publisher={MIT Press}
}

@article{cobbe2021training,
  title={Training verifiers to solve math word problems},
  author={Cobbe, Karl and Kosaraju, Vineet and Bavarian, Mohammad and Chen, Mark and Jun, Heewoo and Kaiser, Lukasz and Plappert, Matthias and Tworek, Jerry and Hilton, Jacob and Nakano, Reiichiro and others},
  journal={arXiv preprint arXiv:2110.14168},
  year={2021}
}

@article{hendrycks2021measuring,
  title={Measuring mathematical problem solving with the {MATH} dataset},
  author={Hendrycks, Dan and Burns, Collin and Kadavath, Saurav and Arora, Akul and Basart, Steven and Tang, Eric and Song, Dawn and Steinhardt, Jacob},
  journal={arXiv preprint arXiv:2103.03874},
  year={2021}
}

@inproceedings{yang2018hotpotqa,
  title={{HotpotQA}: A dataset for diverse, explainable multi-hop question answering},
  author={Yang, Zhilin and Qi, Peng and Zhang, Saizheng and Bengio, Yoshua and Cohen, William and Salakhutdinov, Ruslan and Manning, Christopher D},
  booktitle={Proceedings of the 2018 conference on empirical methods in natural language processing},
  pages={2369--2380},
  year={2018}
}

@article{trivedi2022musique,
  title={{MuSiQue}: Multihop questions via single-hop question composition},
  author={Trivedi, Harsh and Balasubramanian, Niranjan and Khot, Tushar and Sabharwal, Ashish},
  journal={Transactions of the Association for Computational Linguistics},
  volume={10},
  pages={539--554},
  year={2022},
  publisher={MIT Press}
}

@inproceedings{dua2019drop,
  title={{DROP}: A reading comprehension benchmark requiring discrete reasoning over paragraphs},
  author={Dua, Dheeru and Wang, Yizhong and Dasigi, Pradeep and Stanovsky, Gabriel and Singh, Sameer and Gardner, Matt},
  booktitle={Proceedings of the 2019 Conference of the North American Chapter of the Association for Computational Linguistics: Human Language Technologies, Volume 1 (Long and Short Papers)},
  pages={2368--2378},
  year={2019}
}

@article{chen2021evaluating,
  title={Evaluating large language models trained on code},
  author={Chen, Mark and Tworek, Jerry and Jun, Heewoo and Yuan, Qiming and Pinto, Henrique Ponde De Oliveira and Kaplan, Jared and Edwards, Harri and Burda, Yuri and Joseph, Nicholas and Brockman, Greg and others},
  journal={arXiv preprint arXiv:2107.03374},
  year={2021}
}

@article{austin2021program,
  title={Program synthesis with large language models},
  author={Austin, Jacob and Odena, Augustus and Nye, Maxwell and Bosma, Maarten and Michalewski, Henryk and Dohan, David and Jiang, Ellen and Cai, Carrie and Terry, Michael and Le, Quoc and others},
  journal={arXiv preprint arXiv:2108.07732},
  year={2021}
}

@article{jin2021disease,
  title={What disease does this patient have? a large-scale open domain question answering dataset from medical exams},
  author={Jin, Di and Pan, Eileen and Oufattole, Nassim and Weng, Wei-Hung and Fang, Hanyi and Szolovits, Peter},
  journal={Applied Sciences},
  volume={11},
  number={14},
  pages={6421},
  year={2021},
  publisher={MDPI}
}

@inproceedings{jin2019pubmedqa,
  title={{PubMedQA}: A dataset for biomedical research question answering},
  author={Jin, Qiao and Dhingra, Bhuwan and Liu, Zhengping and Cohen, William and Lu, Xinghua},
  booktitle={Proceedings of the 2019 conference on empirical methods in natural language processing and the 9th international joint conference on natural language processing (EMNLP-IJCNLP)},
  pages={2567--2577},
  year={2019}
}

@inproceedings{chen2021finqa,
  title={{FinQA}: A dataset of numerical reasoning over financial data},
  author={Chen, Zhiyu and Chen, Wenhu and Smiley, Charese and Shah, Sameena and Borova, Iana and Langdon, Dylan and Moussa, Reema and Beane, Matt and Huang, Ting-Hao and Routledge, Bryan R and others},
  booktitle={Proceedings of the 2021 Conference on Empirical Methods in Natural Language Processing},
  pages={3697--3711},
  year={2021}
}

@article{islam2023financebench,
  title={{FinanceBench}: A new benchmark for financial question answering},
  author={Islam, Pranab and Kannappan, Anand and Kiela, Douwe and Qian, Rebecca and Scherrer, Nino and Vidgen, Bertie},
  journal={arXiv preprint arXiv:2311.11944},
  year={2023}
}

@inproceedings{welbl2017crowdsourcing,
  title={Crowdsourcing multiple choice science questions},
  author={Welbl, Johannes and Liu, Nelson F and Gardner, Matt},
  booktitle={Proceedings of the 3rd Workshop on Noisy User-generated Text},
  pages={94--106},
  year={2017}
}

@article{wang2023scibench,
  title={{SciBench}: Evaluating college-level scientific problem-solving abilities of large language models},
  author={Wang, Xiaoxuan and Hu, Ziniu and Lu, Pan and Zhu, Yanqiao and Zhang, Jieyu and Subramaniam, Satyen and Loomba, Arjun R and Zhang, Shichang and Sun, Yizhou and Wang, Wei},
  journal={arXiv preprint arXiv:2307.10635},
  year={2023}
}

@inproceedings{zheng2021does,
  title={When does pretraining help? assessing self-supervised learning for law and the {CaseHOLD} dataset of 53,000+ legal holdings},
  author={Zheng, Lucia and Guha, Neel and Anderson, Brandon R and Henderson, Peter and Ho, Daniel E},
  booktitle={Proceedings of the eighteenth international conference on artificial intelligence and law},
  pages={159--168},
  year={2021}
}
